%% file: main.tex
\documentclass[twocolumn]{aastex631}
\usepackage[T1]{fontenc}
\usepackage{newtxtext, newtxmath}
\usepackage{natbib}
\usepackage{graphicx}
\usepackage{amsmath}
\usepackage{amssymb}
\usepackage{tabularx}
\usepackage{hyperref}

\newcommand{\vice}{\textsc{vice}}
\newcommand{\grad}[1]{\ensuremath{\nabla\text{[#1/H]}}}
\newcommand{\msun}{\ensuremath{\text{M}_\odot}}

\newcommand{\flow}{\ensuremath{\text{flow}}}
\newcommand{\eq}{\ensuremath{\text{eq}}}
\newcommand{\sfh}{\ensuremath{\text{sfh}}}
\newcommand{\ddfrac}[2]{\ensuremath{%
    \frac{\displaystyle{#1}}{\displaystyle{#2}}
}}
\newcommand{\partderiv}[2]{\ensuremath{\ddfrac{\partial #1}{\partial #2}}}
\newcommand{\taylorexpand}[2]{\ensuremath{%
    \sum_{i = 1}^\infty \frac{\partial^i #1}{\partial #2^i} \delta #2^i
}}

\newcommand{\ycc}[1]{\ensuremath{
    y_\text{#1}^\text{CC}
}}
\newcommand{\yia}[1]{\ensuremath{
    y_\text{#1}^\text{Ia}
}}
\newcommand{\hi}{H \textsc{i}}
\newcommand{\hii}{H \textsc{ii}}

\defcitealias{Johnson2021}{J21}
\defcitealias{Johnson2025}{J25}

\newcommand{\carnegieaffil}{%
Carnegie Science Observatories, %
813 Santa Barbara Street, Pasadena, CA 91101, USA}

\begin{document}

\title{Constraints on Radial Gas Flows in the Milky Way Disk Revealed by Large Stellar Age Catalogs}
\shorttitle{Radial Gas Flows in the Galactic Disk}

\author[0000-0002-6534-8783]{James W. Johnson}
\affiliation{\carnegieaffil}

\shortauthors{J.W. Johnson}

\input{abstract.tex}

\input{intro.tex}

\input{data.tex}

\input{gce.tex}

\input{results.tex}

\input{discussion.tex}

\input{conclusions.tex}

\input{acknowledgments.tex}

\newpage
\appendix
\input{analytic.tex}

\bibliographystyle{aasjournal}
\bibliography{main}

\end{document}

%% file: abstract.tex
\begin{abstract}
\noindent
Disk galaxies like the Milky Way are expected to experience gas flows carrying matter toward their centers.
This paper investigates the role of these radial gas flows in models of Galactic chemical evolution (GCE).
We follow five different parameterizations of the Galactocentric radial velocity, $v_{r,g}$, of the interstellar medium (ISM).
Relative to the $v_{r,g}=0$ limit, all models predict stellar metallicity to decline less significantly with age in the outer disk and more significantly in the inner disk.
This outcome arises because radial flows cannot remove gas from one region of the Galaxy without placing it elsewhere, leading to opposing effects on enrichment timescales between the inner and outer Galaxy.
This prediction is at odds with recent observational constraints, which indicate remarkably minimal decline in metallicity ($\lesssim$$0.1$ dex) between young ($\sim$$0-2$ Gyr) and old populations ($\sim$$8-10$ Gyr) across the \textit{entire} Galactic disk.
Radial gas flows cannot be the sole explanation of this result at all Galactocentric radii.
Our models reproduce this result at $R\gtrsim6$ kpc if the flow velocity is relatively constant in both radius and time near $v_{r,g}\approx-1$ km/s.
In agreement with previous GCE models, all of our flow prescriptions lead to lower metallicities and steeper radial gradients relative to static models.
Exploiting this universal outcome, we identify mixing effects and the relative rates of star formation and metal-poor accretion as the processes that establish the ISM metallicity at low redshift.
We provide a suite of analytic formulae describing radial metallicity gradient evolution based on this connection.
\end{abstract}

%% file: intro.tex
\section{Introduction}
\label{sec:intro}

Gas moves.
Galaxy formation models predict that the interstellar medium (ISM) should sink toward the Galactic center with speeds of order $\sim$$0.1 - 1$ km/s \citep[e.g.,][]{Vincenzo2020, Barbani2025}.
Galactic chemical evolution (GCE) models generally predict that inward flows at these speeds should significantly impact the timescales of metallicity growth \citep[e.g.,][]{Portinari2000, Spitoni2011, Bilitewski2012}.
Stellar ages provide direct constraints on these timescales.
Age measurements for single stars are challenging (see, e.g., the reviews by \citealt{Soderblom2010} and \citealt{Chaplin2013}), but recent years have seen advancements.
Sample sizes and parameter calibrations have improved in asteroseismology \citep{Boulet2024, Schonhut-Stasik2024, Warfield2024, Pinsonneault2025}, carbon-to-nitrogen ratios \citep{Martig2016, Ho2017, Roberts2025}, gyrochronology \citep{Reinhold2015, Bouma2023, Lu2024a}, and isochrones \citep{Feuillet2016, Sanders2018, Nataf2024}.
Application of machine learning algorithms have also enabled the construction of large catalogs \citep{Anders2023, Leung2023, Stone-Martinez2024, Stone-Martinez2025, Van-Lane2025}.
In this paper, we leverage precise age and metallicity measurements for red giants across the thin disk to constrain radial gas flows in the Milky Way (MW).

\par
GCE models incorporating radial gas flows typically invoke one of a handful of assumptions about the processes that set radial velocities in the ISM.
\citet{Lacey1985} discussed three physically motivated arguments.
In their first scenario, accreting material has a lower angular momentum than the rotation of the disk, inducing an inward flow through phase mixing (see GCE models by \citealt{Mayor1981, Pitts1989, Pitts1996, Chamcham1994, Bilitewski2012, Pezzulli2016}; see also simulation results from \citealt{Trapp2022, Barbani2025}).
Second, the viscosity of the ISM induces inward flows in the inner disk and outward flows in the outer disk, with typical speeds of order $\sim$$0.1$ km/s (see GCE models by \citealt{Clarke1989, Sommer-Larsen1990, Thon1998}).
Third, azimuthal asymmetries such as spiral density waves and central bars induce global torques, driving flows with velocities of order $\sim$$0.1 - 1$ km/s in the ISM.
This scenario arises much more naturally in hydrodynamic simulations than GCE models \citep[e.g.,][]{Grand2016, Orr2023}.
The velocities associated with each of these scenarios are well within observational constraints (see discussion below).

\begin{figure}
\centering
\includegraphics[scale = 1]{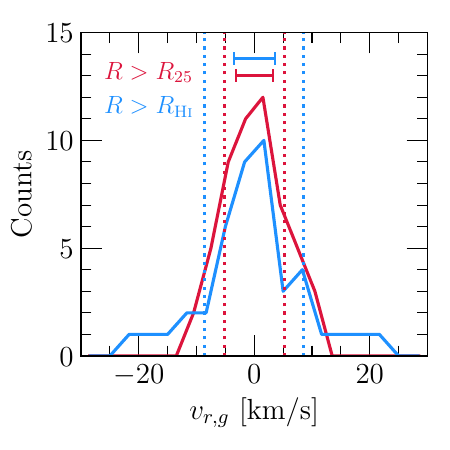}
\caption{
Distribution of observed disk radial velocities, averaged across radius in individual external galaxies, measured using doppler shifts in the \hi\ 21 cm line by \citet{DiTeodoro2021}.
They define $R_{25}$ as the radius corresponding to the 25 mag/arcsec$^2$ isophote in the $z$ band and $R_\text{H \textsc{i}}$ as the radius at which the projected \hi\ surface density drops below 1 \msun/pc$^2$.
\textbf{Summary}: Observed Galactocentric radial velocities are highly variable both between and within galaxies, which simulations suggest is a consequence of variations on short timescales ($\ll 1$ Gyr; \citealt{Trapp2022}).
}
\label{fig:empirical}
\end{figure}

Radial motion in the ISM is difficult to constrain empirically through direct observations of galaxy velocity fields.
Previous observational estimates using CO and \hi\ in nearby spiral galaxies can only rule out the $\left| v_{r,g} \right| \gtrsim 5$ km/s regime \citep{Wong2004}.
Figure \ref{fig:empirical} shows the distribution of galaxy-wide average radial velocities in the sample from \citet[][taken from table 1 therein]{DiTeodoro2021}.
They measured radial motion in the ISM in 54 low-redshift spiral galaxies using doppler shifts of the \hi\ 21 cm line.
Both inward and outward flows are observed with speeds as large as $\sim$$20$ km/s.
Some of these measurements may reflect long-lived flows, but uncertainties in a galaxy's inclination can also lead to spuriously high radial velocities (see discussion in \citealt{Schmidt2016}).
\citet{DiTeodoro2021} also show that $v_{r,g}$ varies substantially both within and between galaxies.
Simulations suggest that this result is a consequence of high-amplitude variability on short timescales ($\lesssim$$100$ Myr), while time-averaged radial velocities are of order $\sim$$1 - 3$ km/s \citep[e.g.,][]{Vincenzo2020, Trapp2022}.
This characteristic speed is within the width of the distribution of observed velocities in Figure \ref{fig:empirical}.
This $\left|v_{r,g}\right| \lesssim -5$ km/s limit is the only clear empirical constraint on long-lived radial gas flows from these direct measurements.
\par
In this paper, we draw on empirical constraints enabled by recent advancements in stellar age measurements.
In particular, old stellar populations are not substantially more metal-poor than young populations.
Stars as old $\sim$$8 - 10$ Gyr are $\lesssim$$0.1$ dex more metal-poor than young stars ($\sim$$0 - 2$ Gyr) at fixed Galactocentric radius \citep[][hereafter \citetalias{Johnson2025}]{Johnson2025}.
This result has been found using color-magnitude diagrams \citep[e.g.,][]{Gallart2024}, isochrone ages for subgiants \citep[e.g.,][]{Xiang2022}, in red giants with asteroseismology \citep[e.g.,][]{Willett2023}, [C/N] ratios \citep[e.g.,][]{Roberts2025}, open clusters \citep[e.g.,][]{Spina2022}, and classical Cepheid variables \citep[e.g.,][]{daSilva2023}.
Significant trends in stellar metallicity with age are a common outcome in GCE models due to ongoing star formation (see discussion in, e.g., the reviews by \citealt{Tinsley1980} and \citealt{Matteucci2021}), but this relatively standard prediction is at odds with these observations.
\par
In \citetalias{Johnson2025}, we argued that this flat age-metallicity relation is indicative of a chemical equilibrium in the MW.
The defining feature of this scenario is that ISM metallicities cease to evolve with time at some point early in the thin disk epoch.
In the equilibrium state, metal production by stars is offset by losses to star formation and ejection\footnote{
    We generally refer to the processes exchanging matter between the MW disk and the CGM as ``accretion'' and ``ejection'' in this paper.
    In previous works \citepalias[e.g.,][]{Johnson2025}, we referred to these processes as ``inflows'' and ``outflows.''
    We adjust our terminology here in order to avoid confusion between these processes and inward or outward radial flows of gas within the disk.
} as well as dilution by metal-poor accretion.
In \citetalias{Johnson2025}, we demonstrated that Galactic winds ejecting ISM gas to the circumgalactic medium (CGM) could explain the observed age-metallicity trend (or lack thereof) if gas is more readily ejected from the outer disk than the inner disk.
Here, we show that radial gas flows also influence chemical equilibria.
The lack of an observed trend in metallicity with stellar age also places limits on sudden events of substantial metal-poor gas accretion, such as the proposed two-infall scenario \citep[e.g.,][]{Chiappini1997, Spitoni2019, Spitoni2020}.
Re-enrichment from the ensuing burst of star formation is rapid ($\lesssim$$1$ Gyr; \citealt{Dalcanton2007, Johnson2020}), which generically leads to trends in metal abundance with stellar age, in tension with recent observations \citep{Dubay2025}.
\par
Inspired by previous work on chemical equilibrium \citep{Larson1972, Weinberg2017}, we also provide a suite of analytic expressions describing the role of radial gas flows in GCE models (see Appendix \ref{sec:analytic}).
Using arguments rooted in chemical equilibration, our framework connects a prescription for the radial velocity profile in the ISM with the low-redshift metallicity and its radial gradient.
These approximations accurately describe the predictions of our numerical GCE models, to which we dedicate most of the discussion in this paper.
The first and foremost difference between our analytic solutions and those by \citet{Sharda2021} is that we do not account for diffusion.
We operate in the limit of instantaneous mixing, allowing for apples-to-apples to comparisons with multi-zone GCE models (see discussion in Section \ref{sec:gce}).
\citet{Sharda2021} also introduced a specific model describing the radial velocities in the ISM, whereas our framework predicts the metallicity gradient based on a radial velocity prescription specified as input.
\par
This paper is organized as follows.
We describe our sample in Section \ref{sec:data} and our GCE models in Section \ref{sec:gce} below.
In Section \ref{sec:results}, we describe their predicted evolutionary histories, comparing them with one another and with broad empirical constraints.
We discuss our results in the context of our analytic solutions in Section \ref{sec:discussion}, with detailed derivations reserved for Appendix \ref{sec:analytic}. 
We summarize our conclusions in Section \ref{sec:conclusions}.

%% file: data.tex
\section{Data}
\label{sec:data}

We use the sample of red clump and red giant stars from \citetalias{Johnson2025}, which are taken from the \textsc{AstroNN} value added catalog\footnote{
\url{https://www.sdss.org/dr18/data_access/value-added-catalogs/?vac_id=85}
} for the seventeenth data release (DR17) of the Apache Point Observatory Galaxy Evolution Experiment \citep[APOGEE;][]{Majewski2017, Abdurrouf2022}.
This catalog provides stellar abundances \citep{Leung2019a} and distances \citep{Leung2019b} using \textit{Gaia}-eDR3 \citep{GaiaCollaboration2021}.
Age estimates in \textsc{AstroNN} are derived from a Bayesian convolutional neural network trained on the APOGEE spectra and asteroseismic ages from APOKASC-2 \citep{Pinsonneault2018}.
We restrict our sample to stars with signal-to-noise ratios of S/N $\geq 80$, surface gravities of $\log g = 1 - 3.8$, and effective temperatures of $T_\text{eff} = 3400 - 5500$ K.
We make additional cuts on age and spatial location to isolate thin disk stellar populations, namely a maximum age of $\tau \leq 10$ Gyr, Galactocentric radius of $R \leq$ 15 kpc, and present-day mid-plane distance of $\left|z\right| \leq 0.5$ kpc.
These criteria yield a sample of $N = 94,387$ red clump and red giant stars, most of which are found in the $R = 5 - 12$ kpc range.
We refer to the discussion in \citetalias{Johnson2025} for further details.
\par
This paper draws primarily qualitative conclusions regarding radial gas flows and their influence on chemical evolution in light of recent age-metallicity trends.
We therefore refrain from fitting our models to this sample quantitatively and instead focus on illustrative comparisons.




%% file: gce.tex
\section{Galactic Chemical Evolution Models}
\label{sec:gce}

\begin{figure}
\centering
\includegraphics[scale = 0.4]{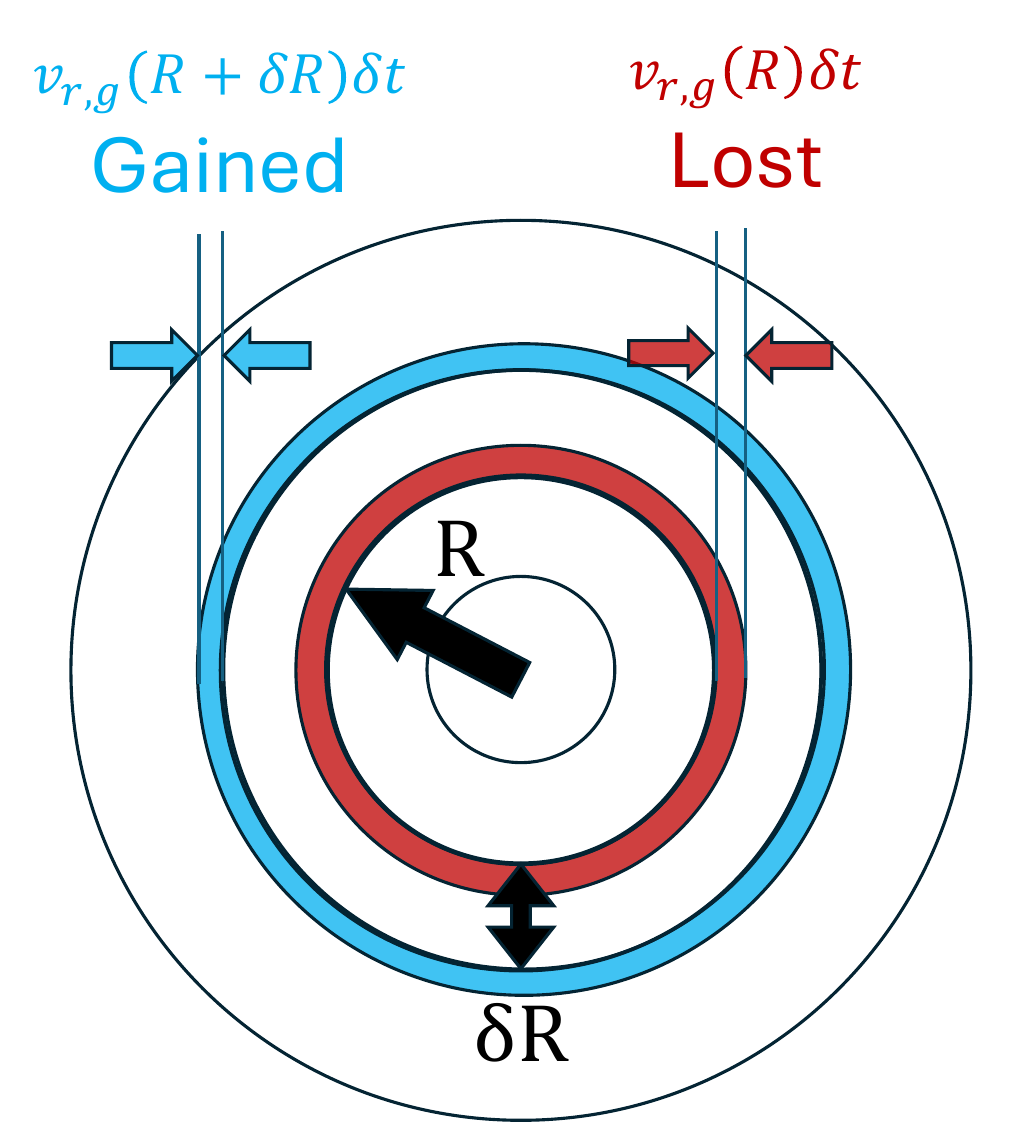}
\caption{
A schematic of the implementation of radial gas flows in our models.
For a given inward flow velocity, $v_{r,g} < 0$, and timestsep size, $\delta t$, a given annulus, $R \rightarrow R + \delta R$, loses the gas within $R \rightarrow R - v_{r,g} \delta t$ to its ``inner neighbor'' at $R - \delta R \rightarrow R$.
Simultaneously, this annulus gains from its ``outer neighbor'' following a similar prescription, though potentially with a different velocity.
The fraction of the mass in each ananulus that is transferred is given by the projected area of the red ring relative to the $R \rightarrow R +\delta R$ annulus.
Our analytic solutions apply the limit as $\delta R, \delta t \rightarrow 0$ to this prescription.
}
\label{fig:schematic}
\end{figure}

We use multi-zone GCE models of the MW disk adapted from \citetalias{Johnson2025}, which were originally developed in \citet[][hereafter \citetalias{Johnson2021}]{Johnson2021}.
We integrate each model numerically using the publicly available \textsc{Versatile Integrator for Chemical Evolution} (\vice; \citealt{Johnson2020}).
Following prior work with similar motivations \citep[e.g.,][]{Matteucci1989, Schoenrich2009, Minchev2013, Minchev2014}, these models discretize the Galactic disk into concentric $\delta R = 100$ pc rings in order to model enrichment in different regions of the MW.
Each ring is coupled to its closest neighbors through the radial migration of stars \citep[e.g.,][]{Sellwood2002, Roskar2008a, Roskar2008b, Loebman2011, Minchev2011, Bird2012, Bird2013, Grand2012, Kubryk2013}.
In this paper, radial gas flows introduce a second source of evolutionary coupling between neighboring annuli \citep[e.g.,][]{Lacey1985, Portinari2000, Spitoni2011, Bilitewski2012}.
With the exception of these processes, the evolution in a given ring is otherwise described by conventional one-zone GCE models (see the review by, e.g., \citealt{Matteucci2021}).
Table \ref{tab:glossary} provides a summary of all of our GCE parameters.
\par
These models compute ISM abundances in each ring at $\delta t = 5$ Myr timesteps forming over a disk lifetime of $\tau_\text{disk} = 13.2$ Gyr, resulting in abundances for $N = 1,056,000$ single stellar populations that form between $R = 0$ and $20$ kpc.
The inclusion of radial gas flows and exclusion of mass-loaded ejection from the Galactic disk is the most important change that we have made from previous versions of these models.
In \citetalias{Johnson2021}, we assumed that the mass loading factor, which describes the rate of ejection relative to star formation according to $\eta \equiv \dot\Sigma_\text{wind} / \dot\Sigma_\star$, increases quasi-exponentially with Galactocentric radius.
This prescription causes the equilibrium metallicity, $Z_\eq$ to decline with radius in a manner that tracks the observed radial metallicity profile.
In \citetalias{Johnson2025}, we demonstrated that this prescription leads to significantly better agreement with recent measurements of the age-metallicity relation in stars relative to $\eta = 0$ models (see discussion in Section \ref{sec:intro}).
In this paper, we are interested in which radial gas flow scenarios, if any, are also able to reproduce this flat age-metallicity trend.
\par
Figure \ref{fig:schematic} provides a qualitative illustration of the numerical implementation of radial gas flows in our models.
Following previous work \citep[e.g.,][]{Portinari2000, Spitoni2011}, we simply move a fraction of each annulus's gas reservoir to one of its neighbors based on the distance the ISM would travel in one timestep for a given velocity, $v_{r,g}$.
In the case of an inward flow (i.e., $v_{r,g} < 0$), which we focus on for most of this paper, each ring loses the gas between its inner boundary $R$ and $R - v_{r,g} \delta t$.
The amount of gas that moves between rings at each timestep is given by the area of the $R \rightarrow R - v_{r,g} \delta t$ annulus relative to the $R \rightarrow R + \delta R$ annulus.
The relative areas for both inward and outward flows are given by
\begin{subequations}
\begin{align}
a_\text{in}(v_{r,g} < 0, R) &= \frac{
    \left(R - v_{r,g} \delta t\right)^2 - R^2
}{
    \left(R + \delta R\right)^2 - R^2
}
\label{eq:areafrac-inward}%
\\
a_\text{out}(v_{r,g} > 0, R) &= \frac{
    (R + \delta R)^2 - (R + \delta R - v_{r,g} \delta t)^2
}{
    (R + \delta R)^2 - R^2
},
\label{eq:areafrac-outward}%
\end{align}
\label{eq:areafracs}%
\end{subequations}
where $a_\text{in}$ is appropriate for inward flows, and $a_\text{out}$ is appropriate for outward flows.
We clarify that these expressions only account for the mass that the $R \rightarrow R + \delta R$ ring loses to either its inner or outer neighbor, depending on the direction of the flow.
Simultaneously, each annulus gains gas from its other neighbor under a similar prescription.
The exchange of metals between neighboring annuli follows the same prescription.

\input{glossary.tablebody.tex}

\subsection{The Gas Supply}
\label{sec:gce:gas}

The rate of change in the surface density of the ISM gas, $\Sigma_g$, can be expressed as a sum of ``source'' and ``sink'' terms, according to
\begin{equation}
\begin{split}
\dot\Sigma_g &= \dot\Sigma_\text{acc} -
\dot\Sigma_\star -
\dot\Sigma_\text{wind} +
\dot\Sigma_{g,\flow} +
\dot\Sigma_r
\\
&= \dot\Sigma_\text{acc} -
\dot\Sigma_\star \left(1 + \eta - \mu_g - r\right),
\label{eq:dot-sigma-gas}
\end{split}
\end{equation}
where $\dot\Sigma_\text{acc}$ is the component for accretion from the CGM, $\dot\Sigma_\star$ is the SFR, $\dot\Sigma_\text{wind}$ is the ejection rate due to Galactic winds, and $\dot\Sigma_\text{r}$ accounts for the return of stellar envelopes back to the ISM.
In the second line of this equality, we make a handful of substitutions following \citet{Weinberg2017}.
In detail, the rate of return of stellar envelopes back to the ISM varies with stellar population age.
We take this subtle evolution in the recycling factor, $r$, into account in our numerical models (see discussion in \citealt{Johnson2020}).
However, for the purposes of analytic solutions, it is sufficiently accurate to approximate envelope return as instantaneous (i.e., $\dot\Sigma_r \approx r\dot\Sigma_\star$, with $r \approx +0.4$ for a \citealt{Kroupa2001} initial mass function, which we use throughout this paper; see discussion in \citealt{Weinberg2017}).
\par
In Equation \ref{eq:dot-sigma-gas} above, we have also substituted in the ``flow coefficient,'' $\mu$.
This coefficient, $\mu_g$ for gas and $\mu_x$ for some chemical element $x$, is defined as the rate of change in the ISM surface density due to the flow relative to the local SFR and metal abundance $Z_x$, according to
\begin{subequations}
\begin{align}
\dot\Sigma_{g,\flow} &\equiv \mu_g \dot\Sigma_\star
\label{eq:mu-g-definition}%
\\
\dot\Sigma_{x,\flow} &\equiv Z_x \mu_x \dot\Sigma_\star.
\label{eq:mu-x-definition}%
\end{align}
\label{eq:mu-definition}%
\end{subequations}
In our numerical models, we simply exchange matter between individual rings by computing the area fractions according to Equation \ref{eq:areafracs} above.
This approach eliminates the requirement to compute $\mu$ for the purposes of integrating our models and instead lets their values arise ``naturally'' through the mass exchange.
However, in Appendix \ref{sec:analytic}, we demonstrate that both $\mu_g$ and $\mu_x$ have well-defined forms for any given radial gas flow prescription in the limit of a thin, axisymmetric disk with efficient mixing.
We discuss analytic solutions to these coefficients in Section \ref{sec:discussion}.
Importantly, both forms of $\mu$ can be compared directly to the mass loading factor, enabling an apples-to-apples comparison with prior versions of these models using ejection from the ISM.
\par
We use the same prescription for star formation efficiency (SFE) as \citetalias{Johnson2025}, which is a classical, single power-law Kennicutt-Schmidt relation, $\dot\Sigma_\star \propto \Sigma_g^N$ with $N = 1.5$ \citep[e.g.,][]{Schmidt1959, Schmidt1963, Kennicutt1998}.
In practice, \vice\ requires user input for the SFE timescale, $\tau_\star \equiv \Sigma_g / \dot \Sigma_\star \propto \Sigma_g^{1 - N}$ (referred to as the ``depletion time'' by some authors).
Following \citetalias{Johnson2021} and \citetalias{Johnson2025}, we postulate that variations in $\tau_\star$ with $\Sigma_g$ reflect shifts in the ISM hydrogen between the atomic and molecular states.
We therefore build in a cutoff to a linear Kennicutt-Schmidt relation (i.e., constant $\tau_\star$ with $N = 1$) at surface densities above $\Sigma_g > 10^8$ \msun\ kpc$^{-2}$.
Above this threshold, we assume the ISM to be dominated by molecular hydrogen, so an increase in surface density does not lead to more efficient star formation.
We use $\tau_\star = 2$ Gyr at the present day as the SFE timescale for a fully molecular gas reservoir \citep{Leroy2008, Blanc2009}, with a $\sqrt{t}$ time-dependence based on variations in the observed $\Sigma_g - \dot\Sigma_\star$ relation with redshift \citep{Tacconi2018}.
\par
Emiprically, molecular gas fractions in the ISM are observed to reach unity near $\Sigma_g \approx 10^6$ \msun\ kpc$^{-2}$ \citep{Bigiel2008, Blanc2009}, which is two orders of magnitude lower than our assumed threshold.
However, our models predict surface densities in the $\Sigma_g = 10^6 - 10^8$ \msun\ kpc$^{-2}$ range across most of the Galactic disk, in broad agreement with observations \citep[e.g.,][]{Kalberla2009}.
The surface density exceeds $\Sigma_g = 10^8$ \msun\ kpc$^{-2}$ only the innermost regions ($R \lesssim 500$ pc).
Adopting the lower cutoff of $\Sigma_g > 10^6$ \msun\ kpc$^{-2}$ places much of the ISM in the molecular phase.
By using this higher threshold, we incorporate variations in SFE with radius, which affects the radial metallicity gradient \citep[e.g.,][]{Palla2020}.
We also explore a variation of our models using this lower cutoff at $\Sigma_g > 10^6$ \msun\ kpc$^{-2}$ in Section \ref{sec:discussion:eq-gradients}, which can be understood as switching to a power-law index of $N = 1$.
\par
We also retain the prescription for the radial migration of stars from \citetalias{Johnson2025}, which in turn is adopted from \citet{Dubay2024}.
For each stellar population, we sample a radial displacement from a normal distribution centered on $\Delta R = 0$ using a width that increase with stellar population age.
This prescription reproduces the distributions of birth and final radius predicted by the \texttt{h277} hydrodynamic simulation \citep[e.g.,][]{Christensen2012, Christensen2016, Governato2012, Zolotov2012, Munshi2013, Brooks2014, Brooks2017, Bird2021}, which we used to drive radial migration in \citetalias{Johnson2021}.
The two prescriptions therefore lead to similar migration patterns, so there are no obvious differences in the effects of radial migration between the models in \citetalias{Johnson2021} and those in either this paper or \citetalias{Johnson2025}.
\par
For simplicity, we focus on models with metal-free accretion throughout this paper.
We relax this assumption in Appendix \ref{sec:analytic:zin}.
Pre-enriched accretion raises overall metal abundances slightly and makes the radial gradient somewhat shallower.
This effect can be quantitatively important for some purposes but is generally irrelevant to our conclusions in this paper.

\subsection{Stellar Yields and Metal Enrichment}
\label{sec:gce:yields}


\par
We integrate our numerical models with O and Fe as representative alpha and iron-peak elements.
O is produced primarily by massive stars \citep[e.g.,][]{Johnson2019, Kobayashi2020}, which have lifetimes much shorter than the age of the Galactic disk \citep[e.g.,][]{Larson1974, Maeder1989, Padovani1993, Kodama1997, Hurley2000}.
Following previous versions of these models, we approximate massive star yields as being released to the ISM instantaneously after a stellar population forms.
In prior GCE models of the MW, this assumption is often combined with the immediate return of stellar envelopes and referred to as ``instantaneous recycling'' \citep[see, e.g., the review by][]{Tinsley1980}.
The rate of change in the O abundance is therefore given by
\begin{equation}
\dot\Sigma_\text{O} = y_\text{O} \dot\Sigma_\star -
Z_\text{O} \dot\Sigma_\star
\left(1 + \eta - \mu_\text{O} - r\right),
\label{eq:dot-sigma-o}
\end{equation}
where $y_\text{O}$ is the stellar yield of O.
The first term in this expression accounts for massive star yields.
The second accounts for the gas processes influencing the ISM (e.g., star formation, ejection) and therefore has a similar appearance as the corresponding term in Equation \ref{eq:dot-sigma-gas}.
There is an additional factor of $Z_\text{O}$ because these processes add and remove mass at the current metallicity of the ISM.
\par
\par
The yield $y_\text{O}$ can be described as a population-averaged, fractional net production factor.
This quantity accounts for nucleosynthesis across the full range of the IMF and marginalizes over, e.g., binary fraction and rotation velocities.
The yield describes only the \textit{newly produced} metal mass and is expressed relative to a single stellar population's initial mass.
For example, a value of $y_x = 0.001$ would imply that a hypothetical $1000$ \msun\ stellar population would produce $1$ \msun\ of some element $x$.
\par
In our numerical models, we set the value of $y_\text{O}$ based on empirical constraints on the scale of stellar yields.
\citet{Rodriguez2023} measured the mean Fe yield from individual Type II supernovae (SNe) using the radioactive tails of their lightcurves.
Based on their measurements, \citet{Weinberg2024} inferred that the total metal yield (i.e., including Type Ia SNe and other sources) of each element is roughly equivalent to its solar abundance.
We therefore adopt $y_\text{O} = Z_{\text{O},\odot}$, for which we use $Z_{\text{O},\odot} = 0.00572$ based on \citet{Asplund2009}.
We refrain from using yields predicted from massive star evolutionary models, since the predictions are affected by substantial uncertainties in, e.g., mass loss \citep{Sukhbold2016} and black hole formation \citep{Griffith2021}.
\par
Our Fe yields follow from the same considerations.
We retain our prescription from \citetalias{Johnson2025}, which attributes 35\% of the Fe in the Sun to massive stars and the remaining 65\% to SNe Ia.
This choice leads to $\ycc{Fe} = 4.51 \times 10^{-4}$ and $\yia{Fe} = 8.38 \times 10^{-4}$.
We retain the delay-time distribution of SN Ia events from prior versions of these models, which imposes a minimum delay of $\Delta t = 150$ Myr and scales with stellar population age as $\tau^{-1.1}$ \citep[e.g.,][]{Maoz2012}.
The choice of DTD impacts the predicted [O/Fe]-[Fe/H] distribution \citep{Dubay2024, Palicio2024}, but these effects are not relevant to this paper.
\par
Although we present Fe abundances predicted by \vice, we only follow O in our analytic models.
The delayed nature of SNe Ia produces a convolution of the DTD and the SFH, which complicates the solution to the abundance evolution.
\citet{Weinberg2017} present analytic solutions incorporating SNe Ia in one-zone models with ejection.
We demonstrate in Appendix \ref{sec:analytic:dtds} that a simple transformation should allow our solutions to account for delayed sources, such as SNe Ia.

\subsection{The Star Formation History}
\label{sec:gce:sfh}

\begin{figure*}
\centering
\includegraphics[scale = 0.75]{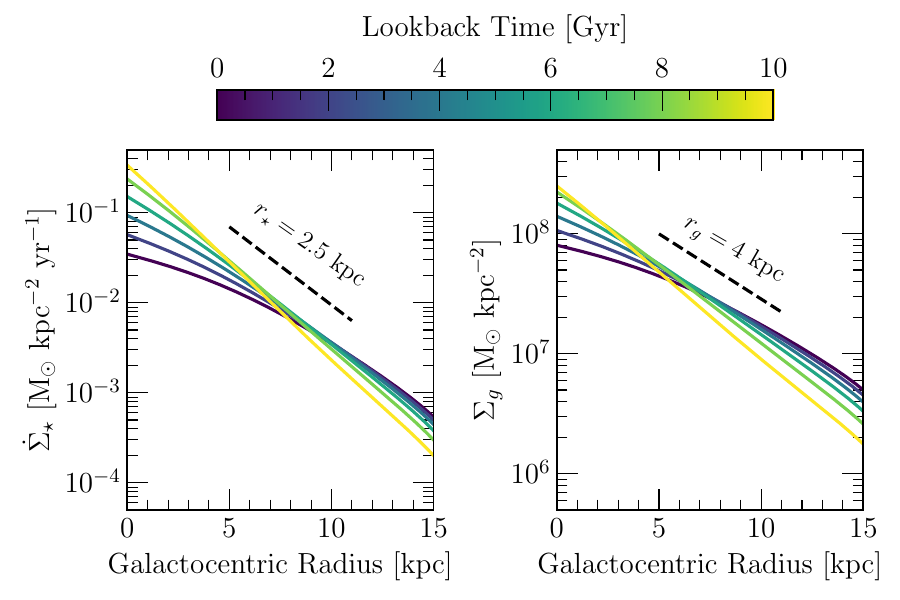}
\includegraphics[scale = 0.77]{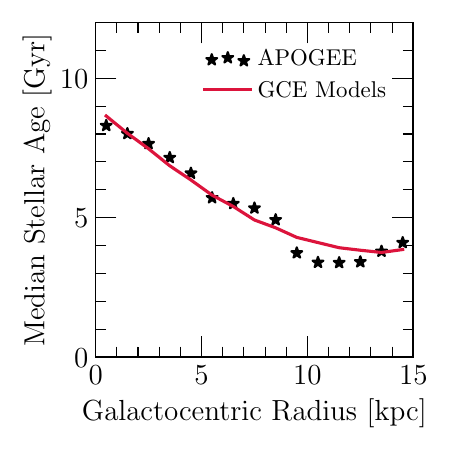}
\caption{
The evolutionary history imposed upon each of our GCE models.
Colored lines show the surface densities of star formation (left) and gas (middle) as functions of Galactocentric radius at lookback times of 0, 2, 4, 6, 8, and 10 Gyr, color coded according to the colorbar at the top.
Dashed black lines show exponential surface density profiles with arbitrary normalizations and scale lengths of $r_\star = 2.5$ kpc (left) and $r_g = 4$ kpc (middle), broadly consistent with the observed surface density gradients \citep[e.g.,][]{Kalberla2009, BlandHawthorn2016}.
The right panel shows the median stellar age predicted by our GCE models after radial migration is taken into account (solid red line) in comparison with our measurements from APOGEE in \citetalias{Johnson2025} (black stars).
\textbf{Summary}: Each model is constrained to form radial profiles in surface density and stellar age in broad agreement with observations.
}
\label{fig:evol}
\end{figure*}

All of our numerical models use the same SFH.
We retain the single-epoch ``rise-fall'' formalism from \citetalias{Johnson2021} and \citetalias{Johnson2025}, given by
\begin{equation}
\dot\Sigma_\star \propto (1 - e^{-t / \tau_\text{rise}})
e^{-t / \tau_\text{sfh}},
\label{eq:sfh}
\end{equation}
where $\tau_\text{rise}$ and $\tau_\text{sfh}$ loosely control the timescales on which the SFR rises at early times and falls late times.
We use smooth exponential scalings of $\tau_\text{rise}$ and $\tau_\text{sfh}$, given by:
\begin{subequations}
\begin{align}
\tau_\text{rise} &= \left(2\text{ Gyr}\right)
e^{R / r_\text{rise}}
\label{eq:taurise}%
\\
\tau_\text{sfh} &= \left(2\text{ Gyr}\right)
\left(1 + e^{R / r_\text{sfh}}\right),
\label{eq:tausfh}%
\end{align}
\label{eq:insideout}%
\end{subequations}
where we use $r_\text{rise} = 6.5$ kpc and $r_\text{sfh}$ = 4.7 kpc throughout this paper.
Tuned by eye, this prescription approximates the radial profile in median stellar age as observed (see discussion below).
The SFH is relatively extended (e.g., $\tau_\text{sfh} \approx 13$ Gyr at $R = 8$ kpc), in line with the original version of these models.
In \citetalias{Johnson2021}, we assumed $\tau_\text{rise} = 2$ Gyr everywhere and determined $\tau_\text{sfh}$ based on the radially resolved SFHs of face-on MW-like galaxies measured with integral field spectroscopy by \citet{Sanchez2020}.
Their data led to similarly large values of $\tau_\text{sfh}$.
\par
In each annulus, the normalization of the SFH is assigned such that the model predicts a total disk mass ($\sim$$(5 \pm 1) \times 10^{10}$ \msun; \citealt{Licquia2015}) and radial density gradient (scale radius $r_\star \sim 2.5$ kpc; \citealt{BlandHawthorn2016}) broadly consistent with observed constraints.
Previous versions of these models imposed a sharp cutoff in the rate of star formation at $R = 15.5$ kpc.
In \citetalias{Johnson2021}, this choice was based on the largest birth radius of all star particles with disk-like kinematics in the \texttt{h277} simulation (see discussion above).
With radial gas flows, this sharp discontinuity can travel inward, leading to numerical artifacts.
We therefore extend the maximum radius of star formation out to $R = 20$ kpc, the outermost edge of the models imposed by \vice's \texttt{milkyway} object \citepalias{Johnson2021}.
We also divide the normalization of the SFH in every annulus by a factor of $1 + e^{(R - 17 \text{ kpc}) / 1 \text{ kpc}}$.
This adjustment softens the cutoff in surface density in the outer Galaxy, preventing the numerical boundary from traveling inward.
\par
The left and middle panels of Figure \ref{fig:evol} show radial profiles in the surface densities of star formation and gas at a handful of representative snapshots.
The inside-out growth of the disk is clearly visible, with both quantities showing more centrally concentrated profiles long ago.
These two panels also show observational constraints for comparison.
In the MW, the stellar surface density follows a scale radius of $r_\star \approx 2.5$ kpc, and the \hi\ follows $r_g \approx 4$ kpc (see, e.g., the reviews by \citealt{BlandHawthorn2016} and \citealt{Kalberla2009}).
Radial migration does not significantly alter the stellar surface density profile in these models \citepalias{Johnson2021}.
Our models predict a stellar surface density gradient broadly consistent with observations primarily because the normalization of the input SFH itself follows a similar gradient.
Our models predict a surface density profile in the ISM that is somewhat shallower than observed, but this offset is inconsequential to our main conclusions.
\par
We specify the SFH as input to \vice\ as opposed to the accretion history because the former is much more directly constrained by observations than the latter.
For a fixed accretion history, different radial gas flow models lead to different surface density profiles due to the gas moving to new locations before it forms stars.
By fixing the SFH, the accretion history is instead constrained by the radial flow prescription, and the desired surface density profile arises regardless.
In the extragalactic context, different disks are most often compared when scaled for their sizes (see the reviews by, e.g., \citealt{Maiolino2019} and \citealt{Sanchez2020}).
In fact, this connection between where gas accretes and where it forms stars due to inward gas flows plays a central role in steepening metallicity gradients relative to static models (see discussion in Section \ref{sec:results}).
We clarify that, with our prescriptions for SFE and radial gas flows, the solution to the evolution of the gas supply, $\dot \Sigma_g$, is mathematically unique.
Therefore, similar predictions arise if we instead parameterize the accretion history and scale each of our models by disk size.
\par
The use of exponential scalings in $\tau_\text{rise}$ and $\tau_\text{sfh}$ is a change from previous versions of these models.
We make this adjustment because the use of radial gas flows requires GCE parameters that are smooth functions of radius not only themselves, but also in their first derivatives.
In \citetalias{Johnson2025}, we determined values of $\tau_\text{rise}$ and $\tau_\text{sfh}$ that lead to ISM abundances and median stellar ages consistent with observations at different Galactocentric radii.
Our procedure therein leads to discontinuities in the effects of radial gas flows due to the presence of breaks in otherwise piece-wise continuous functions of radius.
The radial flow couples the evolution of neighboring annuli, so those sitting near these piece-wise breaks exhibit significant numerical artifacts when input parameters do not vary smoothly with radius.

\subsubsection{Departures on Short Timescales}
\label{sec:gce:sfh:oscil}

We also follow a simple extension of this baseline SFH with oscillations at an amplitude of 50\% and a period of $P = 200$ Myr, according to
\begin{equation}
\dot\Sigma_\star \rightarrow \dot\Sigma_\star \left(
1 + 0.5\sin \left(
\frac{2 \pi t - \phi}{P}
\right)
\right).
\end{equation}
The choices of a 50\% amplitude and a $\sim$$200$ Myr period are motivated by departures from a smooth SFH on similar scales predicted by the FIRE simulations as a result of feedback-regulated star formation \citep{Hopkins2014}.
We explore this SFH in combination with inward flow velocities that vary with a high amplitude ($10$ km/s) and the same period (see discussion in Section \ref{sec:gce:scenarios:constant} below).
We consider cases in which the SFR variations are in-phase ($\phi = 0$) and out-of-phase ($\phi = \pi$) with the flow velocity oscillations.

\subsection{Radial Flow Scenarios}
\label{sec:gce:scenarios}


In this section, we describe our scenarios for setting radial velocities in the ISM.
Our models focus on the effects of angular momentum transport in the ISM and do not explicitly incorporate effects from energy dissipation.
We define a set of three fiducial scenarios, which can be described from the following governing equation
\begin{equation}
v_{r,g} = R\left[
\frac{\dot L}{L} -
\partderiv{\ln v_{\phi,g}}{t} -
\frac{\dot\Sigma_\text{acc}}{\Sigma_g}
\left(1 - \beta_{\phi,\text{acc}}\right)
\right].
\label{eq:vgas-scenarios}
\end{equation}
We derive this expression in Appendix \ref{sec:analytic:velocities}.
Each of our flow scenarios invoke one of the three terms in square brackets.
In the first, non-axisymmetric disk features like spiral arms and central bars induce a torque, which we've written as a fractional rate of change in angular momentum, $\dot L / L$.
In the second, the circular velocity of the ISM gas, $v_{\phi,g}$, increases with time due to the mass growth of the disk and halo.
In the third, gas tends to accrete onto the disk with lower angular momentum than the ISM, which drives radial gas flows through phase mixing \citep[e.g.,][see discussion in Section \ref{sec:intro}]{Lacey1985}.
Although none of these scenarios are mutually exclusive from one another, we explore them in isolation in this paper.
We discuss each scenario further below, and we summarize the key equations in Table \ref{tab:flow-scenarios}.

\input{flowmodels.tablebody.tex}

Following early investigations of radial gas flows in GCE models, we also follow a couple of models in which we simply assert a functional form for the flow velocity \citep[e.g.,][]{Lacey1985, Tosi1988, Goetz1992, Koeppen1994, Edmunds1995}.
This simplest of these direct parameterizations is a constant velocity that we impose at all radii and times.
We also identify a theoretical limit associated with radial gas flows.
In Section \ref{sec:results:dilution-steepening} below, we show that our GCE models with inward gas flows experience metal-poor accretion at systematically larger Galactocentric radii relative to our static model.
The theoretical limit is therefore the scenario in which \textit{all} of the metal-poor accretion is placed at the very outer edge of the star forming disk.
All gas that fuels star formation at intermediate radii is supplied by the inward flow (see discussion in Section \ref{sec:gce:scenarios:ora-limit} below).
\par
In our numerical models, we do not start the radial flow until $t = 1$ Gyr after the onset of star formation.
Some scenarios that we explore would predict substantially large radial velocities during this interval due to, e.g., low ISM surface densities and vigorous accretion (see discussion in Section \ref{sec:gce:scenarios:amd} below).
Due to our prescription for mass transfer between rings (see Figure \ref{fig:schematic}), our models can only resolve speeds as large as the ring width divided by the timestep size (i.e., $\left| v_{r,g} \right| \leq \delta R / \delta t = 50$ km/s).
We circumvent this issue by simply halting the onset of radial flows.
Our models are more directly relevant to disk evolution at lower redshift anyway.

\subsubsection{Global Torque}
\label{sec:gce:scenarios:gt}

In Appendix \ref{sec:analytic:velocities}, we show that any torque, $\dot L/L$, that is present should introduce a radial flow velocity given by
\begin{equation}
v_{r,g} = R \frac{\dot L}{L},
\label{eq:gt-scenario}
\end{equation}
where $L \sim v_\phi R$ is the local angular momentum of the ISM.
Our ``global torque'' (GT) scenario refers to these models with $\dot L / L \neq 0$.
We use $\dot L/L = -0.05$ Gyr$^{-1}$ as a fiducial value, with $\dot L/L = -0.02$ and $-0.08$ Gyr$^{-1}$ as variations thereof.
With $\dot L/L = -0.05$ Gyr$^{-1}$, flow speeds are of order $\sim$$0.5$ km/s across much of the disk.
\par
We interpret this torque as a consequence of azimuthally asymmetric features of the Galactic disk, such as the bar or spiral arms (see discussion in, e.g., \citealt{Portinari2000} and the review by \citealt{Shu2016}).
Observationally, a central bar introduces regions of positive and negative torque in opposing quadrants of the face-on disk (e.g., the torque map of NGC 4579 by \citealt{GarciaBurillo2009}).
Averaged over azimuth, simulations predict net inward flows driven by spiral arms \citep{Grand2016, Orr2023}.
Outside the corotation radius, however, the torque by the bar is expected to drive an outward radial gas flow \citep[e.g.,][]{Lacey1985, Hopkins2011, Beane2023}.
We focus on simple prescriptions for inward flow velocities in this paper, but we plan to investigate bar-driven flow models in future work.
\par
We include a weak central outflow in this model in order to offset the pile-up of matter in the central regions.
This model, along with our constant velocity scenario (see Section \ref{sec:gce:scenarios:constant} below), has faster radial velocities at the present day than our other models.
This feature can lead to unphysical surface densities near $R = 0$, which the weak ejection counteracts.
We use a simple exponential prescription:
\begin{equation}
\eta = \eta_0 e^{-R / r_\eta},
\label{eq:central-ejection}%
\end{equation}
with $\eta_0 = 0.5$ and $r_\eta = 5$ kpc for the GT scenario.
These choices are comfortable within observational constraints on mass loading in MW-like galaxies (see, e.g., the reviews by \citealt{Veilleux2020} and \citealt{Thompson2024}).
We set this normalization because it is high enough to maintain plausible surface densities in the inner regions but still low enough that the effects of radial gas flows dominate chemical enrichment at $R \gtrsim 3$ kpc.
Our models are less applicable in the $R \lesssim 3$ kpc range anyway, since this region corresponds to the Galactic bulge (see also discussion in \citetalias{Johnson2021}).
Even so, our fiducial choice of $\dot L / L = -0.05$ Gyr$^{-1}$ maintains plausible surface densities down to $R \sim 1$ kpc with this choice of $r_\eta$.
We find in practice that radial gas flows on any level generally lead to numerical artifacts in the innermost $\sim$few annuli anyway, since the central zone cannot actually shed any mass through an inward flow.

\subsubsection{Potential Well Deepening}
\label{sec:gce:scenarios:pwd}

In this scenario, which we refer to as ``potential well deepening'' (PWD), the circular velocity of the ISM grows with time as the Galaxy becomes more massive.
This argument is based on the baryonic Tully-Fisher relation, which refers to the empirical result that galaxies with higher baryon mass exhibit faster rotation \citep[e.g.,][]{McGaugh2000, McGaugh2005, Gurovich2004, DeRijcke2007}.
The original Tully-Fisher relation \citep{Tully1977} connected galaxy luminosities with circular velocities.
Here, we simply postulate that this observed relation reflects the circular velocity growing in proportion to mass over time.
As a consequence, the angular momentum of a circular orbit also grows with time (i.e., $\dot v_{\phi,g} > 0$).
Test particles with fixed angular momentum should therefore slowly spiral inward.
\par
Observationally, galaxy circular velocities follow a power-law with stellar mass, $v_{\phi,g} \propto M_\star^\gamma$.
Differentiating this prescription with time, we find an expression relating the tangential acceleration in the disk with the rate of star formation:
\begin{equation}
\partderiv{\ln v_{\phi,g}}{t} = \gamma
\partderiv{\ln M_\star}{t}.
\label{eq:pwd-scenario}
\end{equation}
\citet{Ristea2024} measure a power-law index consistent with $\gamma = 0.19$ based on data from the MaNGA survey \citep{Bundy2015}.
We therefore use $\gamma = 0.2$ as a fiducial value, with $\gamma = 0.1$ and $0.3$ as variations thereof.
\par
The radial flow velocity at each timestep follows from substituting Equation \ref{eq:pwd-scenario} in for the second term in square brackets in Equation \ref{eq:vgas-scenarios}.
The rate of growth in stellar mass can be expressed as the total star formation rate divided by the total mass in stars that have formed up to a given time, according to
\begin{equation}
\frac{\partial \ln M_\star}{\partial t} = \ddfrac{
    \int_0^{R_\text{SF}} \dot\Sigma_\star(R, t)
    2 \pi R dR
}{
    \int_0^t \int_0^{R_\text{SF}}
    \dot\Sigma_\star(R, t') 2 \pi R dR dt'
},
\end{equation}
where $R_\text{SF} = 20$ kpc is the Galactocentric radius beyond which we set $\dot \Sigma_\star = 0$ (see discussion in Section \ref{sec:gce:sfh}).
In this expression, the factor of $1 - r$ accounting for the return of stellar envelopes back to the ISM appears in both the numerator and denominator and therefore cancels.

\subsubsection{Angular Momentum Dilution}
\label{sec:gce:scenarios:amd}

In this scenario, which we refer to as angular momentum dilution (AMD), matter tends to accrete onto the Galactic disk with lower angular momentum than the circular motions of the ISM itself.
Phase mixing then leads to an inward radial flow so that gas can settle at a new orbital radius appropriate for its angular momentum.
Hydrodynamic simulations indeed predict differences in angular momentum between accreting material and the ISM \citep[e.g.,][]{Trapp2022, Barbani2025}, but these differences are not constrained observationally.
We demonstrate in Appendix \ref{sec:analytic:velocities} that the radial flow velocity in this scenario can be expressed as
\begin{equation}
v_{r,g} = -R
\frac{\dot\Sigma_\text{acc}}{\Sigma_g}
\left(1 - \beta_{\phi,\text{acc}}\right),
\label{eq:amd-scenario}
\end{equation}
where $\beta_{\phi,\text{acc}} \equiv v_{\phi,\text{acc}} / v_{\phi,g}$ is the ratio of circular velocities between the accreting material and the ISM.
This scenario corresponds to the third term in square brackets in Equation \ref{eq:vgas-scenarios}.
\citet{Bilitewski2012} found that values of $\beta_{\phi,\text{acc}} \sim 0.7$ led to radial metallicity gradients consistent with their data.
\citet{Pezzulli2016} found similar results.
We therefore use $\beta_{\phi,\text{acc}} = 0.7$ as a fiducial value, with $\beta_{\phi,\text{acc}} = 0.6$ and $0.8$ as variations thereof.
\par
The accretion rate $\dot \Sigma_\text{acc}$ is not known by our models \textit{a priori} because we specify the SFH as input, so Equation \ref{eq:amd-scenario} above does not represent the radial velocity in terms of known input quantities.
In Appendix \ref{sec:analytic:velocities:amd}, we show that the velocity profile is given by the following differential equation
\begin{equation}
\begin{split}
\partderiv{v_{r,g}}{R} &+
v_{r,g} \left[
\frac{1}{R}\left(
1 + \frac{1}{1 - \beta_{\phi,\text{acc}}}
\right) +
\frac{1}{N}
\partderiv{\ln \dot\Sigma_\star}{R}
\right]
\\ &=
\frac{-1}{N}
\partderiv{\ln \dot\Sigma_\star}{t} -
\frac{1 - r}{\tau_\star},
\label{eq:amd-ode}
\end{split}
\end{equation}
which we integrate numerically at each timestep.
We determine the value of the integration constant by setting $v_{r,g} = 0$ at $R = 0$, which follows from Equation \ref{eq:amd-scenario} if $\dot\Sigma_\text{acc}$, $\Sigma_g$, and $\beta_{\phi,\text{acc}}$ are all finite.
\par
Given that previous versions of our GCE models have used ejection as opposed to radial gas flows, we briefly remark that the velocities predicted by the AMD scenario should change in the presence of ejection.
In this case, Equation \ref{eq:vgas-final-appendix} extends to
\begin{equation}
v_{r,g} \rightarrow R \left[
\frac{\dot\Sigma_\text{wind}}{\Sigma_g}
\left(1 - \beta_{\phi,\text{wind}}\right) -
\frac{\dot\Sigma_\text{acc}}{\Sigma_g}
\left(1 - \beta_{\phi,\text{acc}}\right)
\right],
\label{eq:amd-scenario-extended}
\end{equation}
where $\beta_{\phi,\text{wind}} \equiv v_{\phi,\text{wind}} / v_{\phi,g}$ is the circular velocity of ejected material relative to the ISM (see discussion in Appendix \ref{sec:analytic:velocities}).
There is a sign difference between the terms in square brackets in the above expression, since accretion adds material to the ISM while ejection removes it.
If feedback ejects ISM gas by exerting a force perpendicular to the disk midplane, then the circular velocity should be preserved (i.e., $v_{\phi,\text{wind}} = v_{\phi,g}$).
In this limit, ejection has no \textit{direct} influence on the radial flow velocity because $\beta_{\phi,\text{wind}} \rightarrow 1$.
However, ejection necessitates additional accretion in order to fuel star formation and fulfill the stellar mass budget of the MW, thereby increasing $\dot\Sigma_\text{acc} / \Sigma_g$ and leading to faster inward flows (see Equation \ref{eq:ifr-per-sfr} below).
The inclusion of ejection may lead to interesting differences in the AMD scenario predictions, but we do not investigate such models in this paper.

\subsubsection{Constant Velocity}
\label{sec:gce:scenarios:constant}

This scenario is our simplest.
We simply assert a given velocity to use at every Galactocentric radius and time.
We use $v_{r,g} = -1$ km/s as the fiducial value, with $v_{r,g} = -0.5$ km/s and $-1.5$ km/s as variations thereof.
We show that a constant velocity, despite its simplicity, leads to better agreement with trends in metallicity with stellar age across the Galactic disk in Section \ref{sec:results:eq-scenario} below.
\par
Like the GT scenario (see Section \ref{sec:gce:scenarios:gt}), the constant velocity prescription also requires weak ejection at small radii to maintain plausible surface densities down to $R \sim 1$ kpc.
We use the prescription given by Equation \ref{eq:central-ejection}, maintaining the scale radius of $r_\eta = 5$ kpc.
We double the normalization to $\eta_0 = 1$, since this model maintains $R \gg 0$ into the central galaxy.
This feature leads to stronger effects on surface density than in the GT scenario, which uses a linear velocity profile that naturally slows at small $R$.
These choices are based on the same considerations described in Section \ref{sec:gce:scenarios:gt}.
\par
This model leads to slightly lower metal abundances overall than our other radial flow scenarios.
This difference arises because the radial flow, under this prescription, tends to lower the ISM surface density near the Sun while others generally increase surface densities across much of the disk (see discussion in Section \ref{sec:discussion}).
In order to bring the predicted ISM abundances at the present-day back into agreement with observations, we simply increase all of our O and Fe yields by a factor of $10^{0.2}$.
This correction is somewhat larger than statistical uncertainties in the scale of stellar yields ($\sim$$0.1$ dex) but certainly within systematic uncertainties \citep{Weinberg2024}.
We could also make up this difference by using pre-enriched accretion (see discussion in Appendix \ref{sec:analytic:zin}).

\paragraph{Departures on Short Timescales}
We also use this scenario as a base from which we construct variations with high-amplitude, short-timescale fluctuations in the flow velocity, motivated by our discussion in Section \ref{sec:intro}.
In these variations, the mean flow speed remains constant in time and radius, but oscillates sinusoidally with a short timescale according to
\begin{equation}
v_{r,g} = \bar v_{r,g} + A \sin \left(
\frac{2\pi t}{P}
\right),
\end{equation}
where we use $A = 10$ km/s and $P = 200$ Myr.
This amplitude is comparable to the width of the distribution of observed velocities (see Figure \ref{fig:empirical}).
This choice of period allows several full oscillations within $1$ Gyr, broadly consistent with simulation predictions \citep[e.g.,][]{Trapp2022}.
$P = 200$ Myr also places the duration of one cycle in $v_{r,g}$ on a similar order as the rotational period of the Galactic disk.

\subsubsection{The Outer Rim Accretion Limit}
\label{sec:gce:scenarios:ora-limit}

Our final radial gas flow scenario is a theoretical limit, which we refer to as ``outer rim accretion'' (ORA).
We demonstrate in Section \ref{sec:results:dilution-steepening} that radial gas flows shift predicted accretion histories toward large Galactocentric radii relative to static models.
The ORA limit therefore corresponds to the scenario in which \textit{all} of the accretion happens at the very outer edge of the disk, with surface densities in the interior of the disk supplied entirely by the radial flow.
This paper argues that the ORA limit corresponds to the steepest possible metallicity gradient for a given set of choices for GCE parameters.
Our numerical models indeed predict gradients sufficiently steep to be obviously ruled out by available empirical constraints (see discussion in Section \ref{sec:results}).
We explore this limit as a theoretically interesting comparison case but do not dedicate the same level of attention as our other models.
\par
By definition, the ORA limit asserts that $\dot \Sigma_\text{acc} \rightarrow 0$.
The continuity equation for the ISM surface density (Equation \ref{eq:dot-sigma-gas}) therefore reduces, yielding a direct solution for $\mu_g$:
\begin{subequations}
\begin{align}
\dot\Sigma_g &\rightarrow -\dot\Sigma_\star
\left(1 + \eta - \mu_g - r\right)
\\
\implies \mu_g &\rightarrow 1 + \eta - r +
\tau_\star \frac{\dot\Sigma_g}{\Sigma_g}.
\label{eq:ora-limit-mu}
\end{align}
\end{subequations}
This expression indicates that, unsurprisingly, the radial flow simply counteracts all other effects on the ISM surface density in the ORA limit.
A solution to the radial velocity profile in terms of input quantities follows from our solution to $\mu_g$ in the limit that $\delta R, \delta t \rightarrow 0$ (see Equation \ref{eq:mu-g-soln} and discussion in Section \ref{sec:discussion} below).
We demonstrate in Appendix \ref{sec:analytic:velocities:ora} that the velocity profile obeys the following differential equation
\begin{equation}
\partderiv{v_{r,g}}{R} +
v_{r,g} \left(
\frac{1}{R} +
\frac{1}{N} \partderiv{\ln \dot\Sigma_\star}{R}
\right) =
\frac{-1}{N} \partderiv{\ln \dot\Sigma_\star}{t} -
\frac{1 - r}{\tau_\star},
\label{eq:ora-limit-profile}
\end{equation}
which we integrate numerically at each timestep.
We determine the integration constant by computing the velocity in the second innermost ring (i.e., $R = \delta R \rightarrow 2\delta R$) that would fuel the required level of star formation at the next timestep in the innermost ring (i.e., $R = 0 \rightarrow \delta R$) without requiring accretion.
We then proceed with the solution to the above expression as an initial value problem, integrating outward in radius.

%% file: glossary.tablebody.tex
{\renewcommand{\arraystretch}{1.05}
\begin{table*}
\caption{A summary of the GCE parameters appearing throughout this paper.}
\begin{tabularx}{\textwidth}{c @{\extracolsep{\fill}} c c l}
\hline
\hline
Parameter & Section & Fiducial Value & Description
\\
\hline
$R$ & ... & $0 - 20$ kpc & Galactocentric radius.
\\
$\delta R$ & ... & $100$ pc & The width of each annulus.
\\
$\delta t$ & ... & $1$ Myr & The timestep size that we integrate our numerical models with.
\\
$\ycc{O}$ & \ref{sec:gce:yields} & 0.00572 & Mass of O produced by massive stars per solar mass of star formation.
\\
$\yia{O}$ & \ref{sec:gce:yields} & 0 & Mass of O produced by SNe Ia per solar mass of star formation.
\\
$\ycc{Fe}$ & \ref{sec:gce:yields} & $4.51\times10^{-4}$ & Mass of Fe produced by massive stars per solar mass of star formation.
\\
$\yia{Fe}$ & \ref{sec:gce:yields} & $8.38\times10^{-4}$ & Mass of Fe produced by SNe Ia per solar mass of star formation.
\\
$\Sigma_g$ & \ref{sec:gce:gas} & ... & The surface density of the ISM gas.
\\
$\dot\Sigma_\star$ & \ref{sec:gce:sfh} & Equation \ref{eq:sfh} & The surface density of star formation.
\\
$\dot\Sigma_\text{acc}$ & \ref{sec:gce:gas} & ... & The rate of gravitational accretion from the CGM.
\\
$N$ & \ref{sec:gce:gas} & $1.5$ & Power-law index of the relation: $\dot\Sigma_\star \propto \Sigma_g^N$.
\\
$\Sigma_x$ & ... & ... & The surface density of some chemical element $x$ in the ISM.
\\
$Z_x$ & ... & ... & $\equiv \Sigma_x / \Sigma_g$. Abundance by mass of a chemical element $x$.
\\
$Z_{x,\eq}$ & \ref{sec:results} & ... & Equilibrium $Z_x$ (i.e., $\dot Z_x \approx 0$ at $Z_x \approx Z_{x,\eq}$).
\\
$\tau_\eq$ & \ref{sec:discussion:local-enrich} & ... & The timescale on which the local ISM approaches $Z_{x,\eq}$ (see Equation \ref{eq:taueq}).
\\
$\eta$ & \ref{sec:gce:gas} & $\sim$$0$ & $\equiv \dot\Sigma_\text{wind}/\dot\Sigma_\star$. The ejection efficiency. Also known as the mass loading factor.
\\
$\eta_0$ & \ref{sec:gce:scenarios:gt} & 0.5 (GT); 1 (Constant) & The value of $\eta$ at $R = 0$. Only relevant for GT and constant flow velocity scenarios.
\\
$r_\eta$ & \ref{sec:gce:scenarios:gt} & 5 kpc & The scale radius of the exponential decline in $\eta$ with $R$ (see Equation \ref{eq:central-ejection}).
\\
$\mu_g$ & \ref{sec:gce:gas} & ... & $\equiv \dot\Sigma_{g,\flow} / \dot\Sigma_\star$. The flow coefficient for the gas.
\\
$\mu_x$ & \ref{sec:gce:gas} & ... & $\equiv \dot\Sigma_{x,\flow} / Z_x \dot\Sigma_\star$. The flow coefficient for a chemical element $x$.
\\
$r$ & \ref{sec:gce:gas} & $\sim$$0.4$ & $\equiv \dot\Sigma_r / \dot\Sigma_\star$. Fraction of a single stellar population's initial mass returned to the ISM.
\\
$\tau_\star$ & \ref{sec:gce:gas} & ... & $\equiv \Sigma_g / \dot\Sigma_\star$. Star formation efficiency timescale.
\\
$\tau_\text{ISM}$ & \ref{sec:results:eq-scenario} & Equation \ref{eq:tau-ism-def} & $\equiv \tau_\star / (1 + \eta - \mu_g - r)$. The timescale describing turnover in the ISM.
\\
$\tau_\text{rise}$ & \ref{sec:gce:sfh} & Equation \ref{eq:taurise} & Timescale on which the SFR rises at early times.
\\
$\tau_\text{sfh}$ & \ref{sec:gce:sfh} & Equation \ref{eq:tausfh} & Timescale on which the SFR falls at late times.
\\
$v_{r,g}$ & \ref{sec:gce:scenarios} & ... & Mean Galactocentric radial velocity of the ISM.
\\
$a(v_{r,g}, R)$ & \ref{sec:gce} & Equation \ref{eq:areafracs} & Mass fraction of an annulus that moves the next annulus inward for a velocity $v_{r,g}$.
\\
$\dot L/L$ & \ref{sec:gce:scenarios:gt} & $-0.05$ Gyr$^{-1}$ & Torque per unit angular momentum in the ISM.
\\
$\gamma$ & \ref{sec:gce:scenarios:pwd} & $0.2$ & Power-law index of the circular velocity-stellar mass relation.
\\
$\beta_{\phi,\text{acc}}$ & \ref{sec:gce:scenarios:amd} & 0.7 & Angular momentum of accreting matter relative to the disk circular rotation.
\\
$\nabla$[X/H] & \ref{sec:discussion} & ... & $\equiv \partial \log_{10}(Z_x / Z_{x,\odot})/\partial R$. The radial metallicity gradient in some chemical element X.
\\
$\nabla_\eq$[X/H] & \ref{sec:discussion} & ... & The equilibrium metallicity gradient (i.e., $\dot \nabla[\text{X/H}] \approx 0$ at $\nabla[\text{X/H}] \approx \nabla_\eq [\text{X/H}]$).
\\
$\tau_\nabla$ & \ref{sec:discussion} & ... & The timescale on which $\nabla$[X/H] approaches $\nabla_\eq$[X/H].
\\
\hline
\end{tabularx}
\label{tab:glossary}
\end{table*}
}

%% file: flowmodels.tablebody.tex
\begin{table}
\caption{
A summary of our radial gas flow scenarios, which we describe individually in Section \ref{sec:gce:scenarios}.
Our physical scenarios drive radial gas flows based on arguments rooted in angular momentum transport, invoking one of the three terms in the governing equation shown at the top.
Our direct parameterizations instead assert a specific functional form without any explicit assumptions regarding the physical origin.
}
\begin{tabularx}{\columnwidth}{c}
\hline
\hline
\multicolumn{1}{l}{
\textit{Physical Scenarios}
}
\\
\hline
\parbox{\columnwidth}{
\centering
\vspace{2mm}
\(\displaystyle
v_{r,g} = R \left[
\frac{\dot L}{L} -
\partderiv{\ln v_{\phi,g}}{t} -
\frac{\dot\Sigma_\text{acc}}{\dot\Sigma_\star}
\left(1 - \beta_{\phi,\text{acc}}\right)
\right]
\)
\vspace{2mm}
}
\\
\hline
\end{tabularx}
\begin{tabularx}{\columnwidth}{c @{\extracolsep{\fill}} c}
\multicolumn{2}{c}{
\textbf{Global Torque}
}
\\
\parbox{0.55\columnwidth}{
\centering
\(\displaystyle
\frac{\dot L}{L} = \text{Constant}
\)
} & \parbox{0.4\columnwidth}{
\centering
\(\displaystyle
\dot{L}/L = \begin{cases}
-0.02\text{ Gyr}^{-1}
\\
-0.05\text{ Gyr}^{-1}
\\
-0.08\text{ Gyr}^{-1}
\end{cases}
\)
\vspace{2mm}
}
\\
\hline
\multicolumn{2}{c}{
\textbf{Potential Well Deepening}
}
\\
\parbox{0.55\columnwidth}{
\centering
\(\displaystyle
\partderiv{\ln v_{\phi,g}}{t} = \gamma
\partderiv{\ln M_\star}{t}
\)
} & \parbox{0.4\columnwidth}{
\centering
\(\displaystyle
\gamma = \begin{cases}
0.1
\\
0.2
\\
0.3
\end{cases}
\)
\vspace{2mm}
}
\\
\hline
\multicolumn{2}{c}{
\textbf{Angular Momentum Dilution}
}
\\
\parbox{0.55\columnwidth}{
\centering
\(\displaystyle
\frac{\dot \Sigma_\text{acc}}{\Sigma_g}
\left(1 - \beta_{\phi,\text{acc}}\right)
\)
} & \parbox{0.4\columnwidth}{
\centering
\(\displaystyle
\beta_{\phi,\text{acc}} = \begin{cases}
0.8
\\
0.7
\\
0.6
\end{cases}
\)
\vspace{2mm}
}
\\
\hline
\multicolumn{2}{l}{
\textit{Direct Parameterizations}
}
\\
\hline
\multicolumn{2}{c}{
\textbf{Constant}
}
\\
\parbox{0.55\columnwidth}{
\centering
\(\displaystyle
v_{r,g} \rightarrow \text{Constant}
\)
} & \parbox{0.45\columnwidth}{
\centering
\(\displaystyle
v_{r,g} = \begin{cases}
-0.5 \text{ km/s}
\\
-1 \text{ km/s}
\\
-1.5 \text{ km/s}
\end{cases}
\)
\vspace{2mm}
}
\\
\hline
\multicolumn{2}{c}{
\textbf{Outer Rim Accretion Limit}
\vspace{1mm}
}
\\
\multicolumn{2}{c}{
\(\displaystyle
\dot\Sigma_\text{acc} \rightarrow 0
\implies
\mu_g \rightarrow
1 + \eta - r + \tau_\star
\frac{\dot\Sigma_g}{\Sigma_g}
\)
\vspace{2mm}
}
\\
\hline
\end{tabularx}
\label{tab:flow-scenarios}
\end{table}

%% file: results.tex
\section{Results}
\label{sec:results}

\begin{figure*}
\centering
\includegraphics[scale = 0.92]{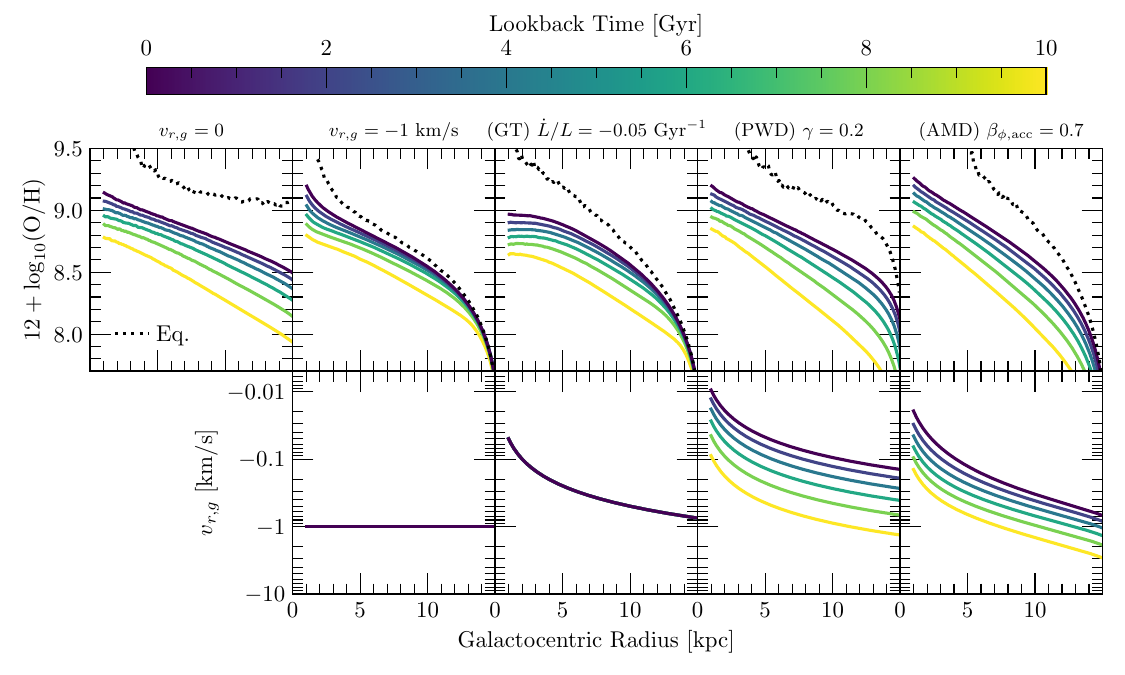}
\caption{
Evolution in the ISM metallicity (top) and radial velocity (bottom) over time in our fiducial GCE models.
From left to right: the base model with $v_{r,g} = 0$ everywhere, a constant velocity of $v_{r,g} = -1$ km/s, the GT scenario with $\dot L/L = -0.1$ Gyr$^{-1}$, the PWD scenario with $\gamma = 0.2$, and the AMD scenario with $\beta_{\phi,\text{acc}} = 0.7$ (see discussion in Section \ref{sec:gce:scenarios}).
Solid lines show snapshots of the radial profiles color coded according to the colorbar at the top.
Dotted black lines in the top row of panels show the equilibrium O abundance as a function of Galactocentric radius at the present day.
\textbf{Summary}: Each radial gas flow scenario has a different impact on the predicted metal enrichment history, with some being more substantial than others.
Inward flows tend to lead to strong variations in the equilibrium metallicity with Galactocentric radius, even if the ISM never reaches the local equilibrium abundance across much of the disk.
}
\label{fig:profile-evol-comp}
\end{figure*}

\begin{figure*}
\centering
\includegraphics[scale = 1]{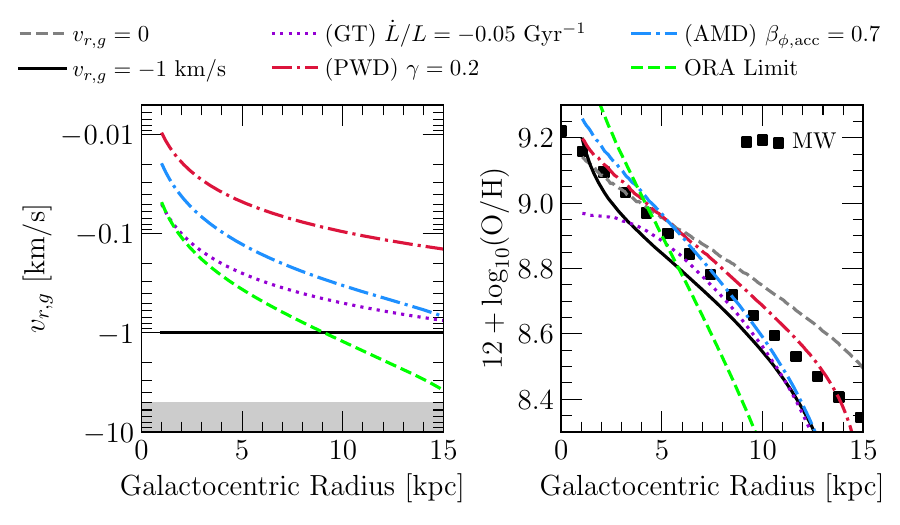}
\caption{
A comparison of the ISM radial velocity (left) and metallicity profiles (right) predicted by our GCE models, marked according to the legend at the top.
We shade the region corresponding to $\left|v_{r,g}\right| > 5$ km/s, which is disfavored by direct observations of galaxy velocity fields \citep[e.g.,][see discussion in Section \ref{sec:intro}]{DiTeodoro2021}.
Black squares in the right panel mark the best-fit profile for \hii\ regions in the MW measured by \citet[][shown as points instead of a line for visual clarity]{MendezDelgado2022}.
\textbf{Summary}: Inward flows with velocities of order $\sim$$0.1 - 1$ km/s at low redshift, within allowed empirical constraints, should have significant effects on abundance growth.
The ORA limit should reflect the steepest possible metallicity gradient for a given choice of GCE parameters (see discussion in Section \ref{sec:results:dilution-steepening}).
}
\label{fig:vgas-oh-comp}
\end{figure*}

Figure \ref{fig:profile-evol-comp} shows the evolution of the metallicity and radial velocities in the ISM with time in our fiducial models.
These models start with simple velocity prescriptions that increase in complexity from left to right.
Our constant and GT models do not evolve with time, but the latter uses a linear as opposed to a flat velocity profile.
The PWD scenario is also a linear velocity profile but evolves with time, and AMD is non-linear.
The PWD and AMD scenarios lead to slower radial flows at low redshift because the velocity is tied to the evolution of the Galaxy.
Under the PWD prescription, the velocity is set by the growth in stellar mass, which fast at high redshift and slow at the present day (see Equations \ref{eq:pwd-scenario} and \ref{eq:sfh}).
With AMD, the rate of gas accretion falls with time, which weakens the effects of phase-mixing low angular momentum material into the ISM (see Equation \ref{eq:amd-scenario}).
\par
These models predict differences in abundance evolution as a consequence of these differences in velocity evolution.
In the base model with $v_{r,g} = 0$, the metal abundance increases steadily with time with no obvious substantial changes in the slope of the radial profile.
The $v_{r,g} = -1$ km/s model predicts noticeably less growth in the abundances over the last $\sim$$8$ Gyr (since redshift $z \sim 1$).
We show that this prediction plays an important role in reproducing trends in metallicity with stellar age in Section \ref{sec:results:eq-scenario} below.
The GT scenario also lessens the abundance growth relative to the $v_{r,g} = 0$ model, though not to the same extent as a constant velocity.
Both the PWD and AMD scenarios lead to more subtle differences in the abundance evolution, primarily due to the slower velocities at low redshifts.
The primary difference between the two is that the PWD scenario ``cuts off'' in metallicity in the outer disk rather suddenly, while this effect is more gradual in the AMD scenario.
\par
For comparison, the top row of panels in Figure \ref{fig:profile-evol-comp} show the equilibrium O abundance, $Z_{\text{O},\eq}$, as a function of Galactocentric radius.
In equilibrium, metal production by stars is balanced by losses to ejection and radial gas flows (if present) as well as the formation of new stars and dilution from metal-poor accretion (i.e., $\dot Z \approx 0$ at $Z \approx Z_\eq$).
In Section \ref{sec:discussion:local-enrich}, we demonstrate that $Z_{\text{O},\eq}$ is given by
\begin{equation}
Z_{\text{O},\eq} = \frac{\ycc{O}}{
    1 + \eta - \mu_\text{O} - r -
    \tau_\star / N\tau_\text{sfh}
}.
\label{eq:zoeq}
\end{equation}
This expression reduces to the form derived in \citet{Weinberg2017} in the absence of radial gas flows ($\mu_\text{O} \rightarrow 0$) with a linear Kennicutt-Schmidt relation ($N \rightarrow 1$; see discussion in Section \ref{sec:gce}).
Each scenario leads to significant variations in the equilibrium abundance with Galactocentric radius, with the exception of the $v_{r,g} = 0$ limit at $R \gtrsim 8$ kpc.
In general, each of our radial flow models lead to lower $Z_{\text{O},\eq}$ and a steeper radial gradient at $R \gtrsim 5 - 8$ kpc.
We discuss this point and its implications further in Section \ref{sec:discussion} below.
\par
Figure \ref{fig:vgas-oh-comp} compares the present day radial profiles in the ISM velocity and metallicity predicted by each model.
Broadly, these models span the $\left|v_{r,g}\right| \sim 0.1 - 1$ km/s range across much of the Galactic disk.
Every model with radial gas flows predicts lower ISM metallicities and a steeper radial gradient at $R \gtrsim 5$ kpc relative to the $v_{r,g} = 0$ base model, consistent with prior work (see discussion in, e.g., \citealt{Portinari2000}).
Despite differences in their radial velocity prescriptions (see Section \ref{sec:gce:scenarios}), all models share this qualitative effect.
We discuss why this outcome should arise in \textit{any} GCE model incorporating inward gas flows in Section \ref{sec:results:dilution-steepening} below.
\par
We compare these predictions to broad empirical constraints in both panels of Figure \ref{fig:vgas-oh-comp}.
In the left panel, we shade the $v_{r,g} < -5$ km/s region, which is disfavored by velocity fields derived from CO and \hi\ emission in external galaxies (\citealt{Wong2004, DiTeodoro2021}; see discussion in Section \ref{sec:intro}).
In the right panel, we show the linear best-fit to the abundance measurements in \hii\ regions by \citet{MendezDelgado2022}.
Each model predicts radial velocities in the ISM that are allowed by currently available direct measurements.
However, the $v_{r,g} = 0$ base model and the ORA limit are in noticeable tension with the ISM metallicities across much of the disk.
With no inward flows, metal abundances are higher than observed, and the radial profile is too shallow.
The opposite arises in the ORA limit, underpredicting metallicities with a radial profile that is too steep.
These two models correspond to opposing limits regarding mass accretion and radial gas flows.
With $v_{r,g} = 0$, every fluid element in the ISM is accreted at its present-day radius, whereas \textit{all} accretion occurs at the outermost edge of the disk in the ORA limit (see discussion in Section \ref{sec:gce:scenarios:ora-limit}).
With neither model in obvious tension with the $v_{r,g} \gtrsim -5$ km/s requirement, our sample of chemical abundances and stellar ages provides sharper constraints on long-lived radial gas flows than direct observations of galaxy velocity fields.

\begin{figure*}
\centering
\includegraphics[scale = 1]{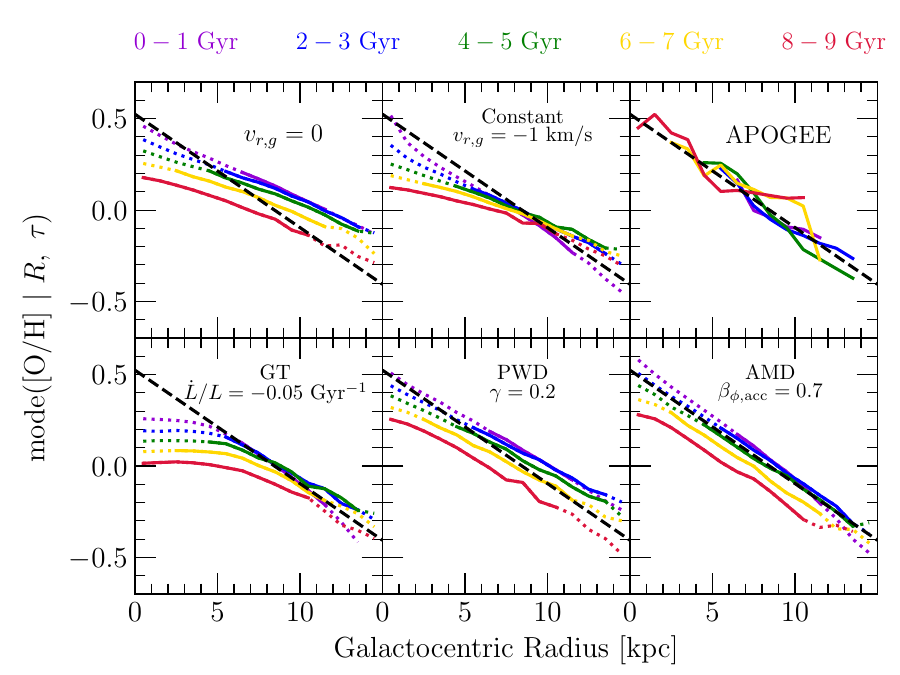}
\caption{
The peak of the [O/H] distribution as a function of Galactocentric radius for mono-age stellar populations near the disk midplane ($\left|z\right| \leq 0.5$ kpc).
Each line shows the trend in a 1-Gyr bin of stellar age, color coded according to the legend at the top.
Individual panels show the predictions from the $v_{r,g} = 0$ base model (top left), the $v_{r,g} = -1$ km/s constant velocity model (top middle), the GT scenario with $\dot L/L = -0.1$ Gyr$^{-1}$ (bottom left), the PWD scenario with $\gamma = 0.2$ (bottom center), and the AMD scenario with $\beta_{\phi,\text{acc}} = 0.7$ (bottom right).
The top-right panel shows the observed trend in our sample.
We show the predictions as dotted as opposed to solid lines at radii and ages where there is no measurement in the top-right panel.
The black dashed line is the same in each panel and shows the radial metallicity profile of our sample across all ages.
\textbf{Summary}: No model is able to reproduce the slope, normalization, and apparent age-independence of the observed trend across the \textit{entire} Galactic disk.
Our constant velocity and GT scenarios are broadly allowed by our sample at $R \gtrsim 6$ kpc and $R \gtrsim 8$ kpc, respectively, but the profile is too shallow at smaller radii.
}
\label{fig:gradoh-agebins}
\end{figure*}

\subsection{The Equilibrium Scenario}
\label{sec:results:eq-scenario}

\begin{figure*}
\centering
\includegraphics[scale = 1]{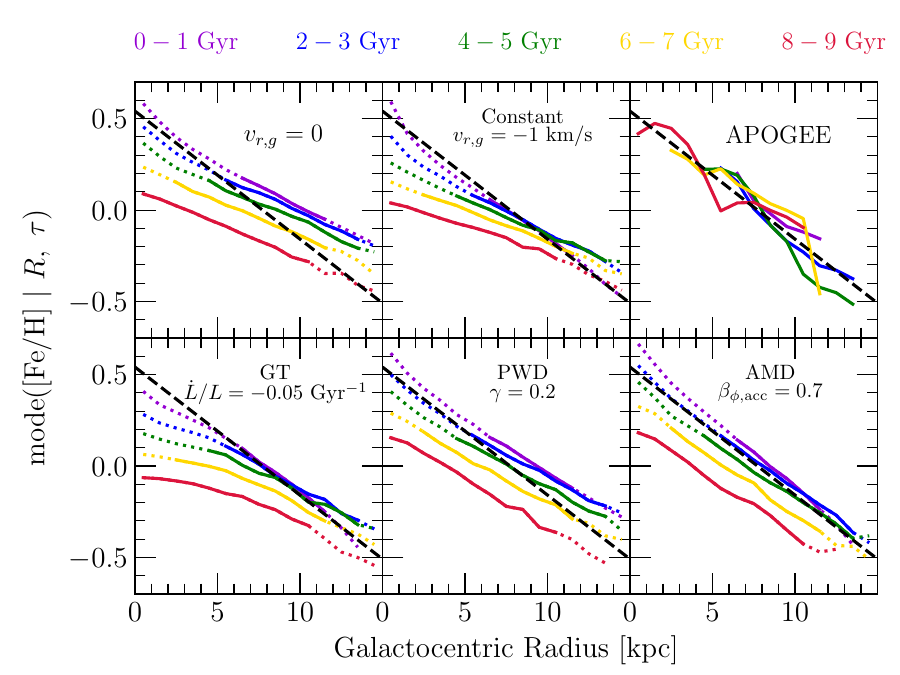}
\caption{
The same as Figure \ref{fig:gradoh-agebins}, but for Fe.
Differences between O and Fe arise due to the delayed nature of SN Ia enrichment (see discussion in Section \ref{sec:gce:yields}).
}
\label{fig:gradfeh-agebins}
\end{figure*}

\begin{figure*}
\centering
\includegraphics[scale = 0.85]{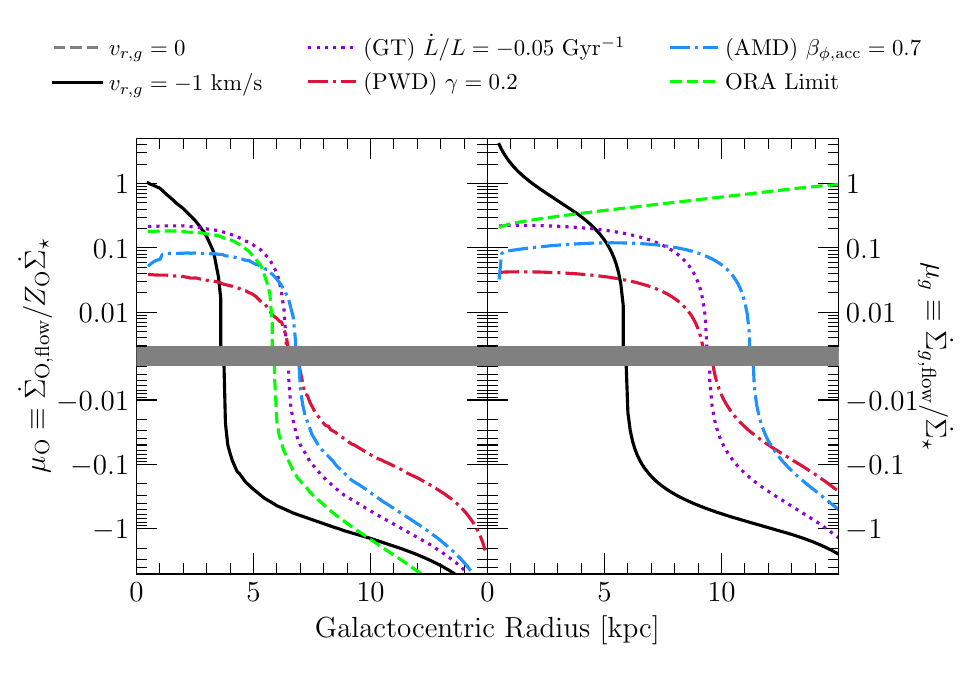}
\includegraphics[scale = 0.9]{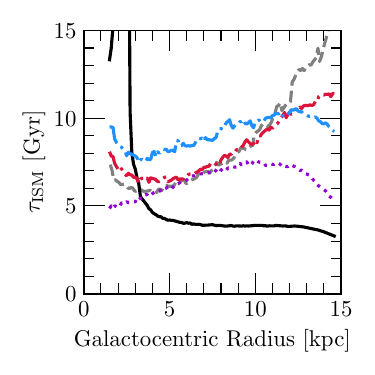}
\includegraphics[scale = 0.9]{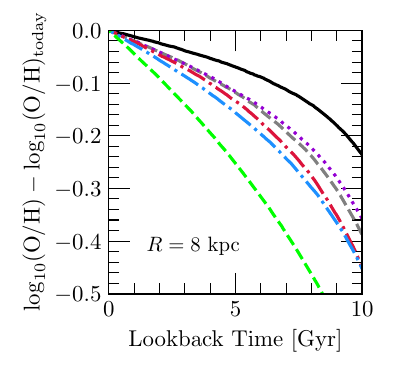}
\includegraphics[scale = 0.9]{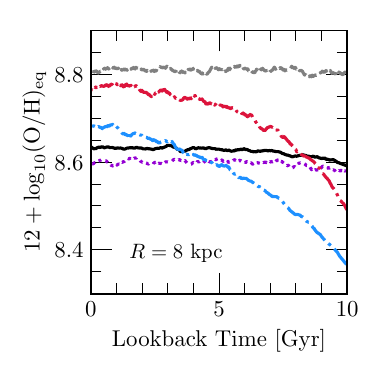}
\caption{
The mechanisms driving chemical equilibrium or non-equilibrium in the Galactic disk in the presence of inward gas flows.
Individual lines show the predictions of our GCE models, marked according to the legend at the top.
\textbf{Top}: Radial profiles in $\mu_\text{O}$ (left) and $\mu_g$ (right) at the present day (see Equations \ref{eq:mu-definition} and \ref{eq:mu-soln} as well as discussion in Sections \ref{sec:gce:gas} and \ref{sec:discussion}).
\textbf{Bottom Middle}: The turnover timescale in the ISM (see Equation \ref{eq:tau-ism-def}) at the present day as a function of Galactocentric radius.
\textbf{Bottom Left}: The growth in the O abundance at $R = 8$ kpc over time.
\textbf{Bottom Right}: The equilibrium O abundance at $R = 8$ kpc as a function of lookback time, relative to the present day abundance.
\textbf{Summary}: The $v_{r,g} = $ constant scenario leads to a clear chemical equilibrium because it acts to lower surface densities ($\mu_g < 0$) across much of the disk, which quickens turnover in the ISM (i.e., $\tau_\text{ISM} \sim $ few Gyr).
The equilibrium abundance is also steady as opposed to unsteady.
}
\label{fig:eq-conditions}
\end{figure*}

\begin{figure*}
\centering
\includegraphics[scale = 1]{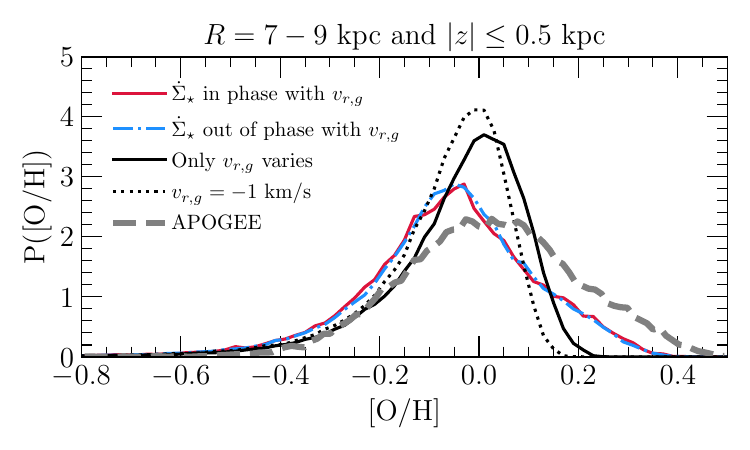}
\includegraphics[scale = 0.9]{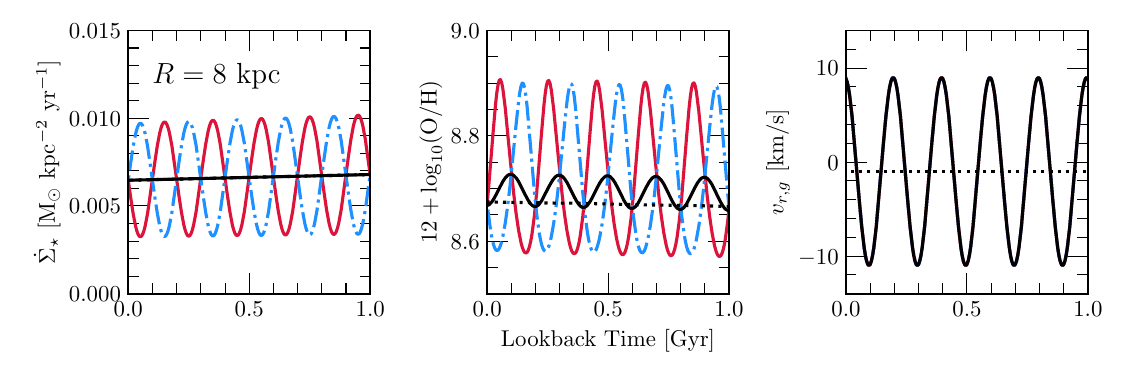}
\caption{
The impact of high-amplitude, short-period variability in ISM radial velocities and SFRs, described in Sections \ref{sec:gce:sfh:oscil} and \ref{sec:gce:scenarios:constant}.
\textbf{Top}: The predicted MDF of stars in the solar annulus (Galactocentric radius $R = 7 - 9$ kpc and mid-plane distance $\left|z\right| \leq 0.5$ kpc) in comparison to our sample (see Section \ref{sec:data}).
\textbf{Bottom Row}: The surface density of star formation (left), the gas-phase O abundance (middle), and the ISM radial velocity (right) at $R = 8$ kpc over the last 1 Gyr of evolution.
\textbf{Summary}: High-amplitude, short-period variability in the radial flow velocity influences the local metallicity only at the $\sim$$0.05$ dex level, whereas departures from a smooth SFH on similarly short timescales lead to variations on the $\sim$$0.3$ dex level.
The primary observational signature is a broadened and more symmetric MDF, in better agreement with observations.
}
\label{fig:oscil}
\end{figure*}

\begin{figure*}
\centering
\includegraphics[scale = 1]{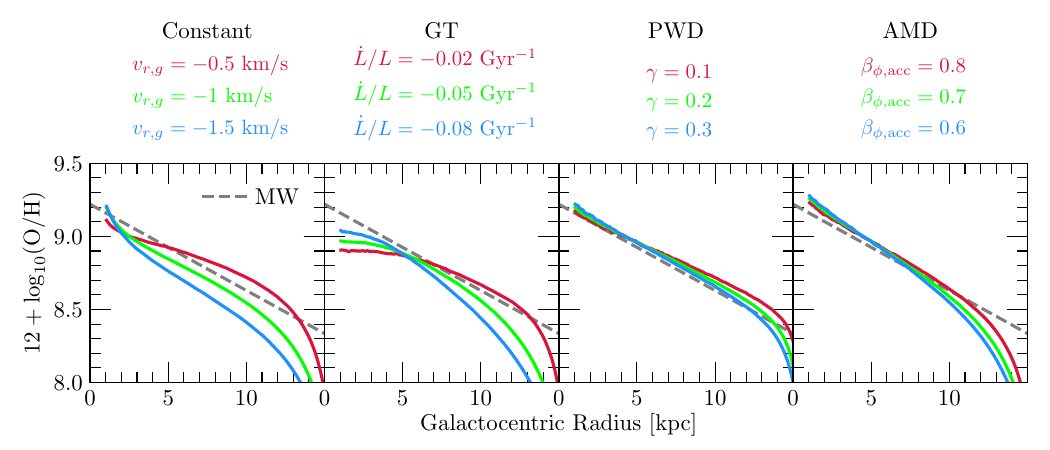}
\caption{
The generic effects of inward gas flows in GCE models.
Each panel shows present day metallicity profiles predicted by our fiducial GCE models: the constant (left), GT (middle left), PWD (middle right), and AMD (right) scenarios (see discussion in Section \ref{sec:gce:scenarios}).
Solid lines show the prediction for a given choice in the corresponding free parameter, color coded according to the legends above each panel.
The grey dashed line shows the profile in \hii\ regions measured by \citet{MendezDelgado2022}.
\textbf{Summary}: Every GCE model using inward gas flows leads to lower metal abundances outside $R \gtrsim 3-5$ kpc and steeper radial metallicity gradients.
The quantitative details and scale of the effect differ between scenarios, but the effect is always present.
}
\label{fig:gradsteepen}
\end{figure*}

In this section, we compare our model predictions to trends in metallicity with stellar population age traced by our sample (see Section \ref{sec:data}).
We sort the stellar populations in each model into bins of age and Galactocentric radius with widths of 1 Gyr and 1 kpc, respectively.
Following our recommendations in \citetalias{Johnson2025}, we compare abundances between our models and our data using the peak of the metallicity distribution function (MDF; i.e., the mode) in each age-radius bin.
This summary statistic is less affected by the radial migration of stars than the mean or median, which are sensitive to the tails of the metallicity distribution.
We fit a skew-normal distribution to the abundances in each age-radius bin and determine the position of the mode through optimization.
We used the same procedure to compute the positions of MDF peaks in \citetalias{Johnson2025}.
We compute the summary statistic at all radii and ages in our models, but for our sample, we only do so if the age-radius bin has at least 200 stars.
\par
Figure \ref{fig:gradoh-agebins} shows the results of this procedure for the [O/H]\footnote{
    We follow conventional notation, where [X/Y] $\equiv \log_{10} (N_X / N_Y) - \log_{10}(N_X / N_Y)_\odot$.
} abundances predicted by each of our models and in our sample.
The empirical trend in the upper-right panel is identical to the result shown in \citetalias{Johnson2025}.
The model trends closely reflect the evolution in the ISM metallicity (see Figure \ref{fig:profile-evol-comp}), but this comparison incorporates the effects of the radial migration of stars (see discussion in Section \ref{sec:gce}).
The $v_{r,g} = 0$ scenario predicts a metallicity profile that is noticeably shallower than observed at all ages.
The PWD and AMD scenarios mitigate this issue at all radii, while the contant velocity and GT scenarios are in agreement with the observed slope at $R \gtrsim 6 - 7$ kpc.
However, the observed metallicity profile exhibits surprisingly minimal decline in its normalization with age.
This realization is the line of evidence for the equilibrium scenario that we identified in \citetalias{Johnson2025} (see discussion in Section \ref{sec:intro}).
In contract, the $v_{r,g} = 0$, PWD, and AMD scenarios exhibit uniform growth by $\sim$$0.2 - 0.3$ dex at all radii across this age range.
The most obvious tensions between our models and our data are generally in the $\tau = 8 - 9$ Gyr age bin, where abundances are often underpredicted.
\par
In Figure \ref{fig:gradoh-agebins}, the constant velocity and GT scenarios are in broad agreement with our sample at $R \gtrsim 6$ kpc but underpredict abundances at smaller radii.
To more directly compare our models to our sample, we visualize the predictions as dotted lines at radii and ages where our sample does not have sufficient size to compute the MDF peak reliably (see discussion above).
The variations in metallicity with age at $R \lesssim 5$ kpc in the constant and GT models would likely not be measured, since these predictions arise in regions where our sample is limited by its size.
However, the predicted metallicity profiles in these regions are noticeably flatter than observed anyway.
Strong age trends at small $R$ did not arise in our previous models in \citetalias{Johnson2025} incorporating ejection from the ISM due to feedback.
Those models predicted no substantial decline in stellar metallicity at all Galactocentric radii up to ages of $\sim$$8$ Gyr.
We demonstrate below that inward gas flows, on their own, should never lead to age-independent stellar metallicities across the entire disk.
\par
Figure \ref{fig:gradfeh-agebins} shows the same comparison as Figure \ref{fig:gradoh-agebins} but for [Fe/H].
Like the O abundances, the empirical trend is also identical to our result from \citetalias{Johnson2025} (see discussion above).
The Fe abundances lead to similar conclusions as O, but the agreement between our models and our sample is noticeably worse.
Obvious tensions now arise not only in the $\tau = 8 - 9$ Gyr age bin but for $\tau = 6 - 7$ Gyr old stars as well.
This difference between O and Fe arises due to the time-delays associated with SNe Ia, which produce most of the Fe in the Universe but negligible amounts of O (see discussion in Section \ref{sec:gce}).
These delays make it challenging to reproduce the solar and super-solar abundances observed at $R \lesssim 8$ kpc in $\tau = 8 - 9$ Gyr old populations.
These discrepancies can be mitigated by, e.g., pre-enriched accretion or an inverse metallicity dependence to the rates of all SNe \citep{Gandhi2022, Johnson2023c, Pessi2023}, both of which act to quicken abundance growth at early times (see discussion in \citetalias{Johnson2025}).
Considering predictions for both O and Fe in Figures \ref{fig:gradoh-agebins} and \ref{fig:gradfeh-agebins}, our $v_{r,g} = -1$ km/s model is in the least amount of tension with our sample from \citetalias{Johnson2025}, followed closely by our GT scenario.
Other models are disfavored for overpredicting the strength of age-metallicity trends in stars.
\par
Figure \ref{fig:eq-conditions} describes the origin of the differences between our fiducial models.
The top panels show the flow coefficients, $\mu_\text{O}$ and $\mu_\text{g}$, as functions of Galactocentric radius at the present day.
These quantities describe the rate of change in the local surface densities of gas and metals due solely to the radial gas flow in terms of the local SFR (see Equation \ref{eq:mu-definition} and discussion in Section \ref{sec:gce}).
The inward flow acts to increase the ISM surface density (i.e., a source) when $\mu > 0$ and to decrease surface density (i.e., a sink) when $\mu < 0$.
There are noticeable differences of detail in $\mu_\text{O}$ as a function of radius in each of our models, but the predictions are qualitatively similar.
The flow acts as a significant source of metals at $R \lesssim 4 - 6$ kpc and a significant sink at larger radii.
Our models exhibit more substantial differences in $\mu_g$ as a function of Galactocentric radius.
The effects of inward flows on the hydrogen surface density therefore influence metallicity evolution much more significantly than the effects on metal surface densities.
\par
In Appendix \ref{sec:analytic:flow-coefficients:mass-conservation}, we demonstrate that mass conservation imposes an important requirement on $\mu_x$ and $\mu_g$.
Namely, the flow cannot lower surface densities in one region of the Galaxy without increasing them elsewhere, and vice versa.
This requirement arises because radial flows, by definition, only affect matter that is already contained within the Galactic disk and are therefore forbidden from changing the total mass of the ISM.
In detail, any model of radial gas flows must obey the following conditions for both forms of $\mu$:
\begin{subequations}
\begin{align}
\int_0^\infty \mu_x Z_x \dot\Sigma_\star R dR &= 0
\label{eq:mu-x-closure}
\\
\int_0^\infty \mu_g \dot\Sigma_\star R dR &= 0.
\label{eq:mu-gas-closure}
\end{align}
\label{eq:mu-closure}%
\end{subequations}
Our numerical models obey these conditions implicitly because the gas exchange algorithm in \vice\ only adds mass to a given zone if it removes the same amount from another.
These integrals extend up to $R \rightarrow \infty$ because, theoretically, the disk is allowed to extend to arbitrarily large radii.
In practice, however, there is an implicit cutoff at the edge of the disk, which is found at $R = 20$ kpc in our numerical models (see discussion in Section \ref{sec:gce:sfh}).
\par
These conditions are the reason why all of our models predict separate regions with $\mu > 0$ and $\mu < 0$ in Figure \ref{fig:eq-conditions}.
The only apparent exception is $\mu_g$ in the ORA limit, in which the radial gas flow replaces accretion as the source of gas everywhere within the Galactic disk, necessitating $\mu_g > 0$.
However, this model also predicts a discontinuity in $\mu_g$ in the outermost annulus at $R = 19.9 \rightarrow 20$ kpc.
By construction, this annulus always loses gas to the radial flow (i.e., $\mu_g< 0$) and gains gas through accretion instead of an ``outer neighbor'' (see Figure \ref{fig:schematic}).
With $\mu_g \ll 0$, this outermost annulus accounts for all other radii.
\par
In the bottom row of Figure \ref{fig:eq-conditions}, we describe how these differences in $\mu$ between models lead to different enrichment timescales.
The bottom-left panel shows the turnover timescale\footnote{
    This timescale has multiple names in the literature.
    In \citetalias{Johnson2021}, we referred to this quantity as the ``depletion time,'' following \citet{Weinberg2017}.
    In \citetalias{Johnson2025}, we switched terminology to use ``processing time,'' since observational work often refers to our definition of $\tau_\star$ as a ``depletion time'' \citep[e.g.,][]{Leroy2008, Tacconi2018}.
} in the ISM, defined as
\begin{equation}
\tau_\text{ISM} \equiv \frac{\tau_\star}{
    1 + \eta - \mu_g - r
},
\label{eq:tau-ism-def}
\end{equation}
as a function of Galactocentric radius.
As the name implies, this timescale describes the interval on which the Galaxy exhausts the local gas supply, replacing it with an entirely new set of baryons through the combination of star formation, accretion, ejection, and radial flows.
Our $v_{r,g} = -1$ km/s model acts as a sink ($\mu_g < 0$) across much of the Galactic disk ($R \gtrsim 6$ kpc), which lowers $\tau_\text{ISM}$.
Relative to the $v_{r,g} = 0$ base model, $\tau_\text{ISM}$ is shorter at $R \gtrsim 3$ kpc.
The weak central ejection included in this model (see discussion in Section \ref{sec:gce:scenarios:constant}) also contributes slightly to lowering $\tau_\text{ISM}$ in the $R \approx 3 - 6$ kpc range.
Our other radial flow models act as weak sources as far out in the disk as $R \approx 10 - 12$ kpc.
As a result, the GT scenario has minimal effect on $\tau_\text{ISM}$ in the $R \approx 5 - 10$ kpc range, while the PWD and AMD scenarios lengthen $\tau_\text{ISM}$ slightly.
Each scenario transitions to the $\mu_g < 0$ regime at sufficiently large $R$, which lowers $\tau_\text{ISM}$ relative to the $v_{r,g} = 0$ limit.
\par
For extended SFHs (such as our own; see discussion in Section \ref{sec:gce:sfh}), $\tau_\text{ISM}$ approximately sets the timescale on which the local metallicity equilibrates (see discussion in Section \ref{sec:discussion:local-enrich}), according to
\begin{equation}
Z_\text{O} \approx Z_{\text{O},\eq} \left(
1 - e^{-t / \tau_\text{ISM}}\right).
\label{eq:eq-approach-turnover}
\end{equation}
Based on this relationship between $\mu_g$ and $\tau_\text{ISM}$, inward gas flows should lead to rapid chemical equilibration when $\mu_g < 0$.
This outcome is similar to ejection, which always acts to lower surface densities \citep{Weinberg2017}.
Conversely, equilibration should be slowed down in cases where $\mu_g > 0$.
\par
The bottom middle panel of Figure \ref{fig:eq-conditions} illustrates this difference between our models by showing the abundance growth at $R = 8$ kpc over time.
Most of the metal enrichment in the $v_{r,g} = -1$ km/s scenario occurs at early times, with minimal growth at late times.
This prediction is a direct consequence of this flow scenario lowering $\tau_\text{ISM}$ near the Sun.
Like our $v_{r,g} = -1$ km/s model, the tendency for the radial flow to increase surface densities slightly is offset by weak ejection at small radii in the GT scenario (see discussion in Section \ref{sec:gce:scenarios:gt}).
This feature leads to similar turnover timescales and therefore similar abundance growth as the $v_{r,g} = 0$ base model.
The PWD and AMD scenarios, as well as the ORA limit, lead to more significant abundance growth with time because the tendency for the flow to increase surface densities is not offset by any centrally concentrated ejection process.
\par
As discussed above, mass conservation dictates that inward gas flows must always lead to $\mu_g > 0$ in the inner Galaxy to offset $\mu_g < 0$ in the outer Galaxy.
These opposing effects therefore lead to both modes of metal enrichment within the disk: rapid equilibration in one region and slow equilibration in another.
This outcome is the reason why our constant and GT flow velocity models predict negligible trends in metallicity with stellar age in the outer Galaxy but significant trends in the inner Galaxy.
In these models, the $R \gtrsim 6$ kpc region is where the radial flow acts as a sufficiently strong sink ($\mu_g \ll 0$) to reach the local equilibrium abundance within the first $\sim$few Gyr of the disk lifetime.
Ongoing enrichment in the inner disk, however, is a necessary secondary effect due the pile-up of matter in those regions, leading to tension with our sample.
\par
The bottom-right panel of Figure \ref{fig:eq-conditions} shows evolution in the equilibrium O abundance itself at $R = 8$ kpc.
The $v_{r,g} = 0$ base model, our constant velocity scenario, and our GT scenario predict $Z_{\text{O},\eq}$ to be steady with time.
There are some small stochastic variations, which are driven by stars releasing their envelopes back to the ISM beyond their birth radius due to radial migration (see Section \ref{sec:gce}).
The PWD and AMD scenarios, however, each lead to $\sim$$0.3$ dex of growth in $Z_{\text{O},\eq}$.
This outcome arises because these models predict significant evolution in the ISM radial velocities with time.
A change in the ISM radial velocity field causes a change in the chemical equilibrium state (see discussion in Section \ref{sec:discussion}).
The unsteady equilibrium in these models contributes to the abundance growth with time shown in the bottom-middle panel, which is in addition to the ongoing growth driven by $\mu_g > 0$.
The observed lack of substantial variation in metallicity with stellar age (see Figures \ref{fig:gradoh-agebins} and \ref{fig:gradfeh-agebins}) disfavors these models, since the ISM metal abundance will never cease to evolve with time.
\par
We acknowledge that the concept of an equilibrium phenomenon is less applicable if the equilibrium itself is evolving.
However, we retain the nomenclature of ``equilibrium'' throughout this paper, since the equilibrium scenario that we proposed in \citetalias{Johnson2025} is a point of discussion.
Whether or not any given model predicts a stable chemical equilibrium on long timescales is a straightforward diagnostic of whether or not the model is compatible with this scenario.
In Section \ref{sec:discussion}, we show that equilibration also provides a reasonably accurate description of evolution in radial metallicity gradients in our numerical models, even if the overall metallicity never reaches equilibrium.

\subsection{Short Timescale Variability}
\label{sec:results:variability}

As discussed in Section \ref{sec:intro}, Galactocentric radial velocities observed in the ISM vary considerably both between and within galaxies \citep{DiTeodoro2021}.
Simulations predict that this result is a consequence of variability at the $\sim$$10$ km/s level on short timescales ($\sim$$100$ Myr), while time-averaged velocities are of order a $\sim$few km/s \citep[e.g.,][]{Vincenzo2020, Trapp2022}.
Simulations also suggest that galaxy SFRs vary at the $\sim$$50$\% level on similar timescales \citep[e.g.,][]{Hopkins2014, Sparre2017, Feldmann2017, Ma2018}.
In this section, we describe the impact of these evolutionary channels in our GCE models.
Each variations uses our $v_{r,g} = -1$ km/s model as a baseline.
The first variation imposes a $10$ km/s amplitude with a period of $200$ Myr (see Section \ref{sec:gce:scenarios:constant}).
The remaining two also impose 50\% variability in the SFH with the same $200$ Myr period, with one in-phase with the $v_{r,g}$ variations and the other out-of-phase (see Section \ref{sec:gce:sfh:oscil}).
\par
Figure \ref{fig:oscil} shows the predictions of these models.
These high-amplitude, short-period variations cause the ISM metallicity to fluctuate.
The effect is minimal ($\sim$$0.05$ dex) with velocity variability and a smooth SFH.
The resulting impact on the MDF near the Sun is proportionally small.
With oscillations in the SFR, however, the ISM metallicity varies more substantially ($\sim$$0.3$ dex).
As a result, the local MDF broadens more significantly.
The models with these departures from a smooth SFH on short timescales do not substantially affect the position of the MDF peak but broaden the distribution by $\sim$$25$\%.
This broadening is an improvement upon our previous models in \citetalias{Johnson2025}, which predicted distributions narrower than observed.
The agreement is not perfect, but this model is not finely tuned to reproduce the observed MDF.
\par
The small effect from the velocity fluctuations can be understood through the distances the ISM moves over one cycle and the corresponding change in metallicity due to the radial gradient.
A $\sim$$10$ km/s flow traverses $\sim$$1$ kpc in $\sim$$100$ Myr.
With a metallicity gradient of order $\nabla[\text{Z/H}] \approx -0.05$ kpc$^{-1}$ \citep[e.g.,][]{MendezDelgado2022, Otto2025}, a distance of $\sim$$1$ kpc corresponds to a change in metallicity of only $\sim$$0.1$ dex.
However, if the SFR oscillates, the enrichment rate itself varies at a similar magnitude.
The factor of $\sim$$2$ range spanned by the SFR variability corresponds directly to the $\sim$$0.3$ dex amplitude of the metallicity oscillations.

\begin{figure}
\centering
\includegraphics[scale = 0.95]{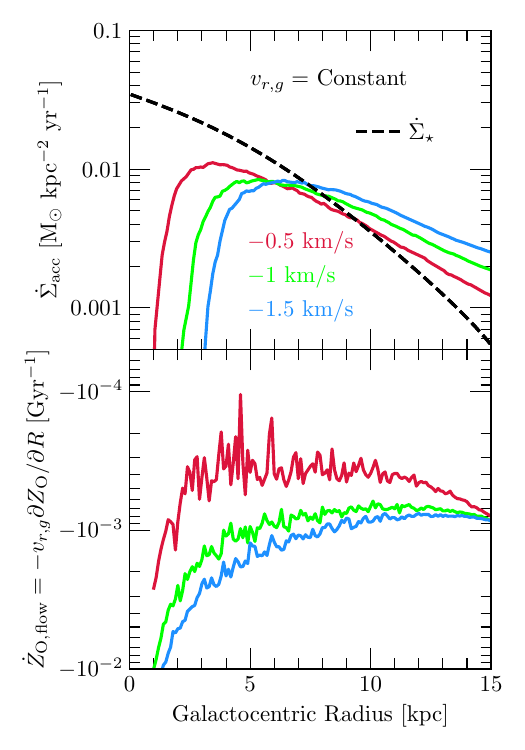}
\caption{
The physical mechanisms behind why the ISM metallicity lowers and why the radial gradient steepens for faster inward flow speeds, showcased for the $v_{r,g} = $ constant scenario.
Lines are color-coded by to the flow velocity according to the legend in the top panel.
\textbf{Top}: The surface density of accreting gas at the present day as a function of Galactocentric radius.
The black dashed line shows the radial profile of the SFR at the present day, which is the same in each model.
\textbf{Bottom}: The instantaneous rate of change in metallicity, $\dot Z_\text{flow}$, as a function of Galactocentric radius due solely to the radial gas flow.
In Section \ref{sec:discussion}, we demonstrate that this rate is given by the product of $v_{r,g}$ and the radial metallicity gradient, $\partial Z_\text{O} / \partial{R}$.
\textbf{Summary}: Inward flows introduce a systematic shift in the Galactocentric radii at which metal-poor gas accretes gravitationally, which raises and lowers metallicity in the inner and outer disk, respectively.
In the presence of a negative metallicity gradient, inward flows drive dilution by carrying low metallicity gas into high metallicity regions.
}
\label{fig:ifrshift-mixing}
\end{figure}

\subsection{Generic Effects: Dilution and Gradient Steepening}
\label{sec:results:dilution-steepening}

In this section, we focus on effects that arise in all of our flow models.
Figure \ref{fig:gradsteepen} shows the present-day radial metallicity profiles in our flow scenarios with different choices of the relevant free parameter.
For comparison, we show the best-fit profile in the observations of \hii\ regions by \citet{MendezDelgado2022}.
In every scenario, the metal abundance outside of $R \sim 3 - 5$ kpc lowers with increasing flow speed, resulting in a steepening of the radial gradient.
The quantitative details vary, but every scenario has this effect, which is in line with previous models of radial gas flows in GCE models (see discussion in, e.g., \citealt{Portinari2000, Spitoni2011}, and \citealt{Bilitewski2012}).
Our constant and GT models predict more substantial differences between parameter choices than PWD and AMD due to faster velocities at low redshift (see Figures \ref{fig:profile-evol-comp} and \ref{fig:vgas-oh-comp}).
We discuss the mathematics behind these outcomes in Section \ref{sec:discussion} below.
\par
Figure \ref{fig:ifrshift-mixing} illustrates the reasons why increasing the flow velocity generically results in lower metallicities and steeper gradients.
We focus on the constant velocity scenario for this comparison, since every model shows similar effects.
The top panel shows the present-day accretion rate as a function of Galactocentric radius for our three different choices of flow speed.
We show the surface density of star formation at the present-day for comparison, which traces the overall stellar suface density (see discussion in Section \ref{sec:gce:sfh}).
With increasing flow speed, metal-poor accretion shifts to large radii.
This effects removes hydrogen from the inner Galaxy and places it in the outer Galaxy, thereby enhancing the differences in chemistry between regions.
This shift in accretion is a consequence of fixing the surface density profiles predicted by each of our models (see Section \ref{sec:gce:sfh}).
If we had instead chosen a fixed accretion history, the SFHs would differ between models, forming more compact systems due to the inward flow.
We would then have to scale our models back up to account for differences in galaxy size.
\par
The bottom panel of Figure \ref{fig:ifrshift-mixing} shows the rate of change in the local metallicity due to the radial gas flow, $\dot Z_\text{flow}$.
We demonstrate in Section \ref{sec:discussion} below that the only direct, \textit{instantaneous} effect of radial gas flows on the local metallicity in our models is a mixing effect, given by
\begin{equation}
\dot Z_{\text{O},\flow} = -v_{r,g}
\partderiv{Z_\text{O}}{R}.
\label{eq:zdotflow}
\end{equation}
The rate of change depends solely on the scale of differences in composition between the gas reservoirs being mixed (i.e., the radial gradient) and how fast the mixing occurs (i.e., the flow velocity).
This functional form arises because the rate goes as $\mu_\text{O} - \mu_g$, since all effects of the radial flow on the gas surface density, $\Sigma_g$, also affect the metal surface density, $\Sigma_\text{O}$.
The relevant terms therefore cancel when computing the effect on the metallicity, $Z_\text{O} \equiv \Sigma_\text{O} / \Sigma_g$.
This result indicates that radial gas flows cannot directly affect the chemistry of the local ISM in the absence of a metallicity gradient ($\dot Z_{\text{O},\flow} = 0$ when $\partial Z_\text{O} / \partial R = 0$).
\par
The bottom panel of Figure \ref{fig:ifrshift-mixing} indicates that $\dot Z_{\text{O},\flow} < 0$ at all radii, as Equation \ref{eq:zdotflow} implies.
This sign uniqueness arises because low metallicity gas funnels into high metallicity regions, which always results in dilution.
The steepening of the radial gradient due the shift in the accretion rates shown in the top panel is an indirect effect in that it only appears in model-to-model comparisons.
This effect therefore does not show up as a term in $\dot Z_\text{flow}$.

%% file: discussion.tex
\section{Analytic Discussion}
\label{sec:discussion}


In this section, we discuss our results in the context of our analytic framework describing abundance evolution under the influence of radial gas flows.
We simply assert the relevant functional forms and focus on qualitative discussion for brevity.
Appendix \ref{sec:analytic} presents detailed mathematical justification.
Table \ref{tab:analytic-summary} provides a summary of the expressions that are particularly useful or insightful.
These analytic solutions are based on the following explicit assumptions:
\begin{enumerate}

    \item The disk is axisymmetric. There are no variations in relevant quantities in the vertical or azimuthal directions, but significant variations occur with Galactocentric radius.

    \item Mixing in the ISM is efficient, occurring on sufficiently short timescales ($\ll 1$ Gyr) to be approximated as instantaneous.
    This assumption applies not only to newly produced metals released to the ISM by stars, but also to gas that is transferred between annuli through the radial flow.

    \item The metals and non-metals follow similar radial velocity distributions at a given location in the disk.

    \item The mean metallicity, local gas supply, and mean ISM radial velocity are accurately described by continuous functions of Galactocentric radius.

\end{enumerate}
Our numerical models rest on the same set of assumptions, allowing an apples-to-apples comparison.
The derivations in Appendix \ref{sec:analytic} start from the same geometric framework as our numerical models, illustrated by Figure \ref{fig:schematic}, but apply the limit that the annulus width and timestep size approach zero (i.e., $\delta R, \delta t \rightarrow 0$).
Consequently, these solutions should accurately describe the evolution of any GCE model with similar geometry.
\par
Under the above assumptions, the flow coefficients $\mu_\text{O}$ and $\mu_g$ have a well-defined form, given by
\begin{subequations}
\begin{align}
\mu_g &= -\tau_\star
\langle v_{r,g} \rangle \left[
\frac{1}{R} +
\partderiv{\ln \Sigma_g}{R} +
\partderiv{\ln \langle v_{r,g} \rangle}{R}
\right]
\label{eq:mu-g-soln}
\\
\mu_\text{O} &= \mu_g - \tau_\star
\langle v_{r,g} \rangle
\partderiv{\ln Z_\text{O}}{R},
\label{eq:mu-o-soln}%
\end{align}
\label{eq:mu-soln}%
\end{subequations}
where $\langle v_{r,g} \rangle$ is the first moment of the Galactocentric radial velocity distribution in the ISM at a given radius.
For brevity, we have referred to this mean radial velocity as simply $v_{r,g}$ throughout this paper, since we do not explore variations in the shape of the radial velocity distribution.
In Appendix \ref{sec:analytic:flow-coefficients}, we introduce $\Sigma_g$, $v_{r,g}$, and $Z_\text{O}$ as arbitrary Taylor series with Galactocentric radius with unknown coefficients.
When taking the limit as $\delta R, \delta t \rightarrow 0$, all higher order derivatives in each quantity vanish.
\par
The direct linear proportionality between $\mu_g$ and $v_{r,g}$ is expected.
As the flow velocity increases, its effect on surface densities increases in proportion.
The factor of $\tau_\star$ arises because the rate of change in surface density due to the flow is tied to local gas supply as opposed to the SFR.
When multiplied by $\dot\Sigma_\star$, the product of the two yields $\Sigma_g$.
The direct radial dependence of $1 / R$ in square brackets arises in combination with the surface density gradient, $\nabla \ln \Sigma_g$.\footnote{
    We use $\nabla \rightarrow \partial / \partial R$ as shorthand.
}
Together, these terms describe the amount of mass present in each annulus, removing the effect of the projected area in determining surface density (i.e., $\nabla \ln M_g = 1 / R + \nabla \ln \Sigma_g$; see Appendix \ref{sec:analytic:flow-coefficients}).
Variations in the radial velocity with disk radius (i.e., $\nabla \ln v_{r,g} \neq 0$) cause gas to enter a given annulus faster or slower than it exits, which in turn raises or lowers the ISM surface density.
$\mu_\text{O}$ depends on each of the same processes as $\mu_g$ but with the additional dependence on the metallicity gradient $\nabla \ln Z_\text{O}$ to account for slight changes in the gas composition between annuli.
\par
In light of our results in Section \ref{sec:results:eq-scenario}, the weak trend observed in metallicity with stellar age arises when $\mu_g \ll 0$ (see Figure \ref{fig:eq-conditions} and discussion in Section \ref{sec:results:eq-scenario}).
Based on Equation \ref{eq:mu-g-soln}, the $\mu_g \ll 0$ regime arises when $\nabla \ln v_{r,g} \ll -\nabla \ln \Sigma_g$.
In Appendix \ref{sec:analytic:velocities:ora}, we show that $\nabla \ln v_{r,g} \approx -\nabla \ln \Sigma_g$ is the regime that arises in the ORA limit (see discussion in Section \ref{sec:analytic:velocities:ora}).
The flat velocity profile, with $\nabla \ln v_{r,g} = 0$ by definition, reaches a chemical quasi-equilibrium at $R \gtrsim 6$ kpc (see Figures \ref{fig:gradoh-agebins} and \ref{fig:gradfeh-agebins}).
No equilibrium arises in the inner Galaxy because the inward flow tends to pile up matter in these regions, which is a necessary consequence of mass conservation (see Equation \ref{eq:mu-closure}).

\input{analytic-summary.tablebody.tex}

Inspection of Equation \ref{eq:mu-soln} indicates that $\mu_g > \mu_\text{O}$ in our models at all radii and times.
This result is a consequence of the term $-\tau_\star v_{r,g} \nabla \ln Z_\text{O}$, which is negative definite if the flow is directed inward ($v_{r,g} < 0$) with a negative metallicity gradient ($\nabla \ln Z_\text{O} < 0$).
This $\mu_g > \mu_\text{O}$ inequality results in the mixing effect that lowers abundances across much of the Galactic disk (see Figures \ref{fig:gradsteepen} and \ref{fig:ifrshift-mixing}).
Inward gas flows are always a stronger source (or equivalently, a weaker sink) of gas than metals.
The instantaneous effect of the radial flow on metallicity can be determined by differentiating $Z_\text{O} = \Sigma_\text{O} / \Sigma_g$ with time and inserting the flow rates, $\dot \Sigma_{\text{O},\flow}$ and $\dot \Sigma_{g,\flow}$, in place of their total time-derivatives, which yields
\begin{equation}
\dot Z_{\text{O},\flow} =
\frac{Z_\text{O}}{\tau_\star}
\left(\mu_\text{O} - \mu_g \right).
\end{equation}
The simple product of two quantities shown in Equation \ref{eq:zdotflow} then arises by plugging in $\mu_\text{O} - \mu_g$ from Equation \ref{eq:mu-o-soln}.
\par
At $R \lesssim 3 - 5$ kpc, our models predict metallicity to increase as opposed to decrease with the radial flow velocity.
This outcome is counterintuitive given the prediction that $\dot Z_{\text{O},\flow} < 0$, as shown in Figure \ref{fig:ifrshift-mixing}.
In these regions, the flow acts as a source of metals ($\mu_\text{O} > 0$), so the flow carries effects similar to metal-rich accretion and acts to raise metal abundances overall (see discussion in Appendix \ref{sec:analytic:zin}).

\subsection{Chemical Enrichment as an Equilibration Process}
\label{sec:discussion:local-enrich}

\begin{figure*}
\centering
\includegraphics[scale = 0.80]{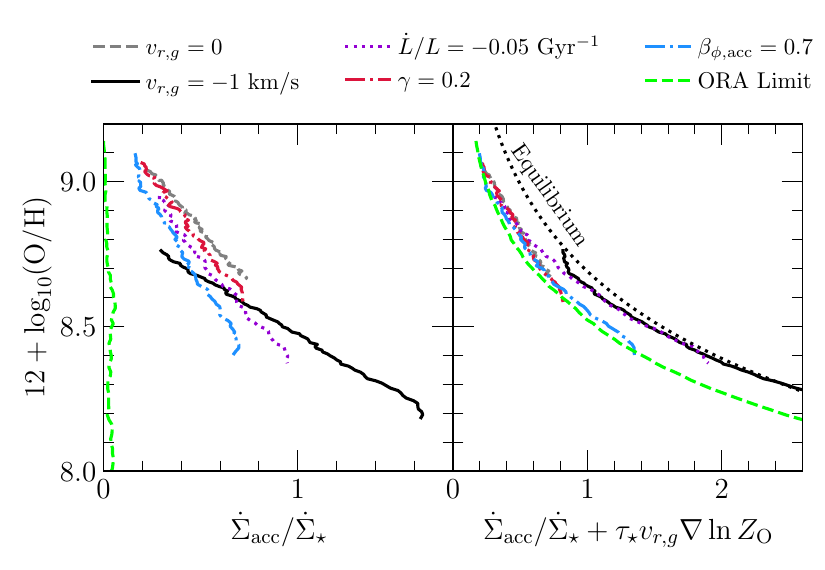}
\includegraphics[scale = 0.82]{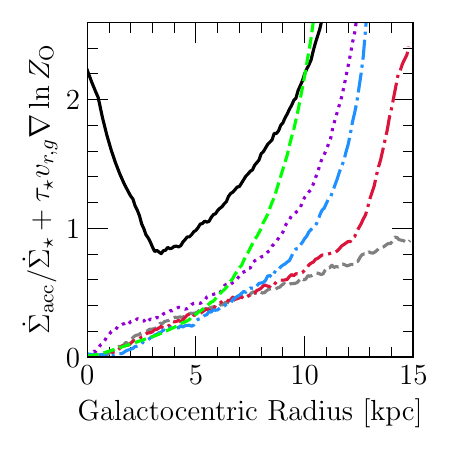}
\caption{
The role of mixing due to inward gas flows in establishing the radial metallicity profile of the Galactic disk.
Individual lines show the predictions of our GCE models, marked according to the legend at the top left.
\textbf{Left}: The gas-phase O abundance as a function of the ratio between star formation and accretion rates, $\dot\Sigma_\text{acc} / \dot\Sigma_\star$.
\textbf{Center}: The same as the left panel, but with an offset of $\tau_\star v_{r,g} \nabla \ln Z_\text{O}$ on the x-axis to account for mixing.
The black dotted line marks the equilibrium state, which is given by the stellar yield divided by the x-coordinate.
\textbf{Right}: The quantity on the x-axis in the center panel as a function of Galactocentric radius.
\textbf{Summary}: Differences in metallicity profiles between models generally reflect differences in predictions regarding $\dot\Sigma_\star / \dot\Sigma_\text{acc}$ and $\tau_\star v_{r,g} \nabla \ln Z_\text{O}$.
}
\label{fig:sfrifr-mixing}
\end{figure*}

The rate of change in the metallicity $\dot Z_\text{O}$ follows from differentiating $Z_\text{O} \equiv \Sigma_\text{O} / \Sigma_g$ with respect to time according to quotient rule.
Substituting in Equation \ref{eq:dot-sigma-gas} for $\dot\Sigma_g$ and Equation \ref{eq:dot-sigma-o} for $\dot\Sigma_\text{O}$ yields the following result:
\begin{equation}
\dot Z_\text{O} =
\frac{\ycc{O}}{\tau_\star} -
\frac{Z_\text{O}}{\tau_\star} \left[
1 + \eta - \mu_\text{O} - r +
\tau_\star \partderiv{\ln \Sigma_g}{t}
\right].
\label{eq:zdot-o}
\end{equation}
At sufficiently late epochs, the SFH across much of the disk declines with time ($R \lesssim 10$ kpc in our models) due to the metal-poor accretion rate slowing down.
With the SFH declining on an e-folding timescale given by $\tau_\sfh$ (see Equation \ref{eq:tausfh}), the gas supply follows a timescale of $\dot \Sigma_g / \Sigma_g = -1 / N\tau_\sfh$.
The equilibrium metallicity then follows from setting $\dot Z_\text{O} = 0$ and solving for $Z_\text{O}$, which yields the form shown by Equation \ref{eq:zoeq} and included in Table \ref{tab:analytic-summary}.
In some of our models, the equilibrium itself is not steady in time, which leads to tension with the weak age-metallicity relation in our sample (see Figure \ref{fig:eq-conditions}).
Inspection of Equation \ref{eq:zoeq} indicates that $\dot Z_{\text{O},\eq} \approx 0$ does not necessarily require $\dot \eta = 0$ or $\dot \mu = 0$, but more generally $\dot \eta \approx \dot \mu$.
A chemical equilibrium could therefore arise if ejection rates and radial flow rates evolve in concert with one another.
\par
A simple approximation for the evolution of $Z_\text{O}$ with time, $t$, follows by assuming that $\tau_\star$ and the linear combination in square brackets in Equation \ref{eq:zdot-o} are constant in time (see Appendix \ref{sec:analytic:eq-gradient:eq-approach}).
This approximation reduces $\dot Z_\text{O}$ to a linear ordinary differential equation, the solution to which is given by
\begin{equation}
Z_\text{O} = Z_{\text{O},\eq} \left(
1 - e^{-t / \tau_\eq}
\right),
\label{eq:z-equilibration}
\end{equation}
where the ``relaxation'' or ``equilibration'' timescale, $\tau_\eq$, is given by
\begin{equation}
\tau_\eq = \left(\frac{
    1 + \eta - \mu_\text{O} - r
}{
    \tau_\star
} - \frac{1}{N \tau_\sfh}
\right)^{-1}.
\label{eq:taueq}
\end{equation}
The approximation shown in Equation \ref{eq:eq-approach-turnover} arises in the limit that $\tau_\text{ISM} \ll N \tau_\sfh$.
Although both $\tau_\star$ and the quantity in square brackets in \ref{eq:zdot-o} generally evolve with time, the assumption that they do not is sufficiently accurate for our analytic approximations in this paper.
The detailed shape of the SFH also influences the approach to equilibrium \citep{Weinberg2017}.
Our \vice\ integrations incorporate these effects, so exact agreement between our analytic and numerical models is not expected.
\par
In \citetalias{Johnson2025}, we showed that the equilibrium metallicity is closely related to the number of stars formed per accreted baryon at a given location in the Galaxy.
The ratio between the rates of accretion and star formation follows from Equation \ref{eq:dot-sigma-gas}:
\begin{equation}
\frac{
    \dot\Sigma_\text{acc}
}{
    \dot\Sigma_\star
} = 1 + \eta - \mu_g - r +
\tau_\star \partderiv{\ln \Sigma_g}{t}.
\label{eq:ifr-per-sfr}
\end{equation}
The right-hand side of this expression becomes the denominator of $Z_{\text{O},\eq}$ by adding $\mu_g - \mu_\text{O}$ to both sides.
Substituting in Equation \ref{eq:mu-o-soln} for $\mu_g - \mu_\text{O}$ and plugging the result into Equation \ref{eq:zoeq} results in the following alternative expression for the equilibrium abundance:
\begin{equation}
Z_{\text{O},\eq} = \ddfrac{\ycc{O}}{
    \frac{
        \dot\Sigma_\text{acc}
    }{
        \dot\Sigma_\star
    } + \tau_\star v_{r,g}
    \partderiv{\ln Z_{\text{O},\eq}}{R}
}.
\label{eq:zoeq-sfrifrmixing}
\end{equation}
We have also applied the substitution $\nabla \ln Z_\text{O} \rightarrow \nabla \ln Z_{\text{O},\eq}$ in the denominator of this expression, since the ISM gradient reflects the equilibrium gradient if the gas is near the equilibrium abundance itself (see discussion in Section \ref{sec:discussion:local-enrich} below).
In the limit that $v_{r,g} \rightarrow 0$, the equilibrium metallicity reduces to the form that we previously derived in \citetalias{Johnson2025}, $Z_{\text{O},\eq} \rightarrow \ycc{O} \dot\Sigma_\star / \dot\Sigma_\text{acc}$.
\par
The left panel of Figure \ref{fig:sfrifr-mixing} shows the present day ISM metallicity between $R = 5$ and $10$ kpc as a function of the ratio $\dot\Sigma_\text{acc} / \dot\Sigma_\star$.
The middle panel adds in the term $\tau_\star v_{r,g} \nabla \ln Z_\text{O}$, which accounts for the effect of mixing due to the radial gas flow, to the x-axis.
We lower the abundances predicted by our $v_{r,g} = -1$ km/s model by $0.2$ dex for the sake of this comparison to account for the slight enhancement in yields in this model (see discussion in Section \ref{sec:gce:scenarios:constant}).
In the left panel, our models follow different trends, but the predictions line up much more noticeably in the middle panel.
The middle panel also shows the relation that arises if the ISM is perfectly in the equilibrium state, which is simply $\ycc{O}$ divided by the x-axis coordinate according to Equation \ref{eq:zoeq-sfrifrmixing}.
The $v_{r,g} = -1$ km/s model is more in line with the equilibrium trend in the outer Galaxy, while other models sit at slightly lower abundances.
This prediction is in line with expectations based on Figure \ref{fig:profile-evol-comp}, which shows minimal change in the ISM metallicity over the last $\sim$$8$ Gyr in the $v_{r,g} = 1$ km/s model due to its proximity to the equilibrium state.
The other models predict abundances that are farther below the equilibrium profile and consequently predict more substantial evolution with time.
This difference plays a key role in the $v_{r,g} = -1$ km/s model reproducing the apparent age-independence of stellar metallicities at $R \gtrsim 8$ kpc (see discussion in Section \ref{sec:results:eq-scenario}), while other models struggle because they predict stronger trends with age.
\par
The consistency between models in the middle panel of Figure \ref{fig:sfrifr-mixing} indicates that metallicity is generically established by mixing and the relative balance of star formation and accretion in our models.
We suspect that this principle should hold for most GCE models, since this comparison already includes six different prescriptions for radial gas flow velocities.
The right panel of Figure \ref{fig:sfrifr-mixing} shows $\dot\Sigma_\text{acc} / \dot\Sigma_\star + \tau_\star v_{r,g} \nabla \ln Z_\text{O}$ as a function of Galactocentric radius.
The models once again differ, indicating that most of the differences in predicted radial abundance profiles between models (see Figure \ref{fig:profile-evol-comp}) arise as a result of differences in $\dot\Sigma_\text{acc} / \dot\Sigma_\star + \tau_\star v_{r,g} \nabla \ln Z_\text{O}$ with radius.
We exploit this relationship mathematically in Section \ref{sec:discussion:eq-gradients} below.
\par
The realization of this link between $\log_{10}$(O/H) and $\dot\Sigma_\text{acc} / \dot\Sigma_\star + \tau_\star v_{r,g} \nabla \ln Z_\text{O}$ highlights the utility of the equilibrium ansatz.
Five of our six models predict present day ISM metallicities to be significantly below $Z_{\text{O},\eq}$ (see Figure \ref{fig:profile-evol-comp}), but these arguments rooted in chemical equilibrium lead to accurate descriptions of the model predictions anyway.
To emphasize this point further, we remark that the notion of chemical equilibrium technically breaks down in the ORA limit, despite its predictions lining up with our other models in the middle panel of Figure \ref{fig:sfrifr-mixing}.
By definition, this scenario asserts $\dot\Sigma_\text{acc} \rightarrow 0$ (see discussion in Section \ref{sec:gce:scenarios:ora-limit}), which reduces Equation \ref{eq:zoeq-sfrifrmixing} to
\begin{equation}
\partderiv{Z_{\text{O},\eq}}{R} \rightarrow \frac{
    \ycc{O}
}{
    \tau_\star v_{r,g}
}.
\end{equation}
$Z_{\text{O},\eq}$ has vanished, so the equilibrium metallicity itself is not clearly defined mathematically, but its radial gradient somehow is (see discussion in Appendix \ref{sec:analytic:scenarios:ora-limit}).
A more accurate description of chemical evolution in the ORA limit requires a new mathematical approach.
This result suggests that accretion is required on some level in order for metal abundances to equilibrate, which is in line with previous GCE models \citep[e.g.,][]{Larson1972}.
In one-zone GCE models omitting metal-poor accretion, all of the hydrogen is eventually fused into metals, so $Z_{\text{O},\eq} \rightarrow \infty$, leading to sufficient numbers of metal-rich stars to create obvious tensions with data (see discussion of so-called ``closed box'' GCE models in, e.g., the review by \citealt{Tinsley1980}).
\par
We emphasize that the equilibrium metallicity is always clearly defined mathematically (see Equation \ref{eq:zoeq}).
However, whether the ISM actually reaches the equilibrium within the disk lifetime, or if it would even be physical to do so, is a different question.
$Z_{\text{O},\eq}$ may be sufficiently large, even infinite, such that metallicity generally increases monotonically over the disk lifetime.
The equilibrium abundance may also be zero or negative.
Such models are not unphysical; close inspection of Equations \ref{eq:zoeq} and \ref{eq:taueq} indicates that $\tau_\eq < 0$ when $Z_{\text{O},\eq} < 0$, which then results in $Z_\text{O}(t) \geq 0$ for all $t \geq 0$ in Equation \ref{eq:z-equilibration} (see discussion in \citealt{Weinberg2017}).

\subsection{Equilibrium Metallicity Gradients}
\label{sec:discussion:eq-gradients}

\begin{figure*}
\centering
\includegraphics[scale = 1]{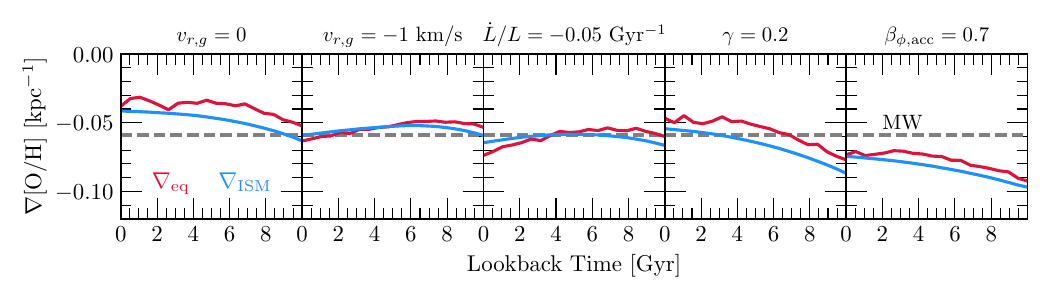}
\caption{
The evolution in the ISM (blue) and equilibrium (red) radial metallicity gradients in our GCE models with fiducial parameter choices (see discussion in Section \ref{sec:gce:scenarios}), shown in individual panels and labeled along the top.
The equilibrium gradient, $\nabla_\eq$[O/H], reflects the slope at which the gradient will not evolve (i.e., $\dot\nabla[\text{O/H}] = 0$) and is established through a decline in the $\dot\Sigma_\star / \dot\Sigma_\text{acc}$ ratio with radius and shaped by mixing effects (see discussion in Section \ref{sec:discussion:eq-gradients}).
The grey dashed line in each panels shows the metallicity gradient in Galactic \hii\ regions measured by \citet{MendezDelgado2022}.
\textbf{Summary}: In all models, the ISM radial metallicity profile tilts with time toward an equilibrium slope, which is generally not stationary.
}
\label{fig:gradism-gradeq-evol}
\end{figure*}

We demonstrate in Appendix \ref{sec:analytic:eq-gradient} that the rate of change in the radial metallicity gradient can be expressed as
\begin{equation}
\begin{split}
\dot\nabla \ln Z_\text{O} &= \frac{
    \ycc{O}
}{
    Z_\text{O} \tau_\star
} \Bigg[
\left(1 - e^{-t / \tau_\eq}\right)
\partderiv{\ln Z_{\text{O},\eq}}{R}
\\
&\qquad -
e^{-t / \tau_\eq}
\partderiv{\ln \tau_\star}{R} -
\partderiv{\ln Z_\text{O}}{R}
\Bigg].
\label{eq:tilt-rate}
\end{split}
\end{equation}
This expression follows from differentiating Equation \ref{eq:zdot-o} with radius and applying the symmetry of second derivatives.
Following the notion that metallicity evolves minimally near the equilibrium abundance, we now define an equilibrium radial gradient, $\nabla \ln Z_\text{O}$, at which the rate of tilt is minimal (i.e., $\dot \nabla \ln Z \approx 0$ at $\nabla \ln Z \approx \nabla_\eq \ln Z$).
Setting the above expression equal to zero yields a straightforward solution to $\nabla \ln Z_\text{O}$, which by definition should equal $\nabla_\eq \ln Z_\text{O}$:
\begin{equation}
\begin{split}
\nabla_\eq[\text{O/H}] &= \frac{1}{\ln 10}
\Bigg[
\frac{Z_\text{O}}{Z_{\text{O},\eq}}
\partderiv{\ln Z_{\text{O},\eq}}{R}
\\
&\qquad -
\left(1 - \frac{Z_\text{O}}{Z_{\text{O},\eq}}\right)
\partderiv{\ln \tau_\star}{R}
\Bigg],
\label{eq:gradeq}
\end{split}
\end{equation}
where the factor of $\ln 10$ arises from transforming between natural logarithms (e.g., $\nabla \ln Z$) to the base-10 logarithms that abundances are typically quantified with (e.g., [O/H]).
\par
At early times, $Z_\text{O} \ll Z_{\text{O},\eq}$; in this limit, $\nabla_\eq \ln Z_\text{O} \rightarrow \nabla \ln \tau_\star$.
High redshift radial metallicity gradients should therefore reflect variations in SFE with radius according to this parameterization.
As expected, the gradient generically evolves toward a slope that reflects the equilibrium metallicity profile (i.e., $\nabla_\eq \ln Z \rightarrow \nabla \ln Z_\eq$ as $Z \rightarrow Z_\eq$).\footnote{
    The term ``equilibrium gradient'' could refer to either $\nabla_\eq \ln Z_\alpha$ or $\nabla \ln Z_{\alpha,\eq}$.
    The distinction is quantitatively important but generally insignificant for interpretive purposes.
    In practice, differentiating between the two is often of little consequence.
    We therefore refrain from introducing new terminology, using ``equilibrium gradient'' interchangeably and referring explicitly to $\nabla_\eq \ln Z$ and $\nabla \ln Z_\eq$ when necessary.
}
A phase transition with significant steepening or shallowing on $\sim$Gyr timescales should therefore arise if $\nabla \ln Z_{\text{O},\eq} \neq \nabla \ln \tau_\star$.
In some hydrodynamic simulations, flat metallicity profiles arise at high redshift not necessarily because $\nabla \ln \tau_\star \approx 0$, but because of a substantial radial velocity dispersion in the ISM, leading to efficient mixing \citep{Graf2024}.
These origins of flat metallicity gradients are not mutually exclusive.
\par
We now demonstrate the concept of the equilibrium metallicity gradient in action in our fiducial numerical models.
We determine the ISM gradients in our \vice\ models by fitting an exponential profile to $Z_\text{O}$ as a function of Galactocentric radius between $R = 5$ and $10$ kpc; the gradient then follows from the scale length of the exponential.
To determine the equilibrium gradients, we first compute $Z_{\text{O},\eq}$ and $\tau_\star$ in each annulus between $R = 5$ and $10$ according to Equation \ref{eq:zoeq} and the definition of the SFE timescale (see Table \ref{tab:glossary}).
We then fit exponential profiles to these quantities and use their scale radii to determine $\nabla_\eq$[O/H] according to Equation \ref{eq:gradeq}.
Figure \ref{fig:gradism-gradeq-evol} shows the evolution of the ISM and equilibrium gradients over time, computed by applying this procedure to snapshots of our models at a range of lookback times.
As expected, all models predict $\nabla_\text{ISM}$[O/H] to evolve toward $\nabla_\eq$[O/H], which is quasi-stationary in some models and significantly evolving in others.
Evolution in the ISM gradient generally reflects evolution in the underlying equilibrium gradient.
The ISM gradient also briefly pauses evolution whenever $\nabla_\text{ISM}[\text{O/H}] \approx \nabla_\eq[\text{O/H}]$, as expected.

\begin{figure*}
\centering
\includegraphics[scale = 1]{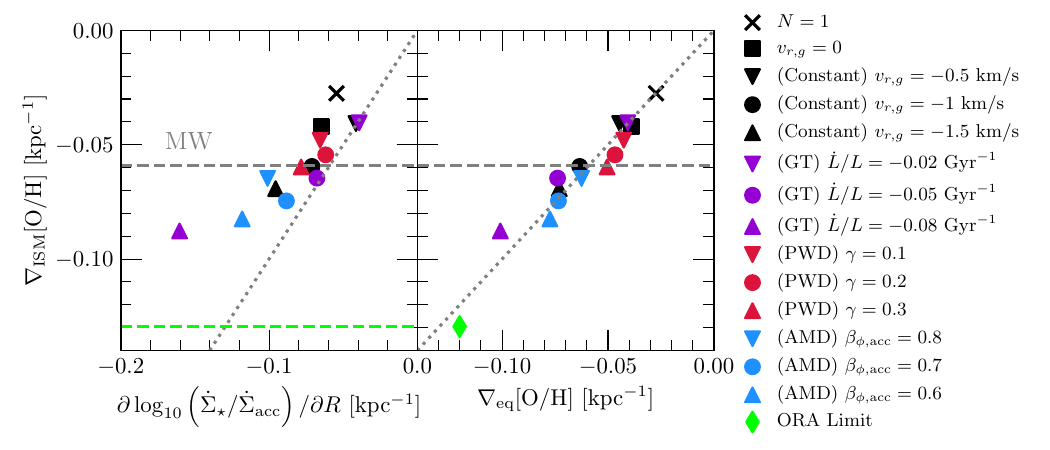}
\caption{
The metallicity gradient in the ISM predicted by each of our models as a function of the radial gradient in the $\dot\Sigma_\star / \dot\Sigma_\text{acc}$ ratio (left) and the equilibrium metallicity gradient (right; see discussion in Section \ref{sec:discussion:eq-gradients}).
Individual radial flow scenarios are color-coded, with our fiducial parameter choices shown as circles and variations as triangles according to the legend on the right.
The ORA limit represents a discontinuity in the left panel ($\dot\Sigma_\text{acc} \rightarrow 0$; see discussion in Section \ref{sec:gce:scenarios:ora-limit}) and is therefore marked as a dashed line instead.
The black square shows the base model with no radial flows, and the black X shows the variation thereof with uniform SFE (i.e., $N = 1$; see discussion in Section \ref{sec:gce}).
Dotted grey lines mark the one-to-one relation in both panels, while dashed grey lines mark the gradient in Galactic \hii\ regions measured by \citet{MendezDelgado2022}.
\textbf{Summary}: In the presence of inward gas flows, radial metallicity gradients in the ISM at the present day closely trace the equilibrium slope, which is established through a decline in the $\dot\Sigma_\star / \dot\Sigma_\text{acc}$ ratio with radius and shaped by mixing effects (see discussion in Section \ref{sec:discussion:eq-gradients}).
}
\label{fig:gradism-vs-gradeq}
\end{figure*}

Figure \ref{fig:gradism-gradeq-evol} indicates that the radial gradient may steepen, remain steady, or shallow with time depending on the prescription for the ISM radial velocity.
In the PWD and AMD scenarios, the ISM gradient shallows significantly.
This outcome is driven by the radial flow slowing down with time, making their low-redshift evolution resemble the $v_{r,g} = 0$ limit more closely than their high-redshift evolution.
Our constant and GT models exhibit minimal tilt in the radial metallicity profile with time.
Evolution at these levels would be difficult to measure empirically, likely resulting in flat trends within uncertainties.
We note that these scenarios lead to gradients that instead steepen significantly with time if the flow is somewhat faster, of order $v_{r,g} \lesssim -1.5$ km/s or $\dot L / L \approx -0.1$ Gyr$^{-1}$.
\par
Based on these results, accurate measurements of the evolution of the ISM metallicity gradient with time in the MW should provide useful constraints on radial gas flows.
However, there are conflicting arguments in the literature.
Some authors argue that the gradient has shallowed significantly with time \citep[e.g.,][]{Anders2023, Ratcliffe2023, Lu2024b}.
Others argue that the gradient has steepened \citep[e.g.,][]{Vickers2021, Lian2023}.
Others find no significant evidence of steepening or shallowing \citep[e.g.,][]{Willett2023, Otto2025}.
Our sample exhibits some slight variations in slope with stellar population age ($\sim$$0.03$ kpc$^{-1}$) but remains near $\nabla[\text{O/H}] \approx -0.06$ kpc$^{-1}$ between ages of $\tau \sim 0$ and $\tau \sim 9$ Gyr \citepalias{Johnson2025}.
\citet{Hu2025} recently presented a comparison of a large suite of measurements based on old open clusters (see figure 7 therein).
Considering all of the measurements collected therein equally, the data appear consistent with a flat trend.
However, there is substantial scatter, indicating that these inferences of $\nabla \ln Z$ with time are subject to substantial systematic uncertainties.
Observations of metallicity gradients across redshift tell a similar precision-limited story (e.g., \citealt{Jones2010, Jones2013, Yuan2011, Leethochawalit2016, Wuyts2016, Ju2025}; see discussion in, e.g., \citealt{Ma2017}).
\par
Figure \ref{fig:gradism-vs-gradeq} shows the close relationship between the ISM metallicity gradient, the gradient in the $\dot\Sigma_\star / \dot\Sigma_\text{acc}$ ratio, and the equilibrium gradient.
All models follow a common trend in both panels.
In qualitative agreement with our arguments in \citetalias{Johnson2025}, the ISM metallicity gradient generally reflects a decline in the number of stars formed per accreted baryon in a given region of the Galaxy.
However, our models therein include ejection of ISM gas to the CGM as opposed to inward gas flows.
In those models, the relationship in the left panel of Figure \ref{fig:gradism-vs-gradeq} is linear, since the equilibrium abundance depends linearly on $\dot\Sigma_\star / \dot\Sigma_\text{acc}$ (see Equation \ref{eq:zoeq-sfrifrmixing} in the limit that $v_{r,g} \rightarrow 0$).
The offset from a linear relation in the left panel is due to the mixing effect induced by the inward flow of gas.
Section \ref{sec:discussion:full-solutions} below offers a natural explanation for the seemingly parabolic shape: $\nabla \ln Z_{\text{O},\eq}$ varies with the square root of $\nabla (\dot\Sigma_\text{acc} / \dot\Sigma_\star)$.
The relationship in the right hand panel is linear because the equilibrium gradient, under our definition, accounts for the effect of mixing.
\par
Figure \ref{fig:gradism-vs-gradeq} also highlights how the ISM metallicity gradients are influenced by variations in SFE with radius and the specific parameter choices regarding radial flows.
The black X shows a base model with no radial gas flows and a constant $\tau_\star$ with radius (i.e., $N = 1$), which we achieve by simply lowering the surface density threshold above which the ISM is entirely in the molecular phase down to $\Sigma_g \geq 10^6$ \msun\ kpc$^{-2}$ (see discussion in Section \ref{sec:gce}).
This model predicts the shallowest metallicity gradient, in line with our analytic arguments.
A slightly steeper gradient arises when the Galaxy follows the observed Kennicutt-Schmidt relation with $N = 1.5$ \citep[e.g.,][]{Schmidt1959, Schmidt1963, Kennicutt1998}.
However, gradients in line with the observations arise in our models with radial gas flows.
Reproducing the observed slope without radial flows would require stronger variations in SFE with radius than in our models or a specialized prescription for gas ejection from the ISM, as in \citetalias{Johnson2025}.
This result is in line with the findings of \citet{Palla2020}.
Our $v_{r,g} = -0.05$ km/s and $\dot L/L = -0.02$ Gyr$^{-1}$ models predict slopes comparable to the $v_{r,g} = 0$ model, which is a consequence of the weak central ejection in these models (see discussion in Section \ref{sec:gce}).
This prescription, necessary for these models to obey mass conservation in the inner disk, flattens the radial gradient slightly in those regions.
\par
Based on the left panel of Figure \ref{fig:gradism-vs-gradeq}, we can infer the scale radius with which matter accretes onto the MW in a relatively model-independent way.
Collectively, our models suggest that a gradient of $\nabla[\text{O/H}] \approx -0.059$ kpc$^{-1}$, as measured in \hii\ regions by \citet{MendezDelgado2022}, should arise when $\nabla \log_{10} (\dot\Sigma_\star / \dot\Sigma_\text{acc}) \approx -0.07$ kpc$^{-1}$.
If $\dot\Sigma_\star$ declines exponentially with a scale radius of $r_\star \sim 2.5$ kpc, based on the observed profile in $\Sigma_\star$ (\citealt{BlandHawthorn2016}; see discussion in Section \ref{sec:gce:sfh}), then this condition is met if $\dot\Sigma_\text{acc}$ follows a scale radius of $r_\text{acc} \sim 4.2$ kpc.
This implication appears to be model-independent.

\subsection{Analytic Predictions}
\label{sec:discussion:full-solutions}

\begin{figure*}
\centering
\includegraphics[scale = 0.78]{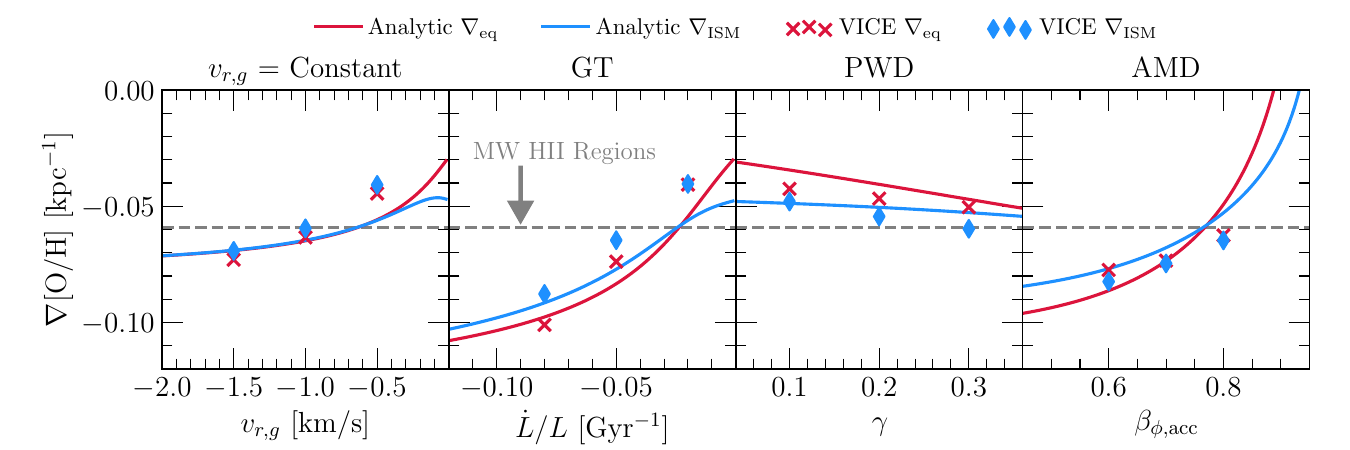}
\caption{
A comparison between our analytic and numerical estimates of metallicity gradients.
Each panel shows the predictions for a given scenario, noted along the top, as a function of the relevant free parameters on the x-axis (see discussion in Section \ref{sec:gce:scenarios}).
Points denote numerical estimates from our VICE integrations, and lines denote analytic estimates.
Equilibrium gradients are shown in red, and ISM gradients are shown in blue.
\textbf{Summary}: Our analytic solutions describe the \vice\ predictions to a typical accuracy of $\nabla[\text{O/H}] \pm \sim$$0.01$ kpc$^{-1}$.
}
\label{fig:gradients-vs-vice}
\end{figure*}

In Appendix \ref{sec:analytic:eq-gradient}, we show that $\nabla \ln Z_{\text{O},\eq}$ can be approximated as a solution to a quadratic equation, according to
\begin{subequations}
\begin{align}
\partderiv{\ln Z_{\text{O},\eq}}{R} &=
\frac{
    -\xi_1 + \sqrt{\xi_1^2 + \xi_2}
}{
    2 \tau_\star v_{r,g}
}
\label{eq:gradeq-quadratic}%
\\
\xi_1 &= \frac{
    \dot\Sigma_\text{acc}
}{
    \dot\Sigma_\star
} + \tau_\star v_{r,g}\left(
\partderiv{\ln \tau_\star}{R} +
\partderiv{\ln v_{r,g}}{R}
\right)
\\
\xi_2 &= -4 \tau_\star v_{r,g}
\partderiv{}{R}\left(\frac{
    \dot\Sigma_\text{acc}
}{
    \dot\Sigma_\star
}\right).
\end{align}
\label{eq:gradeq-quadratic-xi1-xi2}%
\end{subequations}
This form arises by differentiating Equation \ref{eq:zoeq} with Galactocentric radius and assuming that the second-derivative, $\partial^2 \ln Z_{\text{O},\eq} / \partial R^2$, is negligible.
This approximation yields a broadly applicable solution that allows one to simply plug in the functional forms of different radial flow prescriptions.
One could also solve Equation \ref{eq:zoeq-sfrifrmixing} as a differential equation, but this approach requires a unique set of steps to a solution for each individual flow model.
Appendix \ref{sec:analytic:scenarios} provides solutions for $\xi_1$ and $\xi_2$ for each of our radial flow scenarios described in Section \ref{sec:gce:scenarios}.
\par
Appendix \ref{sec:analytic:eq-gradient:eq-approach} demonstrates that Equation \ref{eq:tilt-rate} for $\dot \nabla \ln Z_\text{O}$ can be rewritten as
\begin{equation}
\dot \nabla \ln Z_\text{O} \approx \frac{
    \nabla_\eq \ln Z_\text{O} -
    \nabla \ln Z_\text{O}
}{
    \tau_\nabla
},
\label{eq:tilt-approx}
\end{equation}
where
\begin{equation}
\tau_\nabla \equiv \left(
\frac{1}{\tau_\eq} +
v_{r,g} \partderiv{\ln Z_{\text{O},\eq}}{R}
\right)^{-1}.
\label{eq:tau-nabla-def-orig}
\end{equation}
These expressions describes an equilibration process, wherein the rate of evolution is specified by the displacement from the equilibrium and a characteristic timescale.
Equation \ref{eq:tau-nabla-def-orig} indicates that the radial gradient should equilibrate faster than the ISM ($\tau_\nabla < \tau_\eq$) in the presence of inward gas flows and a negative radial gradient ($v_{r,g} < 0$; $\nabla \ln Z_{\text{O},\eq} < 0$).
Figure \ref{fig:gradism-vs-gradeq} indicates that the ISM gradient generally traces the equilibrium gradient in our models as a consequence, even if the overall metallicity is significantly below $Z_\eq$ (see discussion in Section \ref{sec:discussion:eq-gradients} above).
The closest approximation of present-day radial gradients that we identify applies a simple time offset from the equilibrium.
In principle, $\nabla_\text{ISM}$ should lag behind $\nabla_\eq$ by an interval of order $\sim$$\tau_\nabla$:
\begin{equation}
\nabla_\text{ISM}[\text{O/H}](t \rightarrow \tau_\text{disk}) \approx
\nabla_\eq[\text{O/H}](t - \tau_\nabla),
\label{eq:gradism-taunabla-offset}
\end{equation}
This offset simply accounts for the characteristic lag as the ISM gradient evolves toward its equilibrium slope.
\par
We derive analytic expressions for the coefficients $\xi_1$ and $\xi_2$, defined by Equation \ref{eq:gradeq-quadratic-xi1-xi2}, in Appendix \ref{sec:analytic:scenarios}.
Solutions to the equilibrium and present-day ISM gradients then follow by simply plugging in our parameter values and the relevant flow velocity prescription from Section \ref{sec:gce:scenarios}.
We use $\tau_\star = 4$ Gyr as a fiducial value at $R = 8$ kpc, which is appropriate for an ISM that is split into roughly equal parts molecular H$_2$ and warm \hi.\footnote{
    $\tau_\star = (2 \text{ Gyr}) / f_\text{mol}$.
    $f_\text{mol} = \text{H}_2 / (\text{H}_2 + \text{H \textsc{i}})$.
    See discussion in Section \ref{sec:gce}.
}
We use $\tau_\sfh = 13$ Gyr at $R = 8$ kpc based on Equation \ref{eq:tausfh}.
We also retain the $\eta = 0$ limit from our numerical models.
Although some of our models incorporate ejection in the central regions, the radial flow dominates enrichment across most of the Galactic disk by construction (see discussion in Section \ref{sec:gce:scenarios}).
We reserve further discussion of these analytic solutions for the Appendix, since we have fully described their physics in the subsections above.
The remaining details are primarily mathematical.
\par
Figure \ref{fig:gradients-vs-vice} shows the resulting radial gradients at the present day as functions of the relevant free parameter in our fiducial models.
For comparison, we include the corresponding equilibrium and ISM gradients from our numerical models, identical to the points in Figure \ref{fig:gradism-vs-gradeq}.
Our analytic estimates are in reasonable agreement with the \vice\ predictions at all parameter values.
Small differences arise because we are using a number of parameter simplifications for the sake of analytic solutions, which our \vice\ models relax (see discussion in Appendix \ref{sec:analytic:scenarios}).
For example, we use the instantaneous recycling approximation here, but numerically, we incorporate delayed enrichment from SNe Ia and the continuous return of stellar envelopes back to the ISM.
The key outcome of Figure \ref{fig:gradients-vs-vice} is that our analytic solutions, using simple parameter estimates, accurately describe the numerical predictions to a typical precision of $\pm 0.01$ kpc$^{-1}$.
\par
All of our analytic solutions approach a similar gradient as $v_{r,g} \rightarrow 0$, as expected, with the exception of the AMD scenario.
This solution is not well-behaved near $\beta_{\phi,\text{acc}} = 1$.
This outcome arises due to a factor of $1 / (1 - \beta_{\phi,\text{acc}})$ that arises (see Appendix \ref{sec:analytic:scenarios:amd}).
The AMD and GT scenarios lead to radial profiles in $Z_{\alpha,\eq}$ that are noticeably more concave than the constant and PWD scenarios (see Figure \ref{fig:profile-evol-comp} and discussion in Section \ref{sec:results}).
In these cases, the second derivative, $\partial^2 \ln Z_{\alpha,\eq} / \partial R^2$, is significant.
Our analytic estimates neglect this term.
Relaxing this approximation should provide a more accurate solution in the $\beta_{\phi,\text{acc}} \sim 1$ regime, but we do not have an immediate use for such a solution.

%% file: analytic-summary.tablebody.tex
\begin{table*}
\caption{
A blueprint for analytically modeling metallicity growth and radial gradients incorporating the effects of radial gas flows (for details, see discussion in Section \ref{sec:discussion}).
\textbf{The radial flow coefficients}: $\mu_\text{O}$ and $\mu_g$ describe the effect of the radial gas flow on surface density in units of the local SFR, $\dot\Sigma_\star$, and abundance by mass of some element, $Z_x$.
In the limit of axisymmetry and efficient mixing, both forms of $\mu$ have a well-defined form.
\textbf{Local Enrichment}: The rate of change in the metal abundance, $Z_\text{O}$, follows from the rates of change in the metal and gas surface densities, $\dot\Sigma_\text{O}$ and $\dot\Sigma_g$, respectively.
$Z_\text{O}$ approaches the equilibrium abundance $Z_{\text{O},\eq}$ asymptotically with an e-folding timescale of $\tau_\eq$.
\textbf{Equilibrium Gradients}: The equilibrium gradient, $\nabla_\eq[\text{O/H}]$, corresponds to the value at which the gradient will not evolve with time (i.e., $\dot\nabla[\text{O/H}] \approx 0$ at $ \grad{O} \approx \nabla_\eq[\text{O/H}]$), which is shaped by the number of stars formed per accreted baryon ($\dot \Sigma_\star / \dot\Sigma_\text{acc}$) and mixing effects ($\tau_\star v_{r,g} \nabla \ln Z$; see Figure \ref{fig:sfrifr-mixing}).
We provide solutions to $\xi_1$ and $\xi_2$ for specific radial flow scenarios in Appendix \ref{sec:analytic:scenarios}.
\textbf{ISM Gradients}: The ISM metallicity gradient tilts toward the equilibrium gradient on a timescale given by $\tau_\nabla$.
The gradient at a given moment in time reflects the equilibrium state at a lookback time of order $\tau_\nabla$.
}
\begin{tabularx}{\textwidth}{c @{\extracolsep{\fill}} c}
\hline
\hline
\multicolumn{2}{c}{
\textbf{The Radial Flow Coefficients}
}
\\
\parbox{0.5\textwidth}{
\centering
\vspace{2mm}
\(\displaystyle
\begin{aligned}
\dot\Sigma_{g,\flow} &\equiv
\mu_g \dot\Sigma_\star
\\
&\null
\\
\dot\Sigma_{x,\flow} &\equiv
Z_x \mu_x \dot\Sigma_\star
\end{aligned}
\)
\vspace{2mm}
}
&
\parbox{0.5\textwidth}{
\centering
\vspace{2mm}
\(\displaystyle
\begin{aligned}
\mu_g &= -\tau_\star \langle v_{r,g} \rangle \left(
\frac{1}{R} +
\partderiv{\ln \Sigma_g}{R} +
\partderiv{\ln \langle v_{r,g} \rangle}{R}
\right)
\\
\mu_x &= \mu_g -
\tau_\star \langle v_{r,g} \rangle
\partderiv{\ln Z_x}{R}
\end{aligned}
\)
\vspace{2mm}
}
\\
\hline
\end{tabularx}
\begin{tabularx}{\textwidth}{c  @{\extracolsep{\fill}} c}
\multicolumn{2}{c}{
\textbf{Local Enrichment}
}
\\
\parbox{0.5\textwidth}{
\centering
\vspace{2mm}
\(\displaystyle
\begin{aligned}
\dot\Sigma_g &= \dot\Sigma_\text{acc} -
\dot\Sigma_\star \left(
1 + \eta - \mu_g - r\right)
\\
\dot\Sigma_\text{O} &=
y_\text{O} \dot\Sigma_\star -
Z_\text{O} \dot\Sigma_\star \left(
1 + \eta - \mu_\text{O} - r\right)
\\
\dot Z_\text{O} &= \frac{
    y_\text{O}
}{
    \tau_\star
} - \frac{Z_\text{O}}{\tau_\star}\left(
1 + \eta - \mu_\text{O} - r +
\tau_\star \frac{\dot\Sigma_g}{\Sigma_g}
\right)
\end{aligned}
\)
\vspace{2mm}
} & \parbox{0.5\textwidth}{
\centering
\vspace{2mm}
\(\displaystyle
\begin{aligned}
Z_{\text{O},\eq} &= \frac{\ycc{O}}{
    1 + \eta - \mu_\text{O} - r +
    \tau_\star / N\tau_\sfh
}
\\
Z_\text{O}(t) &\approx Z_{\text{O},\eq}
\left(
1 - e^{-t / \tau_\eq}
\right)
\\
\tau_\text{eq} &\equiv \left(
\frac{
    1 + \eta - \mu_\text{O} - r
}{
    \tau_\star
} - \frac{1}{N \tau_\sfh}\right)^{-1}
\end{aligned}
\)
\vspace{2mm}
}
\\
\hline
\end{tabularx}
\begin{tabularx}{\textwidth}{c @{\extracolsep{\fill}} c}
\multicolumn{2}{c}{
\textbf{Equilibrium Gradients}
}
\\
\parbox{0.5\textwidth}{
\centering
\vspace{2mm}
\(\displaystyle
\begin{aligned}
\nabla_\eq[\text{O}/\text{H}] &= \frac{1}{\ln 10} \left[
\frac{Z_\text{O}}{Z_{\text{O},\eq}}
\partderiv{\ln Z_{\text{O},\eq}}{R} -
\left(1 - \frac{Z_\text{O}}{Z_{\text{O},\eq}}\right)
\partderiv{\ln \tau_\star}{R}
\right]
\\
\partderiv{\ln Z_{\text{O},\eq}}{R} &= \frac{
    -\xi_1 + \sqrt{\xi_1^2 + \xi_2}
}{
    2 \tau_\star v_{r,g}
}
\end{aligned}
\)
} & \parbox{0.5\textwidth}{
\centering
\vspace{2mm}
\(\displaystyle
\begin{aligned}
\xi_1 &= \frac{
    \dot\Sigma_\text{acc}
}{
    \dot\Sigma_\star
} + \tau_\star v_{r,g}\left(
\partderiv{\ln \tau_\star}{R} +
\partderiv{\ln v_{r,g}}{R}
\right)
\\
\xi_2 &= -4 \tau_\star v_{r,g}
\partderiv{}{R}\left(\frac{
    \dot\Sigma_\text{acc}
}{
    \dot\Sigma_\star
}\right)
\\
\frac{
    \dot\Sigma_\text{acc}
}{
    \dot\Sigma_\star
} &= 1 + \eta - \mu_g - r -
\frac{\tau_\star}{N\tau_\sfh}
\end{aligned}
\)
}
\vspace{2mm}
\\
\hline
\end{tabularx}
\begin{tabularx}{\textwidth}{c @{\extracolsep{\fill}} c}
\multicolumn{2}{c}{
\textbf{ISM Gradients}
}
\vspace{2mm}
\\
\parbox{0.6\textwidth}{
\centering
\vspace{2mm}
\(\displaystyle
\begin{aligned}
\nabla_\eq[\text{O/H}] &= \frac{1}{\ln 10}
\left[
\left(1 - e^{-t / \tau_\eq}\right)
\partderiv{\ln Z_{\text{O},\eq}}{R} -
e^{-t / \tau_\eq}
\partderiv{\ln \tau_\star}{R}
\right]
\\
\dot\nabla_\text{ISM} [\text{O/H}] &\approx
\frac{1}{\tau_\nabla}\left(
\nabla_\eq[\text{O/H}] -
\nabla_\text{ISM}[\text{O/H}]
\right)
\\
\nabla_\text{ISM}[\text{O/H}](t \rightarrow \tau_\text{disk}) &\approx
\nabla_\eq[\text{O/H}]\left(
t - \tau_\nabla
\right)
\end{aligned}
\)
\vspace{2mm}
} & \parbox{0.4\textwidth}{
\centering
\vspace{2mm}
\(\displaystyle
\tau_\nabla \equiv \left(
\frac{1}{\tau_\eq} -
\langle v_{r,g} \rangle
\partderiv{\ln Z_{\text{O},\eq}}{R}
\right)^{-1}
\)
\vspace{2mm}
}
\\
\hline
\hline
\end{tabularx}
\label{tab:analytic-summary}
\end{table*}

%% file: conclusions.tex
\section{Conclusions}
\label{sec:conclusions}

Building upon \citetalias{Johnson2025}, this paper has investigated possible origins of little or no decline in metallicity with stellar population age across much of the MW thin disk.
This result has arisen using stars of multiple spectral types and with different measurement techniques (red giants with asteroseismology: \citealt{Willett2023}; red giants with carbon-to-nitrogen ratios: \citealt{Roberts2025}; machine-learning algorithms: \citetalias{Johnson2025}; \citealt{Imig2023, Dubay2025}; open clusters: \citealt{Spina2022, Magrini2023, CarbajoHijarrubia2024}; Cepheid variables: \citealt{daSilva2023}).
\citet{Gallart2024} compared six different catalogs of stellar ages measured with different techniques, finding no significant correlation with metallicity in any of them.
Here, we have considered radial gas flows \citep[e.g.,][]{Lacey1985, Portinari2000, Bilitewski2012} as a possible origin of this apparent age-invariance in stellar metallicity.
We demonstrate that inward gas flows can explain this result at $R \gtrsim 6$ kpc if the Galactocentric radial velocity of the ISM is relatively constant in both radius and time (see Figures \ref{fig:gradoh-agebins} and \ref{fig:gradfeh-agebins}).
Our sample of stellar metallicities and ages are broadly consistent with a flow velocity of order $-1$ km/s in these regions.
\par
In line with our arguments in \citetalias{Johnson2025}, our results in this paper indicate that the best explanation for the lack of an age-metallicity relation in stars is the emergence of a chemical equilibrium early in the disk lifetime.
Without equilibrium, it is challenging to explain the lack of substantial change in metallicity across such a broad range of ages (see also discussion in \citealt{Dubay2025}).
\citet{Palla2024} showed that an episode of recent metal-poor gas accretion can reproduce this result in open clusters, which span a narrower range of ages than our red giants.
These arguments are not mutually exclusive.
However, additional processes are required to hold the ISM metallicity relatively constant on long timescales, since re-enrichment following dilution events is rapid \citep[$\lesssim$$1$ Gyr;][]{Dalcanton2007, Johnson2020}.
\par
A constant velocity of $v_{r,g} \approx -1$ km/s successfully reproduces the weak age-metallicity trends in our sample for stars near the Sun and in the outer disk because it acts to lower the ISM surface density in these regions (see Figure \ref{fig:eq-conditions} and discussion in Section \ref{sec:results:eq-scenario}).
In detail, this outcome arises if the velocity follows a radial profile that is flat in comparison to the ISM surface density (i.e., $\nabla \ln v_{r,g} \ll -\nabla \ln \Sigma_g$; see discussion in Section \ref{sec:discussion}).
By acting to remove gas from the outer disk, these models lower the amount of time that any one baryon typically spends in a given region of the ISM.
As a consequence, the local ISM rapidly approaches an equilibrium metal abundance.
However, at smaller radii ($R \lesssim 6$ kpc, depending on the model), inward gas flows tend to increase the ISM surface density by piling up gas.
In these regions, the flow carries the opposite effect on enrichment timescales and, by extension, trends in metallicity with stellar age.
The ISM metallicity continues to increase until the present day, leading to strong trends in metallicity with age at small Galactocentric radii.
\par
As a consequence of these opposing effects on enrichment timescales due to opposing effects on surface density, radial gas flows cannot be the sole explanation for the weak age-metallicity trend observed across the \textit{entire} Galaxy.
This outcome is a direct consequence of mass conservation (see discussion in Section \ref{sec:results:eq-scenario}).
The radial flow, by definition, only affects matter that is already phase-mixed with the ISM across the Galaxy, which forbids any changes to the total disk mass.
In order to increase or decrease ISM surface densities in one region, \textit{any} prescription for radial gas flows must do the opposite elsewhere.
Weakening age trends in one region of the disk while enhancing them in another is therefore a generic prediction.
Despite these issues with matching observed age trends, all of our flow prescriptions improve agreement relative to static models by steepening the radial gradient into better agreement with our sample (see Figures \ref{fig:gradoh-agebins} and \ref{fig:gradfeh-agebins}).
Based on this improvement, it is likely that radial flows have shaped the disk abundance structure to some significant extent.
\par
Our results nonetheless allow a more finely-tuned model in which inward gas flows tend to remove much of the gas from the ISM at $R \gtrsim 3$ kpc.
Under this prescription, the flow would be the primary driver of the apparent chemical equilibrium in the region of the Galaxy where disk stellar populations dominate and where our models are most applicable.
However, these flow prescriptions require a centralized ejection process to prevent unphysically large surface densities (see discussion in Section \ref{sec:gce:scenarios:gt}).
Mass loading factors of order $\sim$$1$ at $R = 0$ are sufficient for keeping surface densities at plausible levels, which is well within the limits imposed by observations of Galactic winds (see discussion in, e.g., the review by \citealt{Thompson2024}).
The potential requirement for at least one process, in addition to radial gas flows, suggests that some significant level of ISM ejection is required to produce a chemical equilibrium resembling the observations.
\par
The weak age-metallicity trend observed across the Galaxy also disfavors models that predict significant evolution in radial flow velocities with time.
If the speed changes, the equilibrium abundance also changes (see Table \ref{tab:analytic-summary} and Equations \ref{eq:zoeq} and \ref{eq:mu-soln}).
As a consequence, the ISM metallicity never stabilizes, resulting in strong age-metallicity trends (see Figure \ref{fig:eq-conditions} and discussion in Section \ref{sec:results:eq-scenario}).
This outcome generically disfavors models that tie radial flow velocities to the evolution or mass assembly of the Galactic disk.
In particular, our sample disfavors the scenario in which the angular momentum of accreting gas is significantly below the circular motions of the disk, originally described by \citet{Lacey1985}.
In the context of hydrodynamic simulations, this result potentially favors hot-mode accretion over cold-mode accretion.
The former distinguishes itself from the latter in that gas heats to virial temperature of the halo before joining the disk with similar angular momentum as the ISM \citep[e.g.,][]{Keres2005, Grand2019, Hafen2022}.
Cold accretion allows gas to merge with the ISM with kinematics that more closely reflect infall trajectories from filaments and sheets in the cosmic web (see discussion in, e.g., \citealt{vandeVoort2011} and the review by \citealt{SanchezAlmeida2014}).
This feature of cold accretion could lead to significant evolution in the radial flow velocity with time, which our models indicate would lead to strong trends in metallicity with stellar population age, at odds with empirical constraints.
\par
By following multiple scenarios driving radial gas flows based on different physical arguments (see discussion in Section \ref{sec:gce:scenarios}), we are able to identify the predictions that arise regardless of the processes driving the flow.
In line with previous GCE models \citep[e.g.,][]{Portinari2000, Spitoni2011}, all of our radial flow prescriptions lead to lower metal abundances at $R \gtrsim 3-5$ kpc and steeper radial gradients relative to static models (i.e., $v_{r,g} = 0$; see Figure \ref{fig:gradsteepen}).
We identify mixing as the primary reason for lowered metallicities (see Figure \ref{fig:ifrshift-mixing}).
Inward gas flows generically move low metallicity gas into high metallicity regions, leading to dilution.
The steepening of the radial gradient arises because the inward flow systematically shifts metal-poor accretion toward the outer Galaxy, which is a natural consequence of the gas motion.
This shift in accretion places less hydrogen at small radii and more hydrogen at large radii, enhancing the differences in composition between regions (see discussion in Section \ref{sec:results:dilution-steepening}).
We exploit these common outcomes to identify a mathematical relationship establishing metallicity in \textit{all} of our models (see discussion in Section \ref{sec:discussion}).
Metal abundances and radial gradients generically trace the ratio between the rates of star formation and accretion, $\dot\Sigma_\star / \dot\Sigma_\text{acc}$, which is then modified by mixing (see Figure \ref{fig:sfrifr-mixing}).
We derive a suite of analytic formulae based on this relationship, which incorporate the effects of radial gas flows.
The primary use case of these equations is in predicting radial metallicity gradients and their evolution in the MW and in MW-like galaxies at low redshift (see summary in Table \ref{tab:analytic-summary} and derivations in Appendix \ref{sec:analytic}).
\par
This connection bewteen the $\dot\Sigma_\star / \dot\Sigma_\text{acc}$ ratio and the metallicity gradient enables us to constrain the distribution of metal-poor accretion in a relatively model-independent manner (see Figure \ref{fig:gradism-vs-gradeq}).
All of our models are consistent with gas accreted by the MW following an exponential surface density profile with a scale radius of $\sim$$4.2$ kpc (see discussion in Section \ref{sec:discussion:eq-gradients}).
This inference is broadly consistent with our conclusion in \citetalias{Johnson2025} based on GCE models using ejection.
This prediction is useful because accretion has never been directly observed and is generally connected to other galaxy properties based on hydrodynamic simulations \citep[e.g.,][]{LeTiran2011, vandeVoort2011, vandeVoort2017, Hafen2022}.
\par
We also identify a theoretical limit corresponding to the steepest metallicity gradient that can arise for a given set of choices in other GCE parameters.
This so-called ORA limit places all of the metal-poor accretion at the very outer edge of the Galactic disk (see discussion in Section \ref{sec:gce:scenarios:ora-limit}).
The mass that is required to fuel star formation within the disk is supplied entirely by the radial flow.
Our choices of GCE parameters in this paper place the ORA limit at a metallicity gradient near $\nabla[\text{O/H}] \approx -0.13$ kpc$^{-1}$, sufficiently steep to be obviously ruled out by observations (see Figure \ref{fig:vgas-oh-comp}).
This discrepancy indicates that gas accretion must be present on some level across most of the Galactic disk.
Simulations indicate that accretion indeed occurs preferentially at large radii \citep[e.g.,][]{Trapp2022, Barbani2025}, but these predictions are not as extreme as the ORA limit, in line with our models.
\par
Recently, \citet{Wisz2025} showed that face-on galaxies with obvious spiral structure exhibit lower metallicities and steeper gradients than those without.
Based on our models, this result suggests a strong association between inward gas flows and spiral structure, in line with simulation predictions \citep[e.g.,][]{Grand2016, Orr2023}.
Significant radial flows are also expected to arise due to the presence of a central bar \citep[e.g.,][]{Weinberg1985, Debattista1998, Athanassoula2003, Hopkins2011}.
However, this potential connection with spiral structure is based on observations outside of the barred region, where the influence of flow velocities is more subtle.
Understanding the impact of various physically motivated flow velocity prescriptions, such as the influence of the bar, is an interesting direction for future work.
\par
\par
This paper ultimately demonstrates that precise measurements of stellar ages and metallicities can provide useful constraints on gas dynamics in the ISM.
Due to short-timescale effects, direct observations of velocity fields in external galaxies can only rule out the $\left|v_{r,g}\right| \gtrsim 5$ km/s regime (see Figure \ref{fig:empirical} and discussion in Sections \ref{sec:intro} and \ref{sec:results}).
However, chemical abundances are relatively insensitive to high-amplitude variability in ISM flow velocities ($\sim$$10$ km/s) on $\sim$$100$ Myr timescales (see Figure \ref{fig:oscil} and discussion in Section \ref{sec:results:variability}).
Our models lead to obvious discrepancies with observational constraints if the flow velocities are much larger than a $\sim$few km/s (see discussion in Section \ref{sec:results}).
We therefore argue that Galactic archaeology should provide stronger constraints on ISM motion in the radial direction on long timescales than velocity field measurements.
Our results favor the $\sim$$0.5 - 1.5$ km/s range and can confidently rule out the $\left|v_{r,g}\right| \gtrsim 3$ km/s regime.
Future catalogs of the ages and compositions of stars providing good spatial coverage of the Galactic disk, such as those of SDSS-V \citep{Kollmeier2025}, will enable Galactic archaeology to further sharpen the understanding of the baryon cycle in the MW.

%% file: acknowledgments.tex
\section*{Acknowledgments}

We are grateful to David Weinberg, Ana Bonaca, Gwen Rudie, and Fiorenzo Vincenzo for valuable discussions of this project.
JWJ acknowledges support from a Carnegie Theoretical Astrophysics Center postdoctoral fellowship.
\par
Funding for the Sloan Digital Sky 
Survey IV has been provided by the 
Alfred P. Sloan Foundation, the U.S. 
Department of Energy Office of 
Science, and the Participating 
Institutions. 

SDSS-IV acknowledges support and 
resources from the Center for High 
Performance Computing  at the 
University of Utah. The SDSS 
website is www.sdss4.org.

SDSS-IV is managed by the 
Astrophysical Research Consortium 
for the Participating Institutions 
of the SDSS Collaboration including 
the Brazilian Participation Group, 
the Carnegie Institution for Science, 
Carnegie Mellon University, Center for 
Astrophysics | Harvard \& 
Smithsonian, the Chilean Participation 
Group, the French Participation Group, 
Instituto de Astrof\'isica de 
Canarias, The Johns Hopkins 
University, Kavli Institute for the 
Physics and Mathematics of the 
Universe (IPMU) / University of 
Tokyo, the Korean Participation Group, 
Lawrence Berkeley National Laboratory, 
Leibniz Institut f\"ur Astrophysik 
Potsdam (AIP),  Max-Planck-Institut 
f\"ur Astronomie (MPIA Heidelberg), 
Max-Planck-Institut f\"ur 
Astrophysik (MPA Garching), 
Max-Planck-Institut f\"ur 
Extraterrestrische Physik (MPE), 
National Astronomical Observatories of 
China, New Mexico State University, 
New York University, University of 
Notre Dame, Observat\'ario 
Nacional / MCTI, The Ohio State 
University, Pennsylvania State 
University, Shanghai 
Astronomical Observatory, United 
Kingdom Participation Group, 
Universidad Nacional Aut\'onoma 
de M\'exico, University of Arizona, 
University of Colorado Boulder, 
University of Oxford, University of 
Portsmouth, University of Utah, 
University of Virginia, University 
of Washington, University of 
Wisconsin, Vanderbilt University, 
and Yale University.

%% file: analytic.tex
\section{Analytic Derivations}
\label{sec:analytic}

In this Appendix, we present detailed derivations of the mathematical expressions appearing throughout the main body of this paper.
The logic of these derivations rests most directly on the assumptions listed at the beginning of Section \ref{sec:discussion}.
Most of our solutions are enabled by the realization that the flow coefficients, $\mu_x$ and $\mu_g$, have a well-defined form in this limit of a thin, axisymmetric disk (see Appendix \ref{sec:analytic:flow-coefficients}).
We focus on the effects of angular momentum transport in the ISM (see discussion in Section \ref{sec:gce:scenarios}).
Our models do not invoke energy conservation or dissipation explicitly, though some of our scenarios would presumably involve some level of dissipation (e.g., GT and constant velocity scenarios; see Sections \ref{sec:gce:scenarios:gt} and \ref{sec:gce:scenarios:constant}).
We remind the reader that Table \ref{tab:glossary} is available as a reference on individual variable definitions.
\par
We begin with the processes setting radial gas flow velocities in Appendix \ref{sec:analytic:velocities}.
We derive solutions for the flow coefficients $\mu_g$ and $\mu_x$ in Appendix \ref{sec:analytic:flow-coefficients}.
We model the evolution in radial metallicity gradients as an equilibration process in Appendix \ref{sec:analytic:eq-gradient}.
We discuss analytic solutions to specific radial gas flow scenarios in Appendix \ref{sec:analytic:scenarios}.
We describe the impact of delayed enrichment events (e.g., SNe Ia) on gradients and their equilibration in Appendix \ref{sec:analytic:dtds}.
We describe the impact of pre-enriched accretion in Appendix \ref{sec:analytic:zin}.


\subsection{Radial Gas Velocities}
\label{sec:analytic:velocities}

This Appendix concerns the processes setting radial gas flow velocities in the ISM.
The angular momentum of some gas cloud of mass $M_g$, in the plane of disk rotation, is given by
\begin{equation}
L_g = M_g v_{\phi,g} R,
\label{eq:mvr}
\end{equation}
where $v_{\phi,g}$ is the circular velocity at a Galactocentric radius $R$.
In general, this cloud will mix with gas in its immediate surroundings on some timescale.
For some number of ``processes'' $p$, each of which add or remove mass at a rate of $\dot M_p$, the net torque can be expressed as
\begin{equation}
\dot{L}_g = \sum_p \dot{M}_p v_{\phi,p} R + \dot{L}_\text{dyn},
\label{eq:ldotgas}
\end{equation}
where the summation is taken over all processes.
In our models, these processes refer to accretion and ejection.
The circular velocity of the added or removed material, $v_{\phi,p}$, is not necessarily the same as the gas cloud of interest (i.e., $v_{\phi,p} \neq v_{\phi,g}$).
The AMD scenario (see Section \ref{sec:gce:scenarios:amd}) is rooted in this inequality.
The quantity $\dot L_\text{dyn}$ is an unknown term describing dynamical interactions, which may come from, e.g., spiral arms or the central bar.
\par
The new angular momentum, $L_g'$, of the gas cloud one timestep later can be expressed in terms of $L_g$ and $\dot L_g$ as:
\begin{subequations}\begin{align}
L_g' &= L_g + \dot{L}_g\delta t
\\
&=
M_g v_{\phi,g} R + \sum_p \dot{M}_p v_{\phi,p} R \delta t + \dot{L}_\text{dyn} \delta t.
\end{align}\end{subequations}
The second line of this equality follows from plugging in Equations \ref{eq:mvr} and \ref{eq:ldotgas}.
$L_g'$ can also be written in the same form as Equation \ref{eq:mvr}, according to
\begin{equation}
L_g' = \left(M_g + \sum_p \dot{M}_p \delta t\right)
\left(v_{\phi,g} + \sum_{i = 1}^\infty \sum_{j = 0}^i
\ddfrac{\partial^i v_{\phi,g}}{\partial t^{(i - j)} \partial R^j}
\delta t^{(i - j)} \Delta R^j\right)
\left(R + \Delta R\right).
\end{equation}
The first quantity in this product is the new mass of the gas cloud.
The second is the new circular velocity.
We have written the change in $v_{\phi,g}$ as a two-dimensional Taylor expansion in Galactocentric radius, $R$, and time, $t$, since its functional form is not known in terms of quantities defined here.
The third factor on the right hand side is the new Galactocentric radius of the the gas cloud of interest.
We use $\Delta R$ instead of $\delta R$ to distinguish the displacement of this gas cloud of interest from the width of each annulus in our GCE models.
\par
To isolate the radial flow velocity, we set these two expressions for $L_g'$ equal to one another, divide through by the timestep size, $\delta t$, and rearrange terms.
This procedure yields the following intermediate step:
\begin{equation}
\begin{split}
M_g v_{\phi,g} \frac{\Delta R}{\delta t} &=
\sum_p \dot{M}_p v_{\phi,p} R -
M_g \left(R + \Delta R\right)
\sum_{i = 1}^\infty \sum_{j = 0}^i
\ddfrac{\partial^i v_{\phi,g}}{\partial t^{(i - j)} \partial R^j}
\delta t^{(i - j - 1)} \Delta R^j
- \sum_p \dot{M}_p v_{\phi,g} \left(R + \Delta R\right)
\\
&\quad - \left(\sum _p \dot{M}_p\right)
\sum_{i = 1}^\infty \sum_{j = 0}^i
\ddfrac{\partial^i v_{\phi,g}}{\partial t^{(i - j)} \partial R^j}
\delta t^{(i - j)} \Delta R^j + \dot{L}_\text{dyn}.
\end{split}
\end{equation}
We now take the limit as $\delta R, \delta t \rightarrow 0$ to obtain an expression accurate for a disk that is fully resolved in radius and time.
The ratio $\delta R / \delta t$ becomes the ISM radial velocity, $v_{r,g}$.
All terms in the Taylor series vanish with the exception of the $(i, j) = (1, 0)$ combination in the first instance, which leaves
\begin{equation}
M_g v_{\phi,g} v_{r,g} = 
\sum_p \dot{M}_p \left(v_{\phi,p} - v_{\phi,g}\right) R -
M_g \dot{v}_{\phi,g} R + \dot{L}_\text{dyn}.
\end{equation}
Isolating the radial gas velocity yields the following solution:
\begin{equation}
v_{r,g} = R \left[\sum_p \frac{\dot{M}_p}{M_g} \left(\beta_{\phi,p} - 1\right) -
\frac{\dot{v}_{\phi,g}}{v_{\phi,g}} + \frac{\dot{L}_\text{dyn}}{L}\right],
\label{eq:vgas-final-appendix}
\end{equation}
where $\beta_{\phi,p} = v_{\phi,p} / v_{\phi,g}$ is the ratio of effective circular velocities.
\par
For most purposes, the summation over processes reduces to two terms describing accretion and ejection:
\begin{equation}
\sum_p \frac{\dot M_p}{M_g} \left(
\beta_{\phi,p} - 1 \right) \rightarrow
\frac{\dot M_\text{wind}}{M_g} \left(
1 - \beta_{\phi,\text{wind}}\right) -
\frac{\dot M_\text{acc}}{M_g} \left(
1 - \beta_{\phi,\text{acc}}\right).
\label{eq:process-summation-reduction}
\end{equation}
The sign difference arises because Galactic winds remove mass from the ISM, while accretion adds mass.
If the ejecting force is perpendicular to the Galactic disk, then $\beta_{\phi,\text{wind}} = 1$, and this term cancels.
However, the accretion rate depends on the ejection rate (see Equation \ref{eq:ifr-per-sfr}), allowing winds to have a secondary effect on flow velocities.
Our numerical models incorporate only weak winds for the sake of obeying mass conservation (see discussion in Appendix \ref{sec:analytic:flow-coefficients:mass-conservation} below).
After setting $\dot M_\text{wind} = 0$ accordingly and reducing the summation over processes to the single term of accretion, Equation \ref{eq:vgas-final-appendix} becomes the governing expression driving our physical motivated radial flows models (see Equation \ref{eq:vgas-scenarios}).
For brevity, we refer to $\dot L_\text{dyn} / L$ as $\dot L / L$ in the main body of this paper.


\subsubsection{Angular Momentum Dilution}
\label{sec:analytic:velocities:amd}

In this appendix, we derive an expression for the ISM radial velocity profile in the AMD scenario in terms of quantites specified as input to our GCE models.
The velocity follows from expanding Equation \ref{eq:amd-scenario} by plugging in Equation \ref{eq:process-summation-reduction} above:
\begin{equation}
v_{r,g} = R \Bigg[
\frac{\eta}{\tau_\star} \left(1 - \beta_{\phi,\text{wind}}\right) -
\left(\frac{1 + \eta - \mu_g - r}{\tau_\star} +
\frac{\partial \ln \Sigma_g}{\partial t}\right)
\left(1 - \beta_{\phi,\text{acc}}\right)
\Bigg],
\end{equation}
where we have also expanded $\dot\Sigma_\text{acc} / \Sigma_g$ using Equation \ref{eq:ifr-per-sfr} and the SFE timescale.
This expression is not yet a solution to the radial velocity profile, because $\mu_g$ depends on both $v_{r,g}$ and $\nabla \ln v_{r,g}$ (see discussion in Section \ref{sec:discussion}).
Plugging in Equation \ref{eq:mu-g-soln} for $\mu_g$ and isolating the ISM velocity yields the following differential equation
\begin{equation}
\partderiv{v_{r,g}}{R} + v_{r,g} \left[
\frac{1}{R} \left(
1 + \frac{1}{1 - \beta_{\phi,\text{acc}}}
\right) + \partderiv{\ln \Sigma_g}{R}
\right] = \frac{\eta}{\tau_\star}\left(
\frac{
    \beta_{\phi,\text{wind}} -
    \beta_{\phi,\text{acc}}
}{
    \beta_{\phi,\text{acc}} - 1
}
\right) -
\frac{\partial \ln \Sigma_g}{\partial t} -
\frac{1 - r}{\tau_\star}.
\label{eq:dvdr-amd-with-eta}
\end{equation}
This expression then reduces to Equation \ref{eq:amd-ode} when applying the limit that $\eta \rightarrow 0$ and substituting in $\partial \ln \dot\Sigma_\star = N \partial \ln \Sigma_g$ from the Kennicutt-Schmidt relation (see discussion in Section \ref{sec:gce}).
In integrating this equation numerically, we handle the discontinuity at $R = 0$ by using the relation between the radial gradients in the ISM mass and surface density (see Equation \ref{eq:gradmass-gradsigma-relation} and discussion in Appendix \ref{sec:analytic:flow-coefficients} below).
\par
The solution to $v_{r,g}$ as a function of radius does not follow analytically from Equation \ref{eq:dvdr-amd-with-eta}, regardless of the choice of $\eta$.
The factor of $1 / R$ in square brackets results in an integral of the product of a power law and an exponential with radius (akin to a gamma function).
Our analytic solutions for radial metallicity gradients still allow numerical solutions to the velocity profile (see Appendix \ref{sec:analytic:scenarios:amd} below).

\subsubsection{The Outer Rim Accretion Limit}
\label{sec:analytic:velocities:ora}

In this appendix, we describe our solution to the velocity profile in the ORA limit in terms of known input quantities.
By definition, the ORA limit leads to a direct solution to $\mu_g$ (see discussion in Section \ref{sec:gce:scenarios:ora-limit}).
Plugging Equation \ref{eq:mu-g-soln} for $\mu_g$ into Equation \ref{eq:ora-limit-mu} yields the following expression for the velocity profile:
\begin{equation}
\partderiv{v_{r,g}}{R} +
v_g \left(\frac{1}{R} +\partderiv{\ln \Sigma_g}{R} \right) =
\frac{-\partial \ln \Sigma_g}{\partial t} -
\frac{1 + \eta - r}{\tau_\star}.
\label{eq:dvdr-ora-with-eta}
\end{equation}
This expression leads to the form shown by Equation \ref{eq:ora-limit-profile} when substituting in $\partial \ln \dot\Sigma_\star = N\partial \ln \Sigma_g$ from the Kennicutt-Schmidt relation (see discussion in Section \ref{sec:gce}).
\par
Like the AMD scenario, the ORA limit does not have an analytic solution to the radial velocity profile.
However, it is useful to use Equation \ref{eq:dvdr-ora-with-eta} to understand the shape of the velocity profile at large radii.
SFE drops exponentially with radius, so $\tau_\star \rightarrow \infty$ as $R \rightarrow \infty$ (see discussion in Appendix \ref{sec:analytic:scenarios:io-sfe} below).
The SFH also becomes more extended with radius (see discussion in Section \ref{sec:gce:sfh}), so $\dot\Sigma_g / \Sigma_g \rightarrow 0$ as $R \rightarrow \infty$.
The factor of $1 / R$ also quickly becomes negligible beyond the ISM scale radius.
At large radii, Equation \ref{eq:dvdr-ora-with-eta} therefore reduces to
\begin{equation}
\partderiv{\ln v_{r,g}}{R}
\xrightarrow{R \rightarrow \infty}
-\partderiv{\ln \Sigma_g}{R}.
\end{equation}
The velocity gradient approaches the ISM surface density gradient.
Figure \ref{fig:vgas-oh-comp} shows the velocity profile in our numerical models, indicating that the ORA limit is significantly steeper than other flow scenarios.

\subsection{The Flow Coefficients}
\label{sec:analytic:flow-coefficients}

This appendix demonstrates that $\mu_x$ and $\mu_g$ have well-defined forms in the limit of axisymmetry (see Equation \ref{eq:mu-soln} and discussion in Section \ref{sec:discussion}).
Our approach to this derivation is rooted in applying the limit as $\delta R, \delta t \rightarrow 0$ to the concentric ring geometry of our models, shown in Figure \ref{fig:schematic}.
We also marginalize over an arbitrary distribution of radial velocities in the ISM.
The inward and outward components of the velocity distribution must be considered separately, since the area fraction determining how much gas is transferred between annuli has different forms for $v_{r,g} > 0$ and $v_{r,g} < 0$ (see Equation \ref{eq:areafracs}).
We also need to consider both gas and metals separately.
Here, we present a derivation of the metal mass that migrates between annuli for an outward flow ($v_{r,g} > 0$).
Inward flows follow the same procedure; the slight change in the functional form of the area fraction is the sole difference.
The proper solution for the gas also follows by simply setting $Z = 1$.
\par
This derivation parameterizes the rate of mass transfer between neighboring rings instead of surface density.
This choice simplifies our approach because mass is a conserved quantity, while surface density carries an additional dependence on the area of each ring.
The net rate of change in the mass of some element $x$ present in the ISM due to an outward radial flow, $\dot M_{x,\flow}(R | v_{r,g} > 0)$, is given by the gain from the inner neighbor at $R - \delta R \rightarrow R$ and the loss to the outer neighbor at $R + \delta R \rightarrow R + 2\delta R$, divided by the timestep size:
\begin{equation}
\dot{M}_{x,\flow}(R|v_{r,g} > 0) =
Z_x(R - \delta R) M_g(R - \delta R)
\ddfrac{a_\text{out}(v_{r,g}, R - \delta R)}{\delta t}
- Z_x(R) M_g(R) \ddfrac{a_\text{out}(v_{r,g}, R)}{\delta t},
\end{equation}
where $a_\text{out}$ is given by Equation \ref{eq:areafrac-outward}.
The radial velocity $v_{r,g}$ is not necessarily the same at $R - \delta R$ and $R$, which we take into account below.
The total rate of change in the local mass budget from the outward flow follows from weighting this expression by the velocity distribution and integrating over $v_{r,g}$:
\begin{subequations}
\begin{align}
\begin{split}
\dot{M}_{x,\flow}(R|v_{r,g} > 0) &=
Z_x(R - \delta R) M_g(R - \delta R)
\int_0^\infty \ddfrac{a_\text{out}(v_{r,g}, R - \delta R)}{\delta t}
P(v_{r,g} | R - \delta R) dv_{r,g}
\\
&\quad - Z_x(R) M_g(R)
\int_0^\infty \ddfrac{a_\text{out}(v_{r,g}, R)}{\delta t}
P(v_{r,g} | R) dv_{r,g},
\end{split}
\\
\begin{split}
&= Z_x(R) M_g(R) \Bigg[
\ddfrac{Z_x(R - \delta R)}{Z_x(R)}
\ddfrac{M_g(R - \delta R)}{M_g(R)}
\int_0^\infty \ddfrac{a_\text{out}(v_{r,g}, R - \delta R)}{\delta t}
P(v_{r,g} | R - \delta R) dv_{r,g}
\\
&\quad - \int_0^\infty \ddfrac{a_\text{out}(v_{r,g}, R)}{\delta t}
P(v_{r,g} | R) dv_{r,g}
\Bigg],
\end{split}
\end{align}
\end{subequations}
where the limits of integration reflect the outward component of the flow.
\par
In general, the mass present in the ISM, its metallicity, and its velocity distribution could vary as arbitrarily complex functions of Galactocentric radius.
We therefore parameterize each of them as Taylor series with unknown coefficients, which accommodates any smooth functional form.
It turns out to be notationally convenient to write the properties of the inner neighbor as a difference from the annulus of interest to the derivation:
\begin{subequations}
\begin{align}
Z_x(R - \delta R) &= Z_x(R) - \taylorexpand{Z_x}{R}
\\
M_g(R - \delta R) &= M_g(R) - \taylorexpand{M_g}{R}
\\
P(v_{r,g} | R - \delta R) &= P(v_{r,g} | R) -
\taylorexpand{P(v_{r,g} | R)}{R}.
\end{align}
\end{subequations}
\par
We now expand our expression for $\dot M_{x,\flow}$ further by plugging in these Taylor series and Equation \ref{eq:areafrac-outward} for $a_\text{out}$:
\begin{equation}
\begin{split}
\dot{M}_{x,\flow}&(R|v_{r,g} > 0) =
\ddfrac{Z_x M_g}{\delta R}\Bigg[
\\
&\quad
\left(1 - \ddfrac{1}{Z_x}\taylorexpand{Z_x}{R}\right)
\left(1 - \ddfrac{1}{M_g}\taylorexpand{M_g}{R}\right)
\int_0^\infty \ddfrac{
    2Rv_{r,g} - v_{r,g}^2 \delta t
}{
    2 R - \delta R
}
P(v_{r,g} | R) dv_{r,g}
\\
&\quad -
\left(1 - \ddfrac{1}{Z_x}\taylorexpand{Z_x}{R}\right)
\left(1 - \ddfrac{1}{M_g}\taylorexpand{M_g}{R}\right)
\int_0^\infty \ddfrac{
    2Rv_{r,g} - v_{r,g}^2 \delta t
}{
    2R - \delta R
}
\taylorexpand{P(v_{r,g} | R)}{R} dv_{r,g}
\\
&\quad - \int_0^\infty \ddfrac{
    2\left(R + \delta R\right) v_{r,g} - v_{r,g}^2 \delta t
}{
    2R + \delta R
}
P(v_{r,g} | R) dv_{r,g}
\Bigg].
\end{split}
\label{eq:limit-start}%
\end{equation}
We have stopped explicitly referring to $Z_x$ and $M_g$ as functions of radius, since they are always evaluated at $R$ hereafter.
Equation \ref{eq:limit-start} is an important point in this derivation because it exposes every instance of $\delta R$ and $\delta t$.
We can now apply the limit as $\delta R, \delta t \rightarrow 0$.
Each appearance of $-v_{r,g} \delta t$ vanishes.
In the above expression, we have also extracted a factor of $1/\delta R$ out of the quantity in square brackets.
By attempting this derivation other ways, we find that this step is required to reach a physical solution to $\dot M_{x,\flow}$ in the limit that $\delta R \rightarrow 0$.
This approach enables the use of L'H{\^o}pital's rule by producing a $0/0$ indeterminate form.
\par
In applying L'H{\^o}pital's rule, the factor of $\delta R$ outside of square brackets in Equation \ref{eq:limit-start} above simply cancels.
Now, we must take the derivative with respect to $\delta R$ of the linear combination in square brackets, and then apply the limit that $\delta R \rightarrow 0$.
We refer to the three terms in square brackets as $T_1$, $T_2$, and $T_3$, such that Equation \ref{eq:limit-start} can be written as $\dot M_{x,\flow} \sim [T_1 - T_2 - T_3]$.
Differentiating $T_1$ with respect to $\delta R$ and applying the limit:
\begin{subequations}
\begin{align}
\begin{split}
\lim_{\delta R \rightarrow 0} \partderiv{T_1}{\delta R} &=
\lim_{\delta R \rightarrow 0} \Bigg[ \int_0^\infty \ddfrac{
    2Rv_{r,g}
}{
    2 R - \delta R
}
P(v_{r,g} | R) dv_{r,g}
\Bigg[
\left(\frac{-1}{Z_x}
\sum_{i = 1}^\infty i \partderiv{^i Z_x}{R^i}
\delta R^{i - 1}\right)
\left(1 - \frac{1}{M_g}\taylorexpand{M_g}{R}\right)
\\
&\quad\quad +
\left(1 - \frac{1}{Z_x}\taylorexpand{Z_x}{R}\right)
\left(\frac{-1}{M_g}
\sum_{i = 1}^\infty i \partderiv{^i M_g}{R^i}
\delta R^{i - 1}\right)
\Bigg]
\\
&\quad +
\left(1 - \frac{1}{Z_x}\taylorexpand{Z_x}{R}\right)
\left(1 - \frac{1}{M_g}\taylorexpand{M_g}{R}\right)
\int_0^\infty \frac{2Rv_{r,g}}{\left(2R - \delta R\right)^2}
P(v_{r,g} | R) dv_{r,g}
\Bigg]
\end{split}
\\
&=
\int_0^\infty v_{r,g} P(v_{r,g} | R) dv_{r,g} \left[
-\partderiv{\ln Z_x}{R} - \partderiv{\ln M_g}{R}
\right] +
\int_0^\infty \frac{v_{r,g}}{2R} P(v_{r,g} | R) dv_{r,g}.
\end{align}
\end{subequations}
Multiple terms arise due to product rule for derivatives.
Each appearance of $\delta R^{i - 1}$ cancels when $i = 1$, leaving behind a non-zero term that survives the limit as $\delta R \rightarrow 0$.
All higher powers of $\delta R$ vanish, which leaves behind the terms $\nabla \ln Z_x$ and $\nabla \ln M_g$ in square brackets in the second line.
\par
Differentiating $T_2$ with respect to $\delta R$ and applying the limit:
\begin{subequations}
\begin{align}
\begin{split}
\lim_{\delta R \rightarrow 0} \partderiv{T_2}{\delta R} &=
\lim_{\delta R \rightarrow 0} \Bigg[
\int_0^\infty \ddfrac{
    2Rv_{r,g}
}{
    2R - \delta R
}
\sum_{i = 1}^\infty i \partderiv{^i P(v_{r,g} | R)}{R^i}
\delta R^{i - 1} dv_{r,g}
\left(1 - \frac{1}{Z_x}\taylorexpand{Z_x}{R}\right)
\left(1 - \frac{1}{M_g}\taylorexpand{M_g}{R}\right)
\\
&\quad +
\int_0^\infty \ddfrac{
    2Rv_{r,g}
}{
    2R - \delta R
}
\taylorexpand{P(v_{r,g} | R)}{R} dv_{r,g} \Bigg[
\left(
\frac{-1}{Z_x} \sum_{i = 1}^\infty i \partderiv{^i Z_x}{R^i}
\delta R^{i - 1}
\right)
\left(1 - \frac{1}{M_g}\taylorexpand{M_g}{R}\right)
\\
&\quad\quad +
\left(1 - \frac{1}{Z_x}\taylorexpand{Z_x}{R}\right)
\left(
\frac{-1}{M_g} \sum_{i = 1}^\infty i \partderiv{^i M_g}{R^i}
\delta R^{i - 1}
\right)
\Bigg]
\\
&\quad +
\int_0^\infty \frac{2Rv_{r,g}}{
    (2R - \delta R)^2
}\taylorexpand{P(v_{r,g} | R)}{R} dv_{r,g}
\left(1 - \frac{1}{Z_x}\taylorexpand{Z_x}{R}\right)
\left(1 - \frac{1}{M_g}\taylorexpand{M_g}{R}\right)
\Bigg]
\end{split}
\\
&= \int_0^\infty v_{r,g} \partderiv{P(v_{r,g} | R)}{R} dv_{r,g}.
\end{align}
\end{subequations}
Multiple terms arise due to product rule again.
Similar to $T_1$ above, each appearance of $\delta R^{i - 1}$ cancels when $i = 1$.
The second and third terms each vanish due to the Taylor series in $P(v_{r,g} | R)$ and its associated appearance of $\delta R^i$.
The integral on the first line is therefore the only term that survives the limit as $\delta R \rightarrow 0$.
Continuing this procedure for the third term, $T_3$, is much more straightforward:
\begin{equation}
\lim_{\delta R \rightarrow 0} \partderiv{T_3}{\delta R}
= \lim_{\delta R \rightarrow 0} \int_0^\infty
\partderiv{}{\delta R} \left(
\ddfrac{2\left(R + \delta R\right) v_{r,g}}{2R + \delta R}
\right) P(v_{r,g} | R) dv_{r,g}
= \int_0^\infty \ddfrac{v_{r,g}}{2R} P(v_{r,g} | R) dv_{r,g}
\end{equation}
Combining $T_1$, $T_2$, and $T_3$ then yields the following expression for $\dot{M}_{x,\flow}(R|v_{r,g} > 0)$:
\begin{equation}
\dot{M}_{x,\flow}(R|v_{r,g} > 0) = -Z_x M_g \left[
\int_0^\infty v_{r,g} P(v_{r,g} | R) dv_{r,g} \left(
\partderiv{\ln Z_x}{R} + \partderiv{\ln M_g}{R}
\right) +
\int_0^\infty v_{r,g} \partderiv{P(v_{r,g} | R)}{R} dv_{r,g}
\right].
\label{eq:mdotflow-outward}
\end{equation}
\par
An expression for the rate of change in the mass present due to the inward component of the velocity distribution, $\dot{M}_{x,\flow}(R|v_{r,g} < 0)$, follows from a similar derivation.
The intermediate steps have slightly more verbose forms due to the differences between $a_\text{in}$ and $a_\text{out}$ (see Equation \ref{eq:areafracs}).
The result is an expression of the same form as Equation \ref{eq:mdotflow-outward} but with the integration limits on the velocity distribution switched to the $(-\infty,0]$ interval.
The total rate of change in the mass of the element $x$ present at a given Galactocentric radius $R$, $\dot M_{x,\flow}(R)$, is simply the sum of the two components of the velocity distribution:
\begin{equation}
\begin{split}
\dot M_{x,\flow}(R) &=
\dot M_{x,\flow}(R | v_{r,g} < 0) +
\dot M_{x,\flow}(R | v_{R,g} > 0)
\\
&= -Z_x M_g \left[
\left(
\partderiv{\ln Z_x}{R} +
\partderiv{\ln M_g}{R}
\right)
\int_{-\infty}^\infty v_{r,g} P(v_{r,g} | R) dv_{r,g}
+ \partderiv{}{R}
\int_{-\infty}^\infty v_{r,g} P(v_{r,g} | R) dv_{r,g}
\right].
\label{eq:mdotflow-integral}
\end{split}
\end{equation}
In the second line, we have also taken advantage of the fact that $R$ and $v_{r,g}$ are independent variables to move the derivative $\partial / \partial R$ outside of the integral over $v_{r,g}$.
The first moment of the velocity distribution (i.e., the mean velocity),
\begin{equation}
\langle v_{r,g} \rangle \equiv
\int_{-\infty}^\infty v_{r,g} P(v_{r,g} | R) dv_{r,g},
\end{equation}
appears twice in this expression for $\dot{M}_{x,\flow}$.
Substituting this expression into Equation \ref{eq:mdotflow-integral} and pulling one power of $\langle v_{r,g} \rangle$ out of square brackets leads to a more compact form:
\begin{equation}
\dot M_{x,\flow} = -Z_x M_g \langle v_{r,g} \rangle \left[
\partderiv{\ln M_g}{R} +
\partderiv{\ln \langle v_{r,g} \rangle}{R} +
\partderiv{\ln Z_x}{R}
\right].
\label{eq:mdotflow}
\end{equation}
Throughout the rest of this paper, we have referred to $\langle v_{r,g} \rangle$ as $v_{r,g}$ for brevity.
\par
At this point, it is useful to transform from mass, $M_x$, back to surface density, $\Sigma_x$.
In our models, the geometric boundaries of each ring do not move.
Their areas are therefore constant in time, so $\dot M_{x,\flow} \rightarrow \dot \Sigma_{x,\flow}$.
The derivatives of $M_x$ and $\Sigma_x$ with radius, however, must incorporate variations in the area of each annulus ($A = 2\pi R \delta R$):
\begin{equation}
d \ln M_g = d \ln A + d \ln \Sigma_g \implies
\partderiv{\ln M_g}{R} = \frac{1}{R} + \partderiv{\ln \Sigma_g}{R}.
\label{eq:gradmass-gradsigma-relation}
\end{equation}
We now arrive at our final result expressing the rate of change in the ISM surface density of some element $x$ due to radial flows in terms of our GCE parameters:
\begin{equation}
\dot{\Sigma}_{x,\flow} = -Z_x \dot{\Sigma}_\star \tau_\star \langle v_{r,g} \rangle
\left[\frac{1}{R} +
\partderiv{\ln \Sigma_g}{R} +
\partderiv{\ln \langle v_{r,g} \rangle}{R} +
\partderiv{\ln Z_x}{R}
\right],
\label{eq:mu-x-soln-full}
\end{equation}
where we have also substituted in the SFE timescale, $\tau_\star \equiv \Sigma_g / \dot\Sigma_\star$.
The corresponding expression for the rate of change in the gas surface density, $\dot \Sigma_{g,\flow}$, follows a similar derivation.
The result can also be deduced by simply taking the limit that $Z_x \rightarrow 1$:
\begin{equation}
\dot{\Sigma}_{g,\flow} = -\dot{\Sigma}_\star \tau_\star
\langle v_{r,g} \rangle \left[
\frac{1}{R} +
\partderiv{\ln \Sigma_g}{R} +
\partderiv{\ln \langle v_{r,g}\rangle}{R}
\right].
\label{eq:mu-g-soln-full}
\end{equation}
The functional forms of $\mu_\text{O}$ and $\mu_g$ given by Equations \ref{eq:mu-o-soln} and \ref{eq:mu-g-soln}, summarized in Table \ref{tab:analytic-summary}, follow directly from Equations \ref{eq:mu-x-soln-full} and \ref{eq:mu-g-soln-full}, respectively.

\subsubsection{Mass Conservation}
\label{sec:analytic:flow-coefficients:mass-conservation}

This appendix describes an important requirement that the functional forms of $\mu_x$ and $\mu_g$ must satisfy, which originates in mass conservation.
Radial gas flows, by definition, only influence matter that has already phase mixed with the disk ISM.
The flow is therefore forbidden from altering the total mass of the ISM.
This constraint ultimately leads to chemical equilibrium in some regions of the Galaxy and significant disequilibrium in others.
We discuss the impact of this prediction in the context of age-metallicity trends in observations throughout Section \ref{sec:results:eq-scenario}.
\par
A mathematical expression describing this requirement follows by writing two different expressions for the total rate of change in the ISM mass, $\dot M_g$, and setting them equal to one another.
One expression follows from the linear combination of all processes adding or removing mass from the Galaxy:
\begin{equation}
\begin{split}
\dot{M}_g &= \dot{M}_\text{acc} - \dot{M}_\star - \dot{M}_\text{wind} + \dot{M}_r
\\
&= 2 \pi \left[
\int_0^\infty \dot\Sigma_\text{acc}(R) R dR -
(1 - r) \int_0^\infty \dot\Sigma_\star(R) R dR -
\int_0^\infty \eta(R)\dot\Sigma_\star(R) R dR
\right]
\label{eq:total-mdotgas-form1}
\end{split}
\end{equation}
where the second line expresses these quantities as integrals of surface density over radius.
$\dot M_g$ can also be similarly expressed as an integral of $\dot\Sigma_g$ over radius:
\begin{equation}
\begin{split}
\dot{M}_g &= \int_0^\infty
\dot\Sigma_g (R) 2\pi R dR
\\
&= \int_0^\infty \Big(
\dot\Sigma_\text{acc} -
\dot\Sigma_\star \Big(
1 + \eta(R) - \mu_g(R) - r
\Big)\Big) 2\pi R dR.
\label{eq:total-mdotgas-form2}
\end{split}
\end{equation}
Setting Equations \ref{eq:total-mdotgas-form1} and \ref{eq:total-mdotgas-form2} equal to one another results in the condition given by Equation \ref{eq:mu-gas-closure}.
The corresponding requirement based on the metal mass (see Equation \ref{eq:mu-x-closure}) follows similarly.
This constraint enforces the outcome that neither $\mu_g$ nor $\mu_x$ can be negative in one region of the disk without being positive in another, and vice versa.
We discuss the consequences of this implication in Section \ref{sec:results:eq-scenario}.


\subsubsection{Validation Against Numerical Models}
\label{sec:analytic:flow-coefficients:validation}

\begin{figure*}
\centering
\includegraphics[scale = 0.92]{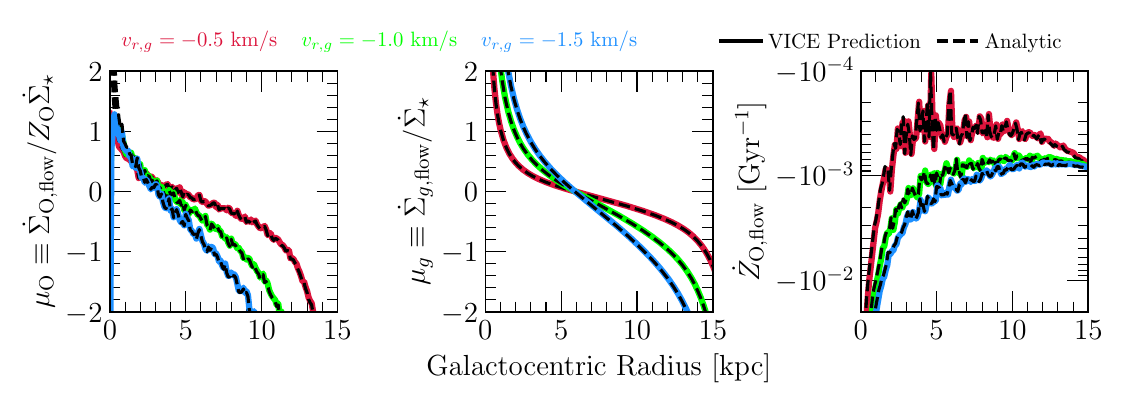}
\caption{
Analytic and numerical calculations of the flow coefficients, $\mu_\text{O}$ (left) and $\mu_g$ (middle), and the mixing effect of the flow on metallicity (right).
Solid lines, color coded according to the legend at the top, show the predictions by our numerical models with VICE.
Black dashed lines show the corresponding analytic expectation based on Equations \ref{eq:mu-g-soln} and \ref{eq:mu-o-soln}.
}
\label{fig:analytic-vs-numeric}
\end{figure*}

Figure \ref{fig:analytic-vs-numeric} compares our analytic solutions to $\mu_\text{O}$, $\mu_g$, and $\dot Z_{\text{O},\flow}$ with our numerical models.
We focus on the constant velocity scenario for this validation, so the colored lines in the right panel are identical to those in the bottom panel of Figure \ref{fig:ifrshift-mixing}.
Analytic solutions come from evaluating Equations \ref{eq:zdotflow}, \ref{eq:mu-g-soln}, and \ref{eq:mu-o-soln} using the present day snapshot of our numerical models directly.
We determine the numerical solutions to these quantities by evaluating the area fractions with Equation \ref{eq:areafracs}, which provides the amount of mass and metals mixed directly.
Our analytic and numerical solutions to $\mu_\text{O}$, $\mu_g$, and $\dot Z_{\text{O},\flow}$ are in excellent agreement.
The only noticeable difference between the two is near $R = 0$, where the analytic solution to $\mu_\text{O}$ diverges to $+\infty$, and the numerical solution diverges to $-\infty$.
This agreement indicates not only that our numerical models have converged, but also that our mathematical expressions for to $\mu_\text{O}$ and $\mu_g$ are accurate.

\subsection{The Equilibrium Gradient}
\label{sec:analytic:eq-gradient}

This appendix describes evolution in the ISM metallicity and the radial gradient thereof as an equilibration process.
For these purposes, it turns out to be notationally convenient to re-write the rate of change the metal abundance, $\dot Z_\alpha$ (see Equation \ref{eq:zdot-o}), in terms of the equilibrium abundance (see Equation \ref{eq:zoeq}):
\begin{equation}
\dot Z_\alpha = \frac{y_\alpha}{\tau_\star}
\left(1 - \frac{Z_\alpha}{Z_{\alpha,\eq}}\right).
\label{eq:zdot-alpha-altform}
\end{equation}
We use the subscript $\alpha$ to refer to alpha elements like O and Mg, whose production is dominated by massive stars.
$Z_\alpha$ can also be interpreted as referring specifically to the products of alpha process nucleosynthesis, with enrichment from other sources (e.g., SNe Ia) requiring additional terms (see discussion in Appendix \ref{sec:analytic:dtds} below).
The instantaneous recycling approximation is more accurate for these elements than others (see discussion in Section \ref{sec:gce:yields}).
\par
The rate of tilt in the radial metallicity profile corresponds to a second-derivative of $Z_\alpha$, with one derivative in time and one in Galactocentric radius:
\begin{equation}
\dot \nabla \ln Z_\alpha = \frac{1}{Z_\alpha} \left(
\frac{\partial^2 Z_\alpha}{\partial t \partial R} -
\partderiv{\ln Z_\alpha}{R}
\partderiv{Z_\alpha}{t}
\right).
\end{equation}
The symmetry of second derivatives allows us to differentiate Equation \ref{eq:zdot-alpha-altform} for $\dot Z_\alpha$ with radius and equate the result to the first term in parentheses above, which yields the following next step:
\begin{equation}
\dot \nabla \ln Z_\alpha = \frac{y_\alpha}{
    \tau_\star Z_\alpha
} \left[
\frac{Z_\alpha}{Z_{\alpha,\eq}}
\partderiv{\ln Z_{\alpha,\eq}}{R} -
\left(1 - \frac{Z_\alpha}{Z_{\alpha,\eq}}\right)
\partderiv{\ln \tau_\star}{R} -
\partderiv{\ln Z_\alpha}{R}
\right].
\label{eq:tilt-rate-alpha}
\end{equation}
We define the equilibrium gradient, $\nabla_\eq$, as the gradient at which the radial metallicity profile will not tilt (i.e., $\dot \nabla \ln Z_\alpha = 0$; see discussion in Section \ref{sec:discussion:eq-gradients}).
The above expression for $\dot \nabla \ln Z_\alpha$ is equal to zero when the linear combination in square brackets is equal to zero.
We can then isolate the slope $\nabla \ln Z_\alpha$ and interpret its value as the equilibrium gradient:
\begin{equation}
\nabla_\eq [\alpha/\text{H}] = \frac{1}{\ln 10}
\left[
\frac{Z_\alpha}{Z_{\alpha,\eq}}
\partderiv{\ln Z_{\alpha,\eq}}{R} -
\left(1 - \frac{Z_\alpha}{Z_{\alpha,\eq}}
\right) \partderiv{\ln \tau_\star}{R}
\right],
\label{eq:gradeq-alpha}
\end{equation}
where the additional factor of $\ln 10$ converts between natural and base-10 logarithms.
The above two expressions have the same forms as Equations \ref{eq:tilt-rate} and \ref{eq:gradeq} for O abundances (see Section \ref{sec:discussion:eq-gradients}).
\par
Equation \ref{eq:gradeq-alpha} becomes useful when every instance of $Z_{\alpha,\eq}$ and $\nabla \ln Z_{\alpha,\eq}$ are expressed in terms of GCE parameters.
It turns out to be notationally convenient to use the expression for the equilibrium abundance, $Z_{\alpha,\eq}$, in terms of $\dot \Sigma_\text{acc} / \dot\Sigma_\star$ and $\tau_\star v_{r,g} \nabla \ln Z_{\alpha,\eq}$ given by Equation \ref{eq:zoeq-sfrifrmixing}.
One possible path forward is to solve Equation \ref{eq:zoeq-sfrifrmixing} as an ordinary differential equation for $Z_{\alpha,\eq}$.
However, this approach requires a different set of steps for different parameter choices.
A more broadly applicable approximation follows from differentiating Equation \ref{eq:zoeq-sfrifrmixing} with radius:
\begin{equation}
\partderiv{\ln Z_{\alpha,\eq}}{R} = \ddfrac{-1}{
\frac{\dot \Sigma_\text{in}}{\dot \Sigma_\star} +
\tau_\star v_{r,g} \partderiv{\ln Z_{\alpha,\eq}}{R}
} \left[
\partderiv{}{R} \left(
\frac{\dot \Sigma_\text{in}}{\dot \Sigma_\star}
\right) +
\tau_\star v_{r,g} \left(
\partderiv{\ln \tau_\star}{R} +
\partderiv{\ln v_{r,g}}{R}\right)
\partderiv{\ln Z_{\alpha,\eq}}{R} +
\tau_\star v_{r,g} \partderiv{^2 \ln Z_{\alpha,\eq}}{R^2}
\right].
\end{equation}
We approximate the second derivative of the equilibrium abundance with radius to be negligible ($\partial^2 \ln Z_{\alpha,\eq} / \partial R^2 \approx 0$), an assumption commonly referred to as ``local linearity.''
The final term in square brackets in the above expression therefore cancels.
Rearranging terms in the above expression then results in the following quadratic equation for $\nabla \ln Z_{\alpha,\eq}$:
\begin{equation}
\tau_\star v_{r,g} \left(
\partderiv{\ln Z_{\alpha,\eq}}{R}\right)^2 +
\left[
\tau_\star v_{r,g} \left(
\partderiv{\ln \tau_\star}{R} +
\partderiv{\ln v_{r,g}}{R} \right) +
\frac{\dot \Sigma_\text{in}}{\dot \Sigma_\star}
\right]\partderiv{\ln Z_{\alpha,\eq}}{R} +
\partderiv{}{R} \left(
\frac{\dot \Sigma_\text{in}}{\dot \Sigma_\star}\right) = 0.
\end{equation}
From here, the solution to $\nabla \ln Z_{\alpha,\eq}$ follows as
\begin{subequations}
\begin{align}
\partderiv{\ln Z_{\alpha,\eq}}{R} &= \ddfrac{
    -\xi_1 + \sqrt{\xi_1^2 + \xi_2}
}{
    2 \tau_\star v_{r,g}
}
\label{eq:quadratic-soln-appendix}
\\
\xi_1 &\equiv \frac{\dot \Sigma_\text{in}}{\dot \Sigma_\star} +
\tau_\star v_{r,g} \left(
\partderiv{\ln \tau_\star}{R} +
\partderiv{\ln v_{r,g}}{R} \right)
\label{eq:xi1-appendix}
\\
\xi_2 &\equiv -4 \tau_\star v_{r,g} \partderiv{}{R} \left(
\frac{\dot \Sigma_\text{in}}{\dot \Sigma_\star}\right).
\label{eq:xi2-appendix}
\end{align}
\end{subequations}
The solution to the quadratic equation is physical with a plus-sign in the numerator of Equation \ref{eq:quadratic-soln-appendix} as opposed to a minus sign.
Reexpanding $\xi_1$ and $\xi_2$, these coefficients can be written more verbosely in terms of input quantities according to
\begin{subequations}
\begin{align}
\xi_1 &\equiv 1 + \eta - r - \frac{\tau_\star}{N \tau_\sfh} +
\tau_\star v_{r,g} \left[
\frac{1}{R} + \partderiv{\ln \Sigma_g}{R} +
\partderiv{\ln \tau_\star}{R} +
2 \partderiv{\ln v_{r,g}}{R}
\right]
\label{eq:xi1}%
\\
\xi_2 &\equiv 4 \tau_\star v_{r,g} \left[
\partderiv{\mu_g}{R} -
\partderiv{\eta}{R} + \frac{\tau_\star}{N \tau_\sfh} \left(
\partderiv{\ln \tau_\star}{R} -
\partderiv{\ln \tau_\sfh}{R}\right)
\right].
\label{eq:xi2}%
\end{align}
\label{eq:xi}%
\end{subequations}
The advantage of this formalism is that one can simply plug in different radial flow prescriptions along with GCE parameters, and the solution to the radial metallicity gradient follows algebraically.
We discuss solutions to $\xi_1$ and $\xi_2$ for individual radial flow scenarios in Appendix \ref{sec:analytic:scenarios} below.

\subsubsection{The Approach to Equilibrium}
\label{sec:analytic:eq-gradient:eq-approach}

This appendix is concerned with the timescale on which the radial metallicity gradient equilibrates.
For these purposes, it is useful to rewrite Equation \ref{eq:gradeq-alpha} for the equilibrium gradient as
\begin{equation}
\nabla_\eq \ln Z_\alpha = \left(1 - e^{-t / \tau_\eq}\right)
\partderiv{\ln Z_{\alpha,\eq}}{R} -
e^{-t / \tau_\eq}
\partderiv{\ln \tau_\star}{R},
\label{eq:gradeq-alpha-alt}
\end{equation}
where we have substituted in $1 - e^{-t / \tau_\eq}$ for $Z_\alpha / Z_{\alpha,\eq}$ from Equation \ref{eq:taueq}.
We have also converted from a base-10 logarithm in Equation \ref{eq:gradeq-alpha} back to a natural logarithm of $Z_\alpha$.
Making the same substitution in Equation \ref{eq:tilt-rate-alpha} for $\dot \nabla \ln Z_\alpha$ yields the following expression for the tilt rate:
\begin{equation}
\dot \nabla \ln Z_\alpha =
\frac{
    y_\alpha
}{
    \tau_\star Z_{\alpha,\eq}
    \left(1 - e^{-t / \tau_\eq}\right)
} \left[
\left(1 - e^{-t / \tau_\eq}\right)
\partderiv{\ln Z_{\alpha,\eq}}{R} -
e^{-t / \tau_\eq} \partderiv{\ln \tau_\star}{R} -
\partderiv{\ln Z_\alpha}{R}
\right],
\label{eq:tilt-rate-alt}
\end{equation}
which can then be re-expressed in terms of $\nabla_\eq \ln Z_\alpha$ as
\begin{equation}
\dot \nabla \ln Z_\alpha = \frac{
    y_\alpha
}{
    \tau_\star Z_{\alpha,\eq}
    \left(1 - e^{-t / \tau_\eq}\right)
} \left(
\nabla_\eq \ln Z_\alpha - \nabla \ln Z_\alpha
\right).
\end{equation}
\par
An identity that is useful at this point is given by
\begin{equation}
\frac{y_\alpha}{Z_{\alpha,\eq}} =
\frac{\tau_\star}{\tau_\eq} +
\tau_\star v_{r,g} \partderiv{\ln Z_\alpha}{R},
\end{equation}
which follows from Equations \ref{eq:taueq}, \ref{eq:ifr-per-sfr}, and \ref{eq:zoeq-sfrifrmixing}.
Substituting this expression into Equation \ref{eq:tilt-rate-alt} yields
\begin{equation}
\dot \nabla \ln Z_\alpha = \frac{1}{\tau_\nabla} \left(
\nabla_\eq \ln Z_\alpha - \nabla \ln Z_{\alpha,\eq}
\right),
\label{eq:tilt-rate-equilibration}
\end{equation}
where
\begin{equation}
\begin{split}
\tau_\nabla &\equiv \left(
1 - e^{-t / \tau_\eq}\right) \left(
\frac{1}{\tau_\eq} +
v_{r,g} \partderiv{\ln Z_{\alpha,\eq}}{R}
\right)^{-1}
\\
&\approx \left(
\frac{1}{\tau_\eq} +
v_{r,g} \partderiv{\ln Z_{\alpha,\eq}}{R}
\right)^{-1}.
\label{eq:tau-nabla-def}
\end{split}
\end{equation}
For our purposes, it is accurate enough to simply neglect the factor of $1 - e^{-t / \tau_\eq}$.
Equation \ref{eq:tilt-rate-equilibration} describes an equilibration process, wherein the rate of evolution is specified by a displacement from the equilibrium state and a characteristic timescale, $\tau_\nabla$.
If the radial metallicity gradient is negative, then the gradient equilibrates faster than the local ISM in the presence of inward gas flows (i.e., $\tau_\nabla < \tau_\eq$).
This outcome is a consequence of coupling the enrichment in neighboring annuli; equilibration in the local abundance gradient is driven by evolution across a narrow range of radii as opposed to local enrichment only.
\par
As discussed in Section \ref{sec:discussion:full-solutions}, the present-day ISM metallicity gradient can be approximated as lagging behind the equilibrium gradient.
We approximate ISM gradients as lagging behind the equilibrium gradient by an interval of $\tau_\nabla$ according to
\begin{equation}
\begin{split}
\nabla_\text{ISM}[\alpha/\text{H}] &\approx
\nabla_\eq[\alpha/\text{H}] (\tau_\text{disk} - \tau_\nabla)
\\
&= \frac{1}{\ln 10} \left[
\left(1 - e^{-(\tau_\text{disk} - \tau_\nabla) / \tau_\eq}\right)
\partderiv{\ln Z_{\alpha,\eq}}{R} -
e^{-(\tau_\text{disk} - \tau_\nabla) / \tau_\eq}
\partderiv{\ln \tau_\star}{R}
\right].
\end{split}
\end{equation}
We use this expression to approximate the radial metallicity gradients that arise in our numerical models in Appendix \ref{sec:analytic:scenarios} below.

\subsection{Radial Gradient Solutions for Specific Scenarios}
\label{sec:analytic:scenarios}

This appendix provides analytic solutions to the equilibrium and ISM metallicity gradients at the present day for each of our radial flow scenarios.
We discuss how these solutions compare to our numerical models in Section \ref{sec:discussion:full-solutions}.
Table \ref{tab:analytic-summary-scenarios} provides a summary of our mathematical expressions, while Table \ref{tab:gradient-values} provides the numerical solutions for specific parameter choices.
For each scenario, we evaluate $\mu_g$ and $\nabla \mu_g$ based on the appropriate prescription for $v_{r,g}$ and $\nabla v_{r,g}$.
We then evaluate $\xi_1$ and $\xi_2$ from Equations \ref{eq:xi1} and \ref{eq:xi2}, respectively.
The equilibrium gradient, $\nabla \ln Z_{\alpha,\eq}$, then follows from the quadratic solution in Equation \ref{eq:gradeq-quadratic}.
We then compute the equilibrium gradient under its alternate meaning, $\nabla_\eq \ln Z_\alpha$, which folds in the disequilibrium in the overall metal abundance according to Equation \ref{eq:gradeq} (equivalent to Equation \ref{eq:gradeq-alpha-alt}).
Finally, we apply a time offset, approximating the present-day ISM gradient as the equilibrium gradient at a lookback time of $\tau_\nabla$ (see discussion in Section \ref{sec:discussion:full-solutions} and in Appendix \ref{sec:analytic:eq-gradient:eq-approach} above).
\par
We follow the mass loading factor $\eta$ describing the rate of ejection through our derivations, even though we focus on the $\eta \rightarrow 0$ limit in our numerical models.
The expressions we provide here reduce to the corresponding forms in \citetalias{Johnson2025}, wherein we explored the $\eta > 0$ regime in the equilibrium context.

\subsubsection{Inside-out Growth and Star Formation Efficiency}
\label{sec:analytic:scenarios:io-sfe}

This appendix computes the radial metallicity gradient that arises due to inside-out disk growth and variations in SFE.
With no ejection or radial gas flows, $\mu, \eta \rightarrow 0$ by construction.
In this limit, our procedure for computing $\nabla \ln Z_{\alpha,\eq}$ described above is overkill; one can simply differentiate Equation \ref{eq:zoeq} for $Z_{\alpha,\eq}$ with radius directly:
\begin{equation}
\partderiv{\ln Z_{\alpha,\eq}}{R} =
\frac{
    \tau_\star / N \tau_\sfh
}{
    1 - r - \tau_\star / N \tau_\sfh
}\left(
\partderiv{\ln \tau_\star}{R} -
\partderiv{\ln \tau_\sfh}{R}
\right).
\end{equation}
In our models, SFE follows the observed Kennicutt-Schmidt relation, with $\dot\Sigma_\star \propto \Sigma_g^N$ (see discussion in Section \ref{sec:gce}).
By definition, it follows that
\begin{equation}
\begin{split}
\partderiv{\ln \tau_\star}{R} &=
(1 - N) \partderiv{\ln \Sigma_g}{R}
\\
&= \frac{N - 1}{r_g},
\label{eq:grad-taustar}
\end{split}
\end{equation}
where the second line is appropriate for an exponential radial profile in $\Sigma_g$ with a scale length of $r_g$, as we have used throughout this paper (see Figure \ref{fig:evol} and discussion in Section \ref{sec:gce}).
Our models also use an exponential scaling of $\tau_\sfh$ with radius (see Equation \ref{eq:tausfh}); therefore,
\begin{equation}
\partderiv{\ln \tau_\sfh}{R} = \frac{1}{r_\sfh},
\label{eq:grad-tausfh}
\end{equation}
where we have used $r_g = 4.7$ kpc throughout this paper (see discussion in Section \ref{sec:gce}).
\par
For our analytic approximations, we use $r_g = 3.75$ kpc \citep{Kalberla2009}.
With $N = 1.5$, we compute an equilibrium gradient of $\nabla_\text{eq}[\alpha/\text{H}] = -0.0288$ kpc$^{-1}$ and an ISM gradient of $\nabla_\text{ISM}[\alpha/\text{H}] = -0.0475$ kpc$^{-1}$.
If we switch to $N = 1$ and use $\tau_\star = 2$ Gyr everywhere, we instead compute $\nabla_\text{eq}[\alpha/\text{H}] = -0.0302$ kpc$^{-1}$ and $\nabla_\text{ISM}[\alpha/\text{H}] = -0.0273$ kpc$^{-1}$.
Each of these estimates are in excellent agreement with our numerical models (see Table \ref{tab:gradient-values}).

\input{analytic-summary-scenarios.tex}
\input{slopes.tablebody.tex}

\subsubsection{Constant Velocity}
\label{sec:analytic:scenarios:constant}

This appendix computes the radial metallicity gradients that arise in our constant velocity scenario for radial flows (see Table \ref{tab:flow-scenarios} and discussion in Section \ref{sec:gce:scenarios:constant}).
The velocity derivative, $\nabla v_{r,g}$ vanishes by construction.
Equation \ref{eq:mu-g-soln} for $\mu_g$ then reduces to
\begin{equation}
\mu_g = -\tau_\star v_{r,g} \left(
\frac{1}{R} + \partderiv{\ln \Sigma_g}{R}
\right).
\end{equation}
Differentiating the above expression with radius yields:
\begin{equation}
\begin{split}
\partderiv{\mu_g}{R} &=
-\tau_\star v_{r,g} \left(
\frac{1}{R} + \partderiv{\ln \Sigma_g}{R}
\right) \partderiv{\ln \tau_\star}{R} -
\tau_\star v_{r,g} \left(
\frac{-1}{R^2} +
\partderiv{^2 \ln \Sigma_g}{R^2}
\right)
\\
&\approx \tau_\star v_{r,g} \left[
\frac{1}{R^2} -
\left(
\frac{1}{R} + \partderiv{\ln \Sigma_g}{R}
\right) \partderiv{\ln \tau_\star}{R}
\right].
\end{split}
\end{equation}
If the surface density of the ISM does not deviate significantly from an exponential radial profile, then the second-derivative of $\ln \Sigma_g$ with Galactocentric radius vanishes, resulting in the second line of this expression.
Plugging these expressions for $\mu_g$ and $\nabla \mu_g$ into Equations \ref{eq:xi1} and \ref{eq:xi2} for $\xi_1$ and $\xi_2$ yields:
\begin{subequations}
\begin{align}
\xi_1 &= 1 + \eta - r - \frac{\tau_\star}{N \tau_\sfh} +
\tau_\star v_{r,g} \left(
\frac{1}{R} +
\partderiv{\ln \Sigma_g}{R} +
\partderiv{\ln \tau_\star}{R}
\right)
\\
\xi_2 &= 4 \tau_\star v_{r,g} \left[
\tau_\star v_{r,g} \left(
\frac{1}{R^2} - \left(
\frac{1}{R} + \partderiv{\ln \Sigma_g}{R}
\right)
\partderiv{\ln \tau_\star}{R} \right) -
\partderiv{\eta}{R} +
\frac{\tau_\star}{N \tau_\sfh} \left(
\partderiv{\ln \tau_\star}{R} -
\partderiv{\ln \tau_\sfh}{R}
\right)
\right].
\end{align}
\end{subequations}
We then arrive at the forms given in Table \ref{tab:analytic-summary-scenarios} by substituting in Equations \ref{eq:grad-taustar} and \ref{eq:grad-tausfh} for $\nabla \ln \tau_\star$ and $\nabla \ln \tau_\sfh$, respectively (see discussion in Appendix \ref{sec:analytic:scenarios:io-sfe}).
\par
From these expressions, we compute $\nabla_\eq[\alpha/\text{H}] = -0.0558$ kpc$^{-1}$, $-0.0649$ kpc$^{-1}$, and $-0.0690$ kpc$^{-1}$ and $\nabla_\text{ISM}[\alpha/\text{H}] = -0.0561$ kpc$^{-1}$, $-0.0646$ kpc$^{-1}$, and $-0.0689$ kpc$^{-1}$ with velocities of $v_{r,g} = -0.5$, $-1$, and $-1.5$ km/s, respectively.
These results are in good agreement with our numerical models (see Table \ref{tab:gradient-values} and Figure \ref{fig:gradients-vs-vice}).

\subsubsection{Global Torque}
\label{sec:analytic:scenarios:gt}

This appendix computes the radial metallicity gradients that arise in our GT scenario for radial flows (see Table \ref{tab:flow-scenarios} and discussion in Section \ref{sec:gce:scenarios:gt}).
The velocity profile is linear with radius, so
\begin{equation}
\partderiv{\ln v_{r,g}}{R} = \frac{1}{R}.
\end{equation}
Equation \ref{eq:mu-g-soln} for $\mu_g$ then reduces to
\begin{equation}
\mu_g = -\tau_\star v_{r,g} \left(
\frac{2}{R} + \partderiv{\ln \Sigma_g}{R}
\right).
\end{equation}
Differentiating with radius yields
\begin{equation}
\begin{split}
\partderiv{\mu_g}{R} &= -\tau_\star v_{r,g}
\left(
\frac{2}{R} + \partderiv{\ln \Sigma_g}{R}
\right)
\left(
\partderiv{\ln \tau_\star}{R} +
\partderiv{\ln v_{r,g}}{R}
\right) -
\tau_\star v_{r,g} \left(
\frac{-2}{R^2} +
\partderiv{^2 \ln \Sigma_g}{R^2}
\right)
\\
&\approx \tau_\star v_{r,g} \left[
\frac{2}{R^2} -
\left(
\frac{2}{R} + \partderiv{\ln \Sigma_g}{R}
\right)
\left(
\frac{1}{R} + \partderiv{\ln \tau_\star}{R}
\right)
\right].
\end{split}
\end{equation}
The second derivative of $\ln \Sigma_g$ with radius vanishes (see discussion in Appendix \ref{sec:analytic:scenarios:constant}), resulting in the second line above.
Plugging these expressions for $\mu_g$ and $\nabla \mu_g$ into Equations \ref{eq:xi1} and \ref{eq:xi2} for $\xi_1$ and $\xi_2$ yields
\begin{subequations}
\begin{align}
\xi_1 &= 1 + \eta - r -
\frac{\tau_\star}{N \tau_\sfh} +
\tau_\star v_{r,g} \left[
\frac{3}{R} +
\partderiv{\ln \Sigma_g}{R} +
\partderiv{\ln \tau_\star}{R}
\right]
\\
\xi_2 &= 4 \tau_\star v_{r,g} \left[
\tau_\star v_{r,g} \left[
\frac{2}{R^2} -
\left(
\frac{2}{R} + \partderiv{\ln \Sigma_g}{R}
\right)
\left(
\frac{1}{R} + \partderiv{\ln \tau_\star}{R}
\right)
\right] -
\partderiv{\eta}{R} +
\frac{\tau_\star}{N \tau_\sfh} \left(
\partderiv{\ln \tau_\star}{R} -
\partderiv{\ln \tau_\sfh}{R}
\right)
\right]
\label{eq:xi2-linear-v}%
\end{align}
\end{subequations}
The forms of $\xi_1$ and $\xi_2$ listed in Table \ref{tab:analytic-summary-scenarios} follow after plugging in Equation \ref{eq:gt-scenario} for the flow velocity in the GT scenario, Equation \ref{eq:grad-taustar} for $\nabla \ln \tau_\star$, and Equation \ref{eq:grad-tausfh} for $\nabla \ln \tau_\sfh$.
\par
In our numerical models, we use $\dot L / L = -0.02$, $-0.05$, and $-0.08$ Gyr$^{-1}$, from which we compute $\nabla_\eq[\alpha/\text{H}] = -0.0545$ kpc$^{-1}$, $-0.0833$ kpc$^{-1}$, and $-0.0976$ kpc$^{-1}$ and $\nabla_\text{ISM}[\alpha/\text{H}] = -0.0561$ kpc$^{-1}$, $-0.0771$ kpc$^{-1}$, and $-0.0915$ kpc$^{-1}$, respectively.
These results are in reasonable agreement with our numerical models (see Table \ref{tab:gradient-values} and Figure \ref{fig:gradients-vs-vice}), but agreement is slightly better in the equilibrium gradients than the ISM gradients.

\subsubsection{Potential Well Deepening}
\label{sec:analytic:scenarios:pwd}

This appendix computes the radial metallicity gradients that arise in our PWD scenario for radial flows (see Table \ref{tab:flow-scenarios} and discussion in Section \ref{sec:gce:scenarios:pwd}).
The velocity profile is linear with Galactocentric radius in the PWD scenario, which is the same form as the GT scenario.
The solution is therefore the same as in Appendix \ref{sec:analytic:scenarios:gt} above.
The difference between the two scenarios does not arise until we plug in Equation \ref{eq:pwd-scenario} for $v_{r,g}$ into Equation \ref{eq:xi2-linear-v}, which results in the form listed in Table \ref{tab:analytic-summary-scenarios}.
\par
Our numerical models predict $\partial \ln M_\star / \partial t = 0.048$ Gyr$^{-1}$ at the present day.
With $\gamma = 0.1$, $0.2$, and $0.3$, we compute $\nabla_\eq[\alpha/\text{H}] = -0.0344$ kpc$^{-1}$, $-0.0406$ kpc$^{-1}$, and $-0.0470$ kpc$^{-1}$ and $\nabla_\text{ISM}[\alpha/\text{H}] = -0.0488$ kpc$^{-1}$, $-0.0505$ kpc$^{-1}$, and $-0.0527$ kpc$^{-1}$, respectively.
These estimates are in good agreement with our numerical models (see Table \ref{tab:gradient-values} and Figure \ref{fig:gradients-vs-vice}).
This prediction of $\partial \ln M_\star / \partial t$ is a factor of $\sim$$2$ higher than current estimates of the mass and SFR of the Galaxy \citep[e.g.,][]{Chomiuk2011, Licquia2015, Elia2022}.
This offset negligibly affects the analytic predictions, resulting in differences of only $\lesssim 0.005$ dex/kpc.


\subsubsection{Angular Momentum Dilution}
\label{sec:analytic:scenarios:amd}

This appendix computes the radial metallicity gradients that arise in our AMD scenario for radial flows (see Table \ref{tab:flow-scenarios} and discussion in Section \ref{sec:gce:scenarios:amd}).
The AMD scenario is the most mathematically complex of all our radial flow models.
Consequently, the same steps for deriving $\xi_1$ and $\xi_2$ outlined in previous sections of this Appendix generally lead to complex expressions for the AMD scenario.
Here, we present the derivation that we have identified with the fewest number of intermediate steps.
\par
The most direct solution to $\xi_1$ arises when starting from the linear ordinary differential equation for the radial velocity profile given by Equation \ref{eq:dvdr-amd-with-eta}.
Dividing through by $v_{r,g}$ and rearranging terms leads to a straightforward expression for $\nabla \ln v_{r,g}$:
\begin{equation}
\partderiv{\ln v_{r,g}}{R} = \frac{1}{v_{r,g}} \left[
\frac{\eta}{\tau_\star} \left(
\frac{
    \beta_{\phi,\text{wind}} - \beta_{\phi,\text{acc}}
}{
    \beta_{\phi,\text{acc}} - 1
}
\right) -
\partderiv{\ln \Sigma_g}{t} -
\frac{1 - r}{\tau_\star}
\right] -
\frac{1}{R} \left(
1 + \frac{1}{ 1 - \beta_{\phi,\text{acc}}}\right) -
\partderiv{\ln \Sigma_g}{R}.
\label{eq:dlnvdr-amd-forxi1}
\end{equation}
This expression can be plugged directly into Equation \ref{eq:xi1} for $\xi_1$.
After combining terms and simplifying, we arrive at the following solution
\begin{equation}
\xi_1 = \eta \left( \frac{
    3 \beta_{\phi,\text{acc}} - 2 \beta_{\phi,\text{wind}} - 1
}{
    \beta_{\phi,\text{acc}} - 1
}\right) - 1 + r + \frac{\tau_\star}{N \tau_\sfh} +
\tau_\star v_{r,g} \left[
\partderiv{\ln \tau_\star}{R} -
\partderiv{\ln \Sigma_g}{R} -
\frac{1}{R} \left(1 + \frac{2}{1 - \beta_{\phi,\text{acc}}}
\right)
\right],
\end{equation}
where we have applied the substitution $\partial \ln \Sigma_g / \partial t \rightarrow -1 / N \tau_\sfh$, appropriate for evolution at late times.
The form of $\xi_1$ listed in Table \ref{tab:analytic-summary-scenarios} follows by plugging in Equations \ref{eq:grad-taustar} and \ref{eq:grad-tausfh} for $\nabla \ln \tau_\star$ and $\nabla \ln \tau_\sfh$, respectively.
\par
The most direct solution to $\xi_2$ arises when starting from the following identity
\begin{equation}
1 + \eta\left(\frac{
    \beta_{\phi,\text{wind}} - \beta_{\phi,\text{acc}}
}{
    \beta_{\phi,\text{acc}} - 1
}\right) - \mu_g - r - \frac{\tau_\star}{N\tau_\sfh} =
\frac{
    -\tau_\star v_{r,g}
}{
    R \left(1 - \beta_{\phi,\text{acc}}\right)
},
\end{equation}
which arises from plugging Equation \ref{eq:ifr-per-sfr} for $\dot\Sigma_\text{acc} / \dot\Sigma_\star$ into Equation \ref{eq:amd-scenario-extended} for $v_{r,g}$ in the AMD scenario in the $\eta > 0$ regime.
We neglect variations in $\beta_{\phi,\text{wind}}$ and $\beta_{\phi,\text{acc}}$ with radius, since we do not explore these prescriptions in this paper.
From here, Equation \ref{eq:dvdr-amd-with-eta} follows from plugging in Equation \ref{eq:mu-g-soln} for $\mu_g$ and isolating $v_{r,g}$ and $\nabla v_{r,g}$.
However, for these purposes, we can simply isolate $\mu_g$ and differentiate with radius:
\begin{equation}
\partderiv{\mu_g}{R} = \partderiv{\eta}{R} \left(
\frac{
    \beta_{\phi,\text{wind}} - \beta_{\phi,\text{acc}}
}{
    \beta_{\phi,\text{acc}} - 1
}\right) -
\frac{\tau_\star}{N \tau_\sfh} \left(
\partderiv{\ln \tau_\star}{R} -
\partderiv{\ln \tau_\sfh}{R}
\right) + \frac{
    \tau_\star v_{r,g}
}{
    R \left(1 - \beta_{\phi,\text{acc}}\right)
}\left(
\partderiv{\ln \tau_\star}{R} +
\partderiv{\ln v_{r,g}}{R} -
\frac{1}{R}
\right).
\end{equation}
The solution to $\xi_2$ then arises from plugging this expression into Equation \ref{eq:xi2} along with Equation \ref{eq:dlnvdr-amd-forxi1} for $\nabla \ln v_{r,g}$:
\begin{equation}
\begin{split}
\xi_2 &= \frac{
    -4 \tau_\star v_{r,g}
}{
    R \left(1 - \beta_{\phi,\text{acc}}\right)
} \Bigg[
1 + \eta \left(\frac{
    \beta_{\phi,\text{wind}} - \beta_{\phi,\text{acc}}
}{
    \beta_{\phi,\text{acc}} - 1
}\right) - r - \frac{\tau_\star}{N \tau_\sfh} -
\tau_\star v_{r,g} \left(
\partderiv{\ln \tau_\star}{R} -
\partderiv{\ln \Sigma_g}{R} -
\frac{1}{R \left(1 - \beta_{\phi,\text{acc}}\right)} -
\frac{2}{R}
\right)
\\
&\qquad -
\partderiv{\eta}{R} \left(
\beta_{\phi,\text{acc}} - \beta_{\phi,\text{wind}} - 1
\right)
\Bigg].
\end{split}
\end{equation}
Plugging in Equations \ref{eq:grad-taustar} and \ref{eq:grad-tausfh} for $\nabla \ln \tau_\star$ and $\nabla \ln \tau_\sfh$, respectively, leads to the form of $\xi_2$ listed in Table \ref{tab:analytic-summary-scenarios}.
\par
We have left $v_{r,g}$ as a free parameter in these expressions for $\xi_1$ and $\xi_2$, since the radial velocity profile does not have an analytic solution in the AMD scenario (see discussion in Appendix \ref{sec:analytic:velocities:amd}).
With $\beta_{\phi,\text{acc}} = 0.8$, $0.7$, and $0.6$, our numerical models predict $v_{r,g} = -0.17$, $-0.24$, and $-0.29$ km/s, respectively.
With these values, we compute $\nabla_\eq[\alpha/\text{H}] = -0.0558$ kpc$^{-1}$, $-0.0736$ kpc$^{-1}$, and $-0.0875$ kpc$^{-1}$ and $\nabla_\text{ISM}[\alpha/\text{H}] = -0.0567$ kpc$^{-1}$, $-0.0679$ kpc$^{-1}$, and $-0.0775$ kpc$^{-1}$, respectively.
These results are in good agreement with our numerical models (see Table \ref{tab:gradient-values} and Figure \ref{fig:gradients-vs-vice}).

\subsubsection{The Outer Rim Accretion Limit}
\label{sec:analytic:scenarios:ora-limit}

In the ORA limit, the equilibrium formalism breaks down, so these analytic arguments do not accurately reproduce the results of our numerical models.
The ORA limit imposes $\dot\Sigma_\text{acc} \rightarrow 0$ by construction, which reduces Equation \ref{eq:zoeq-sfrifrmixing} to
\begin{equation}
\begin{split}
Z_{\alpha,\eq} &\rightarrow \ddfrac{y_\alpha}{
    \tau_\star v_{r,g}
    \partderiv{\ln Z_{\alpha,\eq}}{R}
}
\\
\implies \partderiv{Z_{\alpha,\eq}}{R} &=
\frac{y_\alpha}{\tau_\star v_{r,g}}.
\\
\implies \partderiv{\ln Z_{\alpha,\eq}}{R} &=
-\partderiv{\ln \tau_\star}{R} -
\partderiv{\ln v_{r,g}}{R}.
\label{eq:dlnzeqdr-ora-limit}
\end{split}
\end{equation}
The equilibrium metallicity is not clearly defined, having vanished from the above expression, but its radial gradient somehow is clearly defined.
This mathematical outcome does not make physical sense but is in line with previous work.
Chemical equilibrium arises only in the presence of metal-poor accretion because, in its absence, all of the hydrogen is eventually fused into metals, so the equilibrium diverges ($Z_{\alpha,\eq} \rightarrow 1$; see discussion in, e.g., \citealt{Larson1972} and the review by \citealt{Tinsley1980}).
\par
Nonetheless, the bottom line of Equation \ref{eq:dlnzeqdr-ora-limit} does yield a straightforward solution to the radial gradient that one might expect from the ORA limit.
Like the AMD scenario, the ORA limit does not have an analytic solution to its velocity profile.
However, the expression reduces at large radii to indicate that the velocity profile follows the same scale radius as the ISM surface density (i.e., $\nabla \ln v_{r,g} \rightarrow -\nabla \ln \Sigma_g$; see discussion in Section \ref{sec:analytic:velocities:ora}).
The equilibrium gradient then follows trivially as
\begin{equation}
\partderiv{\ln Z_{\alpha,\eq}}{R} \approx
N \partderiv{\ln \Sigma_g}{R},
\end{equation}
which corresponds to a slope of $\nabla[\alpha/\text{H}]_\eq = -0.174$ kpc$^{-1}$.
This value is indeed considerably steep, in line with our numerical models.
However, our numerical models predict $\nabla_\eq[\alpha/\text{H}] = -0.120$ kpc$^{-1}$ and $\nabla_\text{ISM}[\alpha/\text{H}] = -0.130$ kpc$^{-1}$, which is the largest discrepancy between analytic and numerical gradients of all of our models.
\par
An alternate set of equations is required to describe the evolution of the radial metallicity gradient under the ORA limit.
It follows from Equations \ref{eq:ora-limit-mu}, \ref{eq:mu-soln}, and \ref{eq:zdot-o} that
\begin{equation}
\begin{split}
\dot Z_\alpha &= -\dot Z_{\alpha,\flow}
\\
&= v_{r,g} \partderiv{Z_\alpha}{R}.
\end{split}
\end{equation}
Enrichment in the ORA limit is therefore driven entirely by mixing effects.
This result makes intuitive sense, since the radial flow counteracts the effects of all other processes on the local ISM surface density in this scenario (see Equation \ref{eq:ora-limit-mu}).
From here, one could derive the rate of tilt in $Z_\alpha$ with radius, $\dot \nabla \ln Z_\alpha$, similar to Equation \ref{eq:tilt-rate}.
The equilibrium gradient would then follow by setting $\dot \nabla \ln Z_\alpha = 0$.
We do not pursue this solution any further than we already have, since the ORA limit is in considerable tension with observations anyway (see Figure \ref{fig:vgas-oh-comp} and discussion in Section \ref{sec:results}).

\subsection{Delayed Enrichment Sources}
\label{sec:analytic:dtds}

This appendix is concerned with the effects of delayed enrichment sources, such as SN Ia production of Fe.
Extending Equation \ref{eq:dot-sigma-o}, the rate of change in the surface density of some element $x$ in the ISM can be expressed as
\begin{equation}
\dot \Sigma_x = \sum_i \langle y_{x,i} \dot\Sigma_\star \rangle_i -
\frac{Z_x}{\tau_\star} \left(
1 + \eta - \mu_x - r +
\tau_\star \frac{\dot\Sigma_g}{\Sigma_g}
\right),
\end{equation}
where the term $\langle y_{x,i} \dot\Sigma_\star \rangle_i$ refers to production of $x$ through individual enrichment channels:
\begin{equation}
\langle y_{x,i} \dot\Sigma_\star \rangle_i = \ddfrac{
    y_{x,i} \int_0^t \dot\Sigma_\star (t') R_i (t - t') dt'
}{
    \int_0^\infty R_i (t') dt'
}.
\end{equation}
$y_{x,i}$ is the yield of $x$ through the $i$'th enrichment channel, and $R_i$ is the associated DTD.
The summation simply adds up all of the mass of $x$ produced through each enrichment channel.
This expression describes metal enrichment in a relatively general form, reducing to the prescription for O shown by Equation \ref{eq:dot-sigma-o} under the instantaneous recycling approximation.
The form for Fe includes one additional term describing SN Ia production (i.e., $R_i \rightarrow R_\text{Ia}$), for which we have used the popular $R_\text{Ia} \propto t^{-1}$ single power-law DTD in this paper (see discussion in Section \ref{sec:gce:yields}).
\par
The rate of change in the abundance by mass of $x$ follows by differentiating $Z_x \equiv \Sigma_x / \Sigma_g$ with time and applying the quotient rule for derivatives:
\begin{equation}
\dot Z_x = \frac{1}{\Sigma_g}
\sum_i \langle y_{x,i} \dot\Sigma_\star \rangle_i -
\frac{Z_x}{\tau_\star} \left(
1 + \eta - \mu_x - r +
\tau_\star \frac{\dot \Sigma_g}{\Sigma_g}
\right).
\end{equation}
As in Section \ref{sec:discussion:eq-gradients}, the equilibrium abundance follows from setting $\dot Z_x = 0$.
If the second source of metals is SNe Ia, then
\begin{equation}
Z_{x,\eq} = \frac{
    \ycc{x} + \yia{x} \psi_\text{Ia}
}{
    1 + \eta - \mu_x - r - \tau_\star / N \tau_\sfh
},
\label{eq:zxeq-delayed}
\end{equation}
where the factor $\psi_\text{Ia}$ is a convolution of the SFH and the DTD:
\begin{equation}
\psi_\text{Ia} \equiv \ddfrac{
    \int_0^t \dot\Sigma_\star (t') R_\text{Ia}(t - t') dt'
}{
    \dot\Sigma_\star
    \int_0^\infty R_\text{Ia} (t') dt'
}.
\end{equation}
If other enrichment channels are relevant (e.g., neutron capture nucleosynthesis), then additional terms of the form $y_x^i \psi_i$ would also arise.
$\psi_\text{Ia}$ describes a relative excess or deficit in the SN Ia event rate based on the shape of the SFH.
This factor is zero at the onset of star formation ($\dot \Sigma_\star > 0$, $\dot N_\text{Ia} = 0$) but diverges in a fully quenched environment ($\dot \Sigma_\star = 0$, $\dot N_\text{Ia} > 0$).
Sharply declining SFHs raise the equilibrium abundance of Fe because delayed SNe Ia enrich a rapidly dwindling gas supply.
SFHs that are more extended in time, however, have a more steady supply of hydrogen and therefore reach lower [Fe/H] at low redshift.
\par
Inside-out galaxy growth \citep[e.g.,][]{Bird2013} is implemented in our models through variations in the shape of the SFH with radius (see Equation \ref{eq:insideout}).
Increases in $\tau_\text{sfh}$ with radius not only contribute to lowering $Z_{\alpha,\eq}$ (see the denominator of Equation \ref{eq:zoeq}), but for elements like Fe, $\psi_\text{Ia}$ lowers with radius as well.
The difference in equilibrium gradients follows straightforwardly from Equation \ref{eq:zxeq-delayed} as
\begin{equation}
\partderiv{\ln Z_{x,\eq}}{R} =
\partderiv{\ln Z_{\alpha,\eq}}{R} +
\left(\frac{\ycc{x}}{\yia{x}} + \psi_\text{Ia}\right)^{-1}
\partderiv{\psi_\text{Ia}}{R}.
\end{equation}
This expression indicates that metallicity gradients should be steeper for elements with significant contributions from SNe Ia, or any delayed enrichment source for that matter.
This prediction is consistent with our measurements in \citetalias{Johnson2025}
($\nabla[\text{O/H}] = -0.062$ kpc$^{-1}$; $\nabla[\text{Fe/H}] = -0.070$ kpc$^{-1}$).

\subsection{Metal-Rich Accretion}
\label{sec:analytic:zin}

\begin{figure}
\centering
\includegraphics[scale = 0.9]{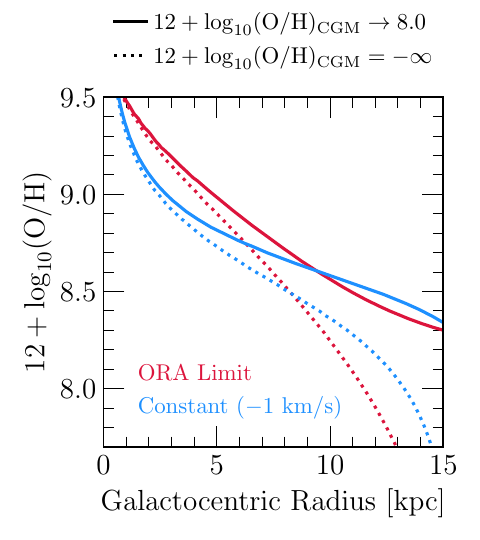}
\caption{
Variations of the ORA limit (red) and constant flow velocity (blue) scenarios with non-zero metallicity accretion (solid lines).
Dotted lines mark the predictions with zero metallicity accretion that appear in previous Figures.
\textbf{Summary}: Non-zero metallicity accretion increases metal abundances overall, but more so in the outer disk, resulting in a slight flattening of the profile.
The flattening is more significant for the ORA limit, which goes from $\grad{O} \approx -0.13$ kpc$^{-1}$ to $\grad{O} \approx -0.084$ kpc$^{-1}$.
}
\label{fig:zin}
\end{figure}

This appendix is concerned with the effects of pre-enriched accretion on the equilibrium metallicity and its radial gradient.
If accreting gas has significant metal content, then Equation \ref{eq:dot-sigma-o} for $\dot\Sigma_\alpha$ picks up an additional term:
\begin{equation}
\dot\Sigma_\alpha \rightarrow
y_\alpha \dot\Sigma_\star -
Z_\alpha \dot\Sigma_\star \left(
1 + \eta - \mu_\alpha - r\right) +
Z_{\alpha,\text{acc}} \dot\Sigma_\text{acc}.
\end{equation}
This transformation affects the rate of change in the metallicity by mass, $\dot Z_\alpha$, given by
\begin{equation}
\dot Z_\alpha \rightarrow
\frac{y_\alpha}{\tau_\star} -
\frac{Z_\alpha}{\tau_\star} \left(
1 + \eta - \mu_\alpha - r +
\tau_\star \frac{\dot \Sigma_g}{\Sigma_g}
\right) +
Z_{\alpha,\text{acc}}
\frac{\dot\Sigma_\text{acc}}{\Sigma_g}.
\end{equation}
As in Section \ref{sec:discussion:local-enrich}, the equilibrium abundance follows from setting $\dot Z_\alpha = 0$ and solving for $Z_\alpha$:
\begin{equation}
Z_{\alpha,\eq} \rightarrow \ddfrac{y_\alpha}{
    \ddfrac{\dot\Sigma_\text{acc}}{\dot\Sigma_\star} +
    \tau_\star v_{r,g} \nabla \ln Z_{\alpha,\eq}
} + \ddfrac{Z_{\alpha,\text{acc}}}{
    1 + \ddfrac{
        \tau_\star v_{r,g} \nabla \ln Z_{\alpha,\eq}
    }{
        \dot\Sigma_\text{acc} / \dot\Sigma_\star
    }
}
\end{equation}
In the limit that $\tau_\star v_{r,g} \nabla \ln Z_{\alpha,\eq} \ll \dot\Sigma_\text{acc} / \dot\Sigma_\star$, the metallicity of accretion is divided by a factor of nearly unity in the above expression.
For these models, pre-enriched accretion simply shifts the equilibrium metallicity upward.
Beyond the radius of star formation, $\dot\Sigma_\star \rightarrow 0$ by definition, and $Z_{\alpha,\eq} \rightarrow Z_{\alpha,\text{acc}}$ as expected.
In the ORA limit, $\dot \Sigma_\text{acc} \rightarrow 0$ by definition, so the role of metal-rich accretion vanishes, also as expected.
\par
Figure \ref{fig:zin} shows the results of a brief investigation of the effects of metal-rich accretion in our numerical GCE models.
We use the simple prescription for the growth of the CGM metallicity from \citetalias{Johnson2025}:
\begin{equation}
12 + \log_{10}(\text{O/H})_\text{CGM} =
8.0 \left(1 - e^{-t / 3 \text{ Gyr}}\right).
\end{equation}
In \citetalias{Johnson2025}, we describe how this prescription increases the metallicities of old stellar populations relative to models with metal-free accretion.
Here, we are interested in the effect on the radial metallicity profile at the present day.
Figure \ref{fig:zin} shows the impact of pre-enriched accretion in our constant velocity model with $v_{r,g} = -1$ km/s and in the ORA limit.
As expected, the overall metal abundance increases across the entire Galaxy when pre-enriched accretion is added to the model.
This effect is stronger in the outer Galaxy, where local abundances are closer to the metallicity of the accreting material, as large as $\sim$$0.3$ dex at $R \sim 12$ kpc ($\sim$$0.5$ dex in the ORA limit).
The result is a slight flattening in the metallicity gradient.
The impact of pre-enriched accretion is small in any region where the local abundance is significantly larger than the metallicity of the accreting material.
These effects therefore do not have any significant impact on the conclusions of this paper.

%% file: analytic-summary-scenarios.tex
\begin{table*}
\caption{
Analytic formula for computing radial metallicity gradients at low redshift based on our radial gas flow scenarios (see derivations in Appendix \ref{sec:analytic:scenarios}).
}
\begin{tabularx}{\textwidth}{c}
\hline
\hline
\parbox{\textwidth}{
\centering
\vspace{2mm}
\(\displaystyle
\begin{aligned}
\nabla_\eq[\alpha/\text{H}] &= \frac{1}{\ln 10}
\left[
\left(1 - e^{-t / \tau_\eq}\right)
\partderiv{\ln Z_{\alpha,\eq}}{R} -
e^{-t / \tau_\eq} \partderiv{\ln \tau_\star}{R}
\right]
\\
\partderiv{\ln Z_{\alpha,\eq}}{R} &=
\frac{
    -\xi_1 + \sqrt{\xi_1^2 + \xi_2}
}{
    2 \tau_\star v_{r,g}
}
\\
\xi_1 &\equiv 1 + \eta - r - \frac{\tau_\star}{N \tau_\sfh} +
\tau_\star v_{r,g} \left[
\frac{1}{R} + \partderiv{\ln \Sigma_g}{R} +
\partderiv{\ln \tau_\star}{R} +
2 \partderiv{\ln v_{r,g}}{R}
\right]
\\
\xi_2 &\equiv 4 \tau_\star v_{r,g} \left[
\partderiv{\mu_g}{R} -
\partderiv{\eta}{R} + \frac{\tau_\star}{N \tau_\sfh} \left(
\partderiv{\ln \tau_\star}{R} -
\partderiv{\ln \tau_\sfh}{R}\right)
\right]
\end{aligned}
\)
\vspace{2mm}
}
\\
\hline
\end{tabularx}
\begin{tabularx}{\textwidth}{c @{\extracolsep{\fill}} c c}
\multicolumn{3}{l}{
\textit{Inside-out Growth and Star Formation Efficiency}
}
\\
\parbox{0.4\textwidth}{
\centering
\vspace{2mm}
\(\displaystyle
\partderiv{\ln Z_{\alpha,\eq}}{R} = \frac{
    \tau_\star / N\tau_\text{sfh}
}{
    1 - r - \tau_\star / N\tau_\text{sfh}
} \left(
\frac{N - 1}{r_g} -
\frac{1}{r_\sfh}
\right)
\)
\vspace{2mm}
}
&
\parbox{0.25\textwidth}{
\centering
\vspace{2mm}
\(\displaystyle
\partderiv{\ln\tau_\star}{R} =
(1 - N) \partderiv{\ln \Sigma_g}{R} =
\frac{N - 1}{r_g}
\)
\vspace{2mm}
}
&
\parbox{0.25\textwidth}{
\centering
\vspace{2mm}
\(\displaystyle
\partderiv{\ln \tau_\sfh}{R} = \frac{1}{r_\sfh}
\)
\vspace{2mm}
}
\\
\hline
\end{tabularx}
\begin{tabularx}{\textwidth}{l}
\textit{Constant Velocity}
\\
\parbox{\textwidth}{
\centering
\vspace{2mm}
\(\displaystyle
\begin{aligned}
\xi_1 &= 1 + \eta - r -
\frac{\tau_\star}{N\tau_\text{sfh}} +
\tau_\star v_{r,g} \left(
\frac{1}{R} +
\frac{N - 2}{r_g}
\right)
\\
\xi_2 &= 4 \tau_\star v_{r,g} \left[
\tau_\star v_{r,g} \left(
\frac{1}{R^2} +
\frac{N - 1}{r_g^2} -
\frac{N - 1}{R r_g}
\right) -
\partderiv{\eta}{R} +
\frac{\tau_\star}{N\tau_\text{sfh}} \left(
\frac{N - 1}{r_g} - \frac{1}{r_\text{sfh}}
\right)
\right]
\end{aligned}
\)
\vspace{2mm}
}
\\
\hline
\textit{Global Torque}
\\
\parbox{\textwidth}{
\centering
\vspace{2mm}
\(\displaystyle
\begin{aligned}
\xi_1 &= 1 + \eta - r - \frac{\tau_\star}{N\tau_\text{sfh}} +
\tau_\star \frac{\dot L}{L} \left(
3 + (N - 2) \frac{R}{r_g}
\right)
\\
\xi_2 &= 4 \tau_\star \frac{\dot L}{L} \left[
\tau_\star \frac{\dot L}{L} \left(
2 - \left((N - 1)\frac{R}{r_g} - 1\right)
\left(2 - \frac{R}{r_g}\right)
\right) -
R \partderiv{\eta}{R} +
R \frac{\tau_\star}{N \tau_\sfh} \left(
\frac{N - 1}{r_g} - \frac{1}{r_\sfh}
\right)
\right]
\end{aligned}
\)
\vspace{2mm}
}
\\
\hline
\textit{Potential Well Deepening}
\\
\parbox{\textwidth}{
\centering
\vspace{2mm}
\(\displaystyle
\begin{aligned}
\xi_1 &= 1 + \eta - r - \frac{\tau_\star}{N\tau_\text{sfh}} +
\tau_\star \gamma \partderiv{\ln M_\star}{t} \left(
\frac{3}{R} + \frac{N - 2}{r_g}
\right)
\\
\xi_2 &= 4 \tau_\star
\gamma \partderiv{\ln M_\star}{t} \left[
\tau_\star \gamma \partderiv{\ln M_\star}{t} \left(
\frac{2}{R^2} - \left(
\frac{2}{R} - \frac{1}{r_g}
\right)\left(
\frac{1}{R} + \frac{N - 1}{r_g}
\right)\right) -
\partderiv{\eta}{R} +
\frac{\tau_\star}{N \tau_\sfh} \left(
\frac{N - 1}{r_g} - \frac{1}{r_\sfh}
\right)
\right]
\end{aligned}
\)
\vspace{2mm}
}
\\
\hline
\textit{Angular Momentum Dilution}
\\
\parbox{\textwidth}{
\centering
\vspace{2mm}
\(\displaystyle
\begin{aligned}
\xi_1 &= \tau_\star v_{r,g} \left[
\frac{N}{r_g} -
\frac{1}{R}\left(
1 + \frac{2}{1 - \beta_{\phi,\text{acc}}}
\right)
\right] -
\left(
1 - r - \frac{\tau_\star}{N\tau_\sfh}
\right)
+ \eta \left(
\frac{
    3\beta_{\phi,\text{acc}} -
    2\beta_{\phi,\text{wind}} - 1
}{
    \beta_{\phi,\text{acc}} - 1
}\right).
\\
\xi_2 &= \frac{
    -4 \tau_\star v_{r,g}
}{
    R \left(1 - \beta_{\phi,\text{acc}}\right)
} \left[
1 + \eta\left(
\frac{
    \beta_{\phi,\text{wind}} - \beta_{\phi,\text{acc}}
}{
    \beta_{\phi,\text{acc}} - 1
}
\right)
- r - \frac{\tau_\star}{N\tau_\sfh} -
\tau_\star v_{r,g} \left(
\frac{N}{r_g} -
\frac{1}{R\left(1 - \beta_{\phi,\text{acc}}\right)} -
\frac{2}{R}
\right)
- \partderiv{\eta}{R} \left(
\beta_{\phi,\text{acc}} -
\beta_{\phi,\text{wind}} - 1\right)
\right]
\end{aligned}
\)
\vspace{2mm}
}
\\
\hline
\hline
\end{tabularx}
\label{tab:analytic-summary-scenarios}
\end{table*}

%% file: slopes.tablebody.tex
\begin{table*}
\caption{
Comparing Equilibrium and ISM gradients averaged between $R = 5 - 10$ kpc predicted by \vice\ and estimated analytically.
\\
$^{*}$The equilibrium formalism breaks down in the ORA Limit (see discussion in Section \ref{sec:discussion:local-enrich} and Appendix \ref{sec:analytic:scenarios:ora-limit}).
}
\begin{tabularx}{\textwidth}{c @{\extracolsep{\fill}} c c c c}
\hline
Parameter Value & Analytic $\nabla_\text{eq}[\alpha/\text{H}]$ & \vice\ $\nabla_\text{eq} [\alpha/\text{H}]$ & Analytic $\nabla_\text{ISM}[\alpha/\text{H}]$ & \vice\ $\nabla_\text{ISM}[\alpha/\text{H}]$
\\
\hline
\hline
\multicolumn{3}{l}{
\textit{No Radial Flow}
}
\\
$N = 1$ & $-0.0302$ kpc$^{-1}$ & $-0.0273$ kpc$^{-1}$ & $-0.0273$ kpc$^{-1}$ & $-0.0275$ kpc$^{-1}$
\\
$N = 1.5$ & $-0.0288$ kpc$^{-1}$ & $-0.0391$ kpc$^{-1}$ & $-0.0475$ kpc$^{-1}$ & $-0.0419$ kpc$^{-1}$
\\
\hline
\multicolumn{3}{l}{
\textit{Constant Velocity}
}
\\
$v_{r,g} = -0.5$ km/s & $-0.0558$ kpc$^{-1}$ & $-0.0445$ 
kpc$^{-1}$ & $-0.0561$ kpc$^{-1}$ & $-0.0408$ kpc$^{-1}$
\\
$v_{r,g} = -1.0$ km/s & $-0.0649$ kpc$^{-1}$ & $-0.0634$ kpc$^{-1}$ & $-0.0646$ kpc$^{-1}$ & $-0.0595$ kpc$^{-1}$
\\
$v_{r,g} = -1.5$ km/s & $-0.0690$ kpc$^{-1}$ & $-0.0730$ kpc$^{-1}$ & $-0.0689$ kpc$^{-1}$ & $-0.0692$ kpc$^{-1}$
\\
\hline
\multicolumn{3}{l}{
\textit{Global Torque}
}
\\
$\dot L / L = -0.02$ Gyr$^{-1}$ & $-0.0545$ kpc$^{-1}$ & $-0.0407$ kpc$^{-1}$ & $-0.0561$ kpc$^{-1}$ & $-0.0404$ kpc$^{-1}$
\\
$\dot L / L = -0.05$ Gyr$^{-1}$ & $-0.0833$ kpc$^{-1}$ & $-0.0738$ kpc$^{-1}$ & $-0.0771$ kpc$^{-1}$ & $-0.0646$ kpc$^{-1}$
\\
$\dot L / L = -0.08$ Gyr$^{-1}$ & $-0.0976$ kpc$^{-1}$ & $-0.101$ kpc$^{-1}$ & $-0.0915$ kpc$^{-1}$ & $-0.0877$ kpc$^{-1}$
\\
\hline
\multicolumn{3}{l}{
\textit{Potential Well Deepening}
}
\\
$\gamma = 0.1$ & $-0.0344$ kpc$^{-1}$ & $-0.0425$ kpc$^{-1}$ & $-0.0488$ kpc$^{-1}$ & $-0.0480$ kpc$^{-1}$
\\
$\gamma = 0.2$ & $-0.0406$ kpc$^{-1}$ & $-0.0467$ kpc$^{-1}$ & $-0.0505$ kpc$^{-1}$ & $-0.0544$ kpc$^{-1}$
\\
$\gamma = 0.3$ & $-0.0470$ kpc$^{-1}$ & $-0.0504$ kpc$^{-1}$ & $-0.0527$ kpc$^{-1}$ & $-0.0597$ kpc$^{-1}$
\\
\hline
\multicolumn{3}{l}{
\textit{Angular Momentum Dilution}
}
\\
$\beta_{\phi,\text{in}} = 0.8$ & $-0.0558$ kpc$^{-1}$ & $-0.0625$ kpc$^{-1}$ & $-0.0567$ kpc$^{-1}$ & $-0.0647$ kpc$^{-1}$
\\
$\beta_{\phi,\text{in}} = 0.7$ & $-0.0736$ kpc$^{-1}$ & $-0.0734$ kpc$^{-1}$ & $-0.0679$ kpc$^{-1}$ & $-0.0746$ kpc$^{-1}$
\\
$\beta_{\phi,\text{in}} = 0.6$ & $-0.0875$ kpc$^{-1}$ & $-0.0774$ kpc$^{-1}$ & $-0.0775$ kpc$^{-1}$ & $-0.0824$ kpc$^{-1}$
\\
\hline
\multicolumn{3}{l}{
\textit{Outer Rim Accretion Limit}$^{*}$
}
\\
N/A & $-0.174$ kpc$^{-1}$ & $-0.120$ kpc$^{-1}$ & -- & $-0.130$ kpc$^{-1}$
\\
\hline
\hline
\end{tabularx}
\label{tab:gradient-values}
\end{table*}